\documentclass{aa}  

\usepackage{graphicx}
\usepackage[colorlinks=true,citecolor=blue]{hyperref}
\usepackage{txfonts}
\usepackage{newtxtext, newtxmath}
\usepackage[T1]{fontenc}
\usepackage{amsmath}
\usepackage{amssymb}
\usepackage{color}
\usepackage{xcolor}
\usepackage{hyperref}
\usepackage{pdflscape}
\usepackage{multirow}
\usepackage{threeparttable}
\usepackage{makecell}
\usepackage{bm}
\usepackage{arydshln}

\newcommand{\Msun}{M$_\odot$}

\newcommand{\Zsun}{Z$_\odot$}

\newcommand{\naid}{\ion{Na}{i}\,D\,}

\newcommand{\caii}{\ion{Ca}{ii}\,}
\newcommand{\ki}{\ion{K}{i}\,}

\newcommand{\mgi}{\ion{Mg}{i}\,}

\usepackage{hyperref}
\definecolor{yaleblue}{rgb}{0.1,0.3,0.9}
\definecolor{lava}{rgb}{0.81, 0.06, 0.13}
\definecolor{forestgreen}{rgb}{0.0, 0.45, 0.13}
\hypersetup{colorlinks=true, linkcolor=lava, urlcolor=forestgreen, citecolor=yaleblue}

\newcommand{\orcid}[1]{\href{https://orcid.org/#1}{\includegraphics[width=10pt]{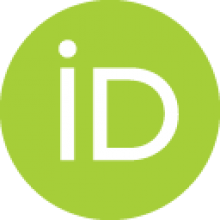}}}

\defcitealias{GG24}{Paper I}
\defcitealias{Gutierrez24}{Paper III}
\begin{document}

   \title{Narrow absorption lines from intervening material in supernovae}

   \subtitle{II. Galaxy properties}

   \author{Santiago Gonz\'alez-Gait\'an
          \inst{1,2,3}\orcid{0000-0001-9541-0317}
          \and
          Claudia P. Guti\'errez
          \inst{4,5}\orcid{0000-0003-2375-2064}
          \and
          Gon\c{c}alo Martins
          \inst{2}
          \and
          Tom\'as E. M\"uller-Bravo
          \inst{6}\orcid{0000-0003-3939-7167}
          \and
          Jo\~ao Duarte
          \inst{2}         
          \and
          Joseph P. Anderson
          \inst{3,7}\orcid{0000-0003-0227-3451}
          \and
          Lluis Galbany
          \inst{5,4}\orcid{0000-0002-1296-6887}
          \and 
          Mark Sullivan
          \inst{8}\orcid{0000-0001-9053-4820}
          \and{Jo\~ao Rino-Silvestre}
          \inst{2}
           \and
          Mariona Caixach
          \inst{9}\orcid{0000-0001-6332-7527}
          \and
          Antonia Morales-Garoffolo
          \inst{9}
          \and
          Sabyasachi Goswami
          \inst{10}
          \and
          Ana M. Mour\~ao
          \inst{2}\orcid{0000-0002-0855-1849}
          \and 
          Seppo Mattila
          \inst{11,12}\orcid{0000-0001-7497-2994}
          }

    \institute{
    Instituto de Astrof\'isica e Ci\^encias do Espaço, Faculdade de Ci\^encias, Universidade de Lisboa, Ed. C8, Campo Grande, 1749-016 Lisbon, Portugal\\
    \email{gongsale@gmail.com}    
    \and
    CENTRA, Instituto Superior T\'ecnico, Universidade de   Lisboa, Av. Rovisco Pais 1, 1049-001 Lisboa, Portugal
    \and
        European Southern Observatory, Alonso de C\'ordova 3107, Casilla 19, Santiago, Chile
    \and
    Institut d'Estudis Espacials de Catalunya (IEEC), Edifici RDIT, Campus UPC, 08860 Castelldefels (Barcelona), Spain\\
    \email{cgutierrez@ice.csic.es}
    \and
    Institute of Space Sciences (ICE, CSIC), Campus UAB, Carrer de Can Magrans, s/n, E-08193 Barcelona, Spain
    \and
    School of Physics, Trinity College Dublin, The University of Dublin, Dublin 2, Ireland
    \and
    Millennium Institute of Astrophysics MAS, Nuncio Monsenor Sotero Sanz 100, Off. 104, Providencia, Santiago, Chile
    \and
    School of Physics and Astronomy, University of Southampton, Southampton, SO17 1BJ, UK
    \and
    Department of Applied Physics, School of Engineering, University of C\'adiz, Campus of Puerto Real, E-11519 C\'adiz, Spain
    \and
     Instituto de Astrof\'isica de Andaluc\'ia – CSIC, Apdo 3004, 18080 Granada, Spain
    \and
    Tuorla Observatory, Department of Physics and Astronomy, University of Turku, FI-20014 Turku, Finland
    \and
    European University Cyprus, Diogenes Street, Engomi, 1516 Nicosia, Cyprus
    }
   \date{}

 
\abstract
{The interstellar medium (ISM) has a number of tracers such as the \naid $\lambda\lambda$ 5890, 5896 absorption lines that are evident in the spectra of galaxies but also in those of individual astrophysical sources such as stars, novae or quasars. Here, we investigate narrow absorption features in the spectra of nearby supernovae (SNe) and compare them to local ($<0.5$ kpc) and global host galaxy properties. With a large and heterogeneous sample of spectra, we are able to recover the known relations of ISM with galaxy properties: larger columns of ISM gas are found in environments that are more massive, more actively star-forming, younger and viewed from a more inclined angle. Most trends are stronger for local than global properties, and we find that the ISM column density decreases exponentially with the offset from the host galaxy centre, as expected for a gas distribution following an exponential radial profile. We also confirm trends for the velocity of galactic outflows increasing with radius. The current study demonstrates the capability of individual light sources to serve as ubiquitous tracers of ISM properties across various environments and galaxies.}

\keywords{supernovae: general, ISM: lines and bands, dust}
\authorrunning{Gonz\'alez-Gait\'an, Guti\'errez et al.}
\titlerunning{SN narrow lines and galaxy properties}
\maketitle
%

\section{Introduction}

The interstellar medium (ISM) of galaxies plays a crucial role in providing the constituents for forming new stars, the reservoir for dying stars and other galactic processes such as feedback and fuelling of active galactic nuclei. The ISM thus regulates the galactic chemical enrichment and the star formation history of galaxies. The cold ISM responsible for star formation represents approximately 10-20\% of the total baryon content (gas and stars) of a late-type galaxy like the Milky Way \citep{Draine11}. Most of this cold ISM ($\sim75$\%) consists of atomic gas, essentially hydrogen, with the rest in molecular form. Dust, responsible for the absorption and scattering of light bluewards of the optical regime and re-emission redwards of the infrared, constitutes 1\% of the gas mass but is an excellent tracer of the atomic and molecular ISM. For early-type galaxies, the ISM content is rather low at $\sim$3\% of the total baryonic mass \citep{Saintonge22}.

Heavier elements are also mixed together with the atomic hydrogen and absorb the light of background sources at characteristic wavelengths. Therefore, it is possible to study the ISM gas content and its kinematics with absorption-line spectroscopy of atomic ions. The sodium doublet lines (\naid $\lambda\lambda$5890, 5896) are among the strongest absorption lines in galaxy spectra. Other atomic lines include ionized calcium (\caii H \& K $\lambda\lambda$ 3970, 3935) and neutral potassium (\ki $\lambda\lambda$ 7665, 7669). The strength of these lines, particularly sodium, strongly correlates with gas phase and dust attenuation, together with galaxy inclination, stellar mass and star formation rate \citep{Chen10}. The kinematics of the lines, on the other hand, can reveal important information on galactic outflows and their relation to galactic properties \citep{Rupke05b, Chen10, Park15, Cazzoli22}.

Studies of ISM absorption lines have traditionally relied on galaxy spectra with the galaxy continuum as the background source of light. This has the complication of having a contribution of stellar atmospheres in the absorption lines. Such contamination can even reach $\sim$80\% in cool stars \citep{Chen10}. There are mechanisms to correct for the stellar component, e.g. through the correlation of stellar sodium with magnesium lines (\mgi triplet $\lambda\lambda$ 5167, 5173, 5184) only present in stars \citep{Martin05, Rupke05}; or through stellar population fits to the continuum to obtain the stellar part \citep{Chen10}. However, both methods have drawbacks and large uncertainty.  

Absorption lines of the atomic ISM are also present in the spectra of other background sources such as resolved single stars in molecular clouds \citep[e.g.][]{Pascucci15}, novae \citep[e.g.][]{Jack19} or quasars \citep[e.g.][]{Boisse15, Kacprzak15}.  Supernovae (SNe), bright stellar explosions, are particularly suitable lighthouses, as they occur in various environments and are so bright that they may outshine their neighbourhood or even their entire host galaxy. Their spectra are known to present absorption lines from the ISM, notably sodium but also potassium, calcium and diffuse interstellar bands \citep[DIBs; e.g.][]{Sollerman05, Gutierrez16}, which have been used as tracers of gas and as proxies of dust extinction \citep[e.g.][]{Welty14, Ritchey15}, even in extreme environments \citep[e.g. tidal tails;][]{Ferretti17}. Therefore, absorption lines in SN spectra can serve as probes of the local ISM content at individual sight lines with the advantage that there is a negligible contribution from stellar populations in their spectra (although see \citealt{Kangas16}). Furthermore, the ISM lines are narrow and easily distinguishable from the intrinsic SN ejecta component, which has a characteristic broad P-Cygni profile. Nevertheless, care should be taken when using low-resolution spectroscopy as this broad profile may affect the measurement of the ISM lines \citep[][hereafter Paper I]{GG24}. 

A disadvantage of using SNe to study the ISM is the influence on the lines of very nearby circumstellar material (CSM) ejected by the progenitor system before the explosion, which is not representative of the ISM. Indeed, CSM signatures are clearly seen in the strength and evolution of Balmer emission lines in interacting SNe, particularly H$\alpha$ \citep[e.g.][]{Chugai94, Chevalier94, Fransson02, Dessart23a}, and also in the absorption line strengths of gas atoms such as sodium in intermediate-luminosity red transients \citep{Byrne23}. However, for most SNe these lines are essentially constant throughout their evolution \citepalias{GG24}. There is a handful of well-studied exceptions that show variation \citep{Patat07, Blondin09, Simon09, Graham15, Ferretti16}, and there is a fraction of SNe with excess sodium absorption strength that is also blue-shifted \citep{Sternberg11, Phillips13, Maguire13, Hachinger17}. This may indicate outflows from the progenitor system or nearby ISM clouds accelerated by the SN radiation \citep{Hoang19, Bulla18}. 

In \citetalias{GG24}, we developed a robust automated technique to measure the narrow lines of ISM tracers from SN spectra of various resolutions, showing the limitations of low-resolution spectra and how to circumvent these constraints. We also applied the methodology to a large sample of SNe to investigate the evolution of the lines throughout the SN lifetime, finding no statistically significant change in those lines. In this paper, we use the same large heterogeneous sample presented in \citetalias{GG24} of nearby SNe occurring at multiple locations and in different galaxy types to study the spectroscopic absorption lines of ISM tracers as a function of various galaxy properties. To our knowledge, this is the first time that individual point sources are used as possible generic large-scale tracers of resolved ISM properties within galaxies and across multiple galaxies of different types and characteristics. We aim to gauge if SNe can truly be used as universal tracers of ISM properties.
The paper is organised as follows. Section~\ref{sec:meth} describes our methodology.  The analysis and results are presented in Section~\ref{sec:res} and discussed in Section~\ref{sec:discussion}, while our conclusions are in Section~\ref{sec:conc}.

In a companion paper \citep[Paper III;][]{Gutierrez24}, we put constraints on the nature of the progenitor systems of the different SN types by using the known galaxy relations on the ISM gas composition, distribution and kinematics.

\section{Methodology}
\label{sec:meth}

This section outlines the methodology and analysis and introduces the data used for this study. However, a detailed description of the data and SN measurements can be found in \citetalias{GG24}.

\subsection{SN EW measurements}
\label{sec:EWmeas}

The public SN sample used in the present study consists of a large heterogeneous nearby spectroscopic dataset obtained by multiple surveys. It initially consists of thousands of spectra of nearly 1300 SNe of different types. Table~\ref{table:sne} summarises the number of SNe used. For a full list of the sample, see Table A.1 of \citetalias{GG24}.

\begin{table}
\centering
\caption{Number of SNe used in this study}
\label{table:sne}
\renewcommand{\arraystretch}{1.4}
\begin{tabular}{c|ccc}

Type & N$_{\mathrm{total}}$ & N$_{\mathrm{global}}$ &   N$_{\mathrm{local}}$ \\
\hline
\hline
SNe~Ia & 701 & 357 & 167\\
SNe~II & 214 & 113 & 84\\
SE-SNe & 233 & 107 & 83 \\
SNe-int & 98 & 36 & 26 \\
\hline
Total & 1246 & 613 & 357 \\
\end{tabular}
\tablefoot{Number of SNe with at least one spectrum passing the continuum and S/N cuts of Paper I for the \naid\ line. Left, centre, and right columns indicate the total number, the number of SNe with global galaxy parameters, and the number of SNe with local galaxy parameters (see \S~\ref{sec:galmeas}). Note that numbers may change for each spectral line.}
\end{table}

To characterise the narrow interstellar absorption lines observed in SNe spectra, we measure their equivalent widths (EWs) and velocities (VELs). 
The algorithm to measure the EW first finds the continuum by smoothing the flux with a cosine kernel; then, the EW is obtained by calculating the area under/over the continuum within a given window. The velocity is determined from the weighted average wavelength shift from the rest frame according to the area under/over the continuum. In the case of multiple spectra for a given SN, a stacked flux-to-continuum spectrum is obtained from which the EW and VEL are measured, and a bootstrap of 100 realisations gives the statistical uncertainty. We make sure that the signal-to-noise (S/N) of the individual spectra is large enough (S/N $>15$) and that the continuum slope is not too steep to avoid the underlying P-Cygni profile of the SN ejecta affecting the narrow line. We refer the reader to \citetalias{GG24} for more details. In Appendix~\ref{ap:lines}, we compare the different lines to each other, whereas in Appendix~\ref{ap:EW-VEL}, the EW and VEL of the lines are compared.   

\subsection{Galaxy properties}\label{sec:galmeas}

\begin{figure}
\centering
\includegraphics[width=\columnwidth]{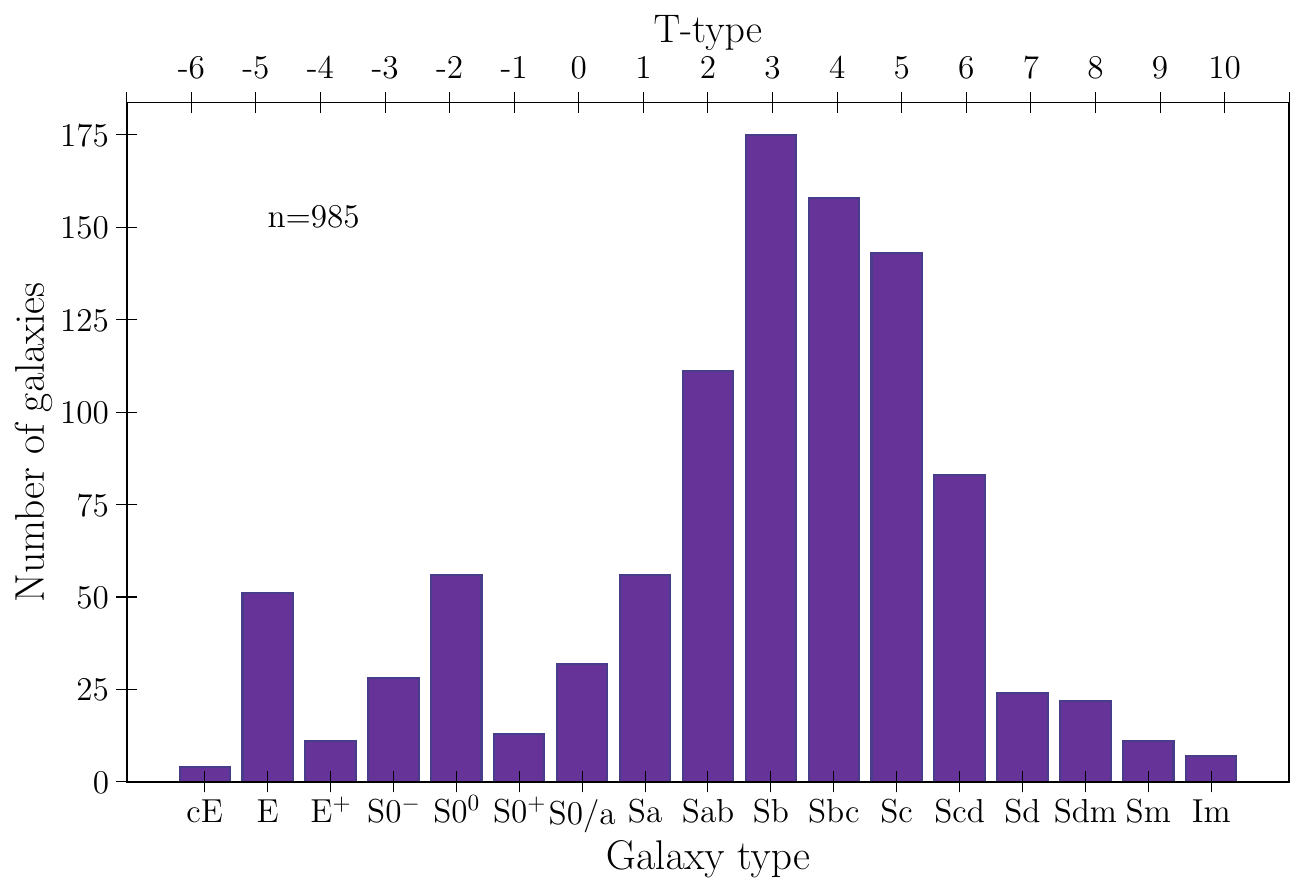}
\caption{Galaxy type distribution of our SN sample. 
$n=985$ is the number of SN host galaxies.}
\label{fig:ttype}
\end{figure}

The host galaxies of the SN sample are nearby ($z\lesssim0.2$) and have abundant public information. We obtain the spectroscopic redshift, the right ascension and declination, the morphological type and the semi-major/minor axes, $a$ and $b$, from the NASA/IPAC Extragalactic Database (NED) database\footnote{\url{https://ned.ipac.caltech.edu/}}. The galaxy type is complemented with the Asiago catalogue\footnote{\url{http://graspa.oapd.inaf.it/asnc/}} information \citep{Barbon10} and transformed to a T-type \citep{deVaucouleurs59} number between $-6$ and $10$. Figure~\ref{fig:ttype} shows the galaxy T-type distribution for our sample. Our SNe cover a large diversity of environments, going from passive to star-forming galaxies. The peak of the distribution is around the Sb type, corresponding to the T-type of 3.

From the SN and host galaxy (GAL) coordinates, we obtain the angular offset:

\begin{equation}  
\label{eq-offset}
\Delta\alpha(^{\circ}) = \sqrt{\Delta\mathrm{Ra}^2+\Delta\mathrm{Dec}^2}, 
\end{equation} 
with 
\begin{eqnarray*}
\Delta\mathrm{Ra}  &=& (\mathrm{Ra}_\mathrm{SN}-\mathrm{Ra}_\mathrm{GAL})\cos{\left[0.5(\mathrm{Dec}_\mathrm{SN}+\mathrm{Dec}_\mathrm{GAL})\right]} \\
\Delta\mathrm{Dec}  &=& \mathrm{Dec}_\mathrm{SN}-\mathrm{Dec}_\mathrm{GAL}.
\end{eqnarray*} 

\begin{figure}
\centering
\includegraphics[width=\columnwidth]{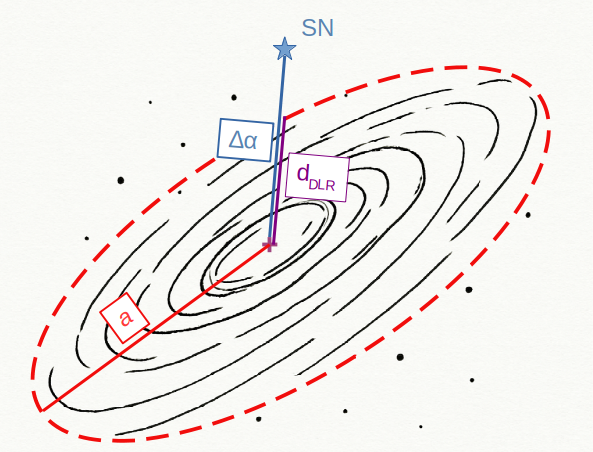}
\caption{Sketch of a spiral galaxy\textsuperscript{$\dagger$} with the SN (blue star) indicating as a blue line the angular offset with the host centre $\Delta\alpha$, in red the semi-major axis $a$ of the ellipse, and in purple the directional light radius $d_{\mathrm{DLR}}$.}
\vspace{0.3cm}
\label{fig:dlr}
\small\textsuperscript{$\dagger$ Using OpenAI, (2025), ChatGPT (Apr version), \url{https://chat.openai.com}.}

\end{figure}

We define the unit-less normalised offset, $\overline{\Delta\alpha}$, by dividing the angular offset by the semi-major axis ($a$) of the galaxy (see Fig.~\ref{fig:dlr}). This normalisation considers the galaxy's size but also reduces the effect of the distance of the galaxy:

\begin{equation} \label{eq-normoffset}
\overline{\Delta\alpha} = \frac{\Delta\alpha(^{\circ})}{a(^{\circ})} . 
\end{equation} 



A more accurate normalisation of the angular separation takes into account the host ellipse axis size in the direction of the SN, the so-called directional light radius, $d_{\mathrm{DLR}}$ \citep{Sullivan06,Gupta16,Gagliano21} given by:

\begin{equation} \label{eq-dlr}
d_{\mathrm{DLR}} = \frac{a}{\sqrt{(a/b\cdot\sin{\beta})^2+(\cos{\beta})^2}}, 
\end{equation} 
where $a$ and $b$ are the semi-major and semi-minor axes, while $\beta$, the angle subtended between the galaxy semi-major axis and the line connecting the SN to the galactic centre, is given by:

\begin{equation*} 
\beta = \phi - \tan^{-1}\left(\frac{\Delta\mathrm{Dec}}{\Delta{\mathrm{Ra}}}\right) 
\end{equation*}
with $\phi$ being the angular tilt of the ellipse with respect to the celestial north. The final normalised separation is:

\begin{equation} \label{eq-dlroffset}
\Delta \alpha_{\mathrm{DLR}} = \frac{\Delta\alpha(^{\circ})}{d_{\mathrm{DLR}}(^{\circ})}. 
\end{equation} 

The inclination of spiral galaxies is estimated following \citet{Hubble26}, and assuming that the disk is circular when viewed face-on and with a finite thickness:

\begin{equation} \label{eq-inclination}
i(^{\circ}) = \arccos{\left(\sqrt{\frac{q^2-q_0^2}{1-q_0^2}}\right)},
\end{equation} 
where $q=b/a$ is the ratio of the semi-minor to semi-major axes and $q_0$ is the intrinsic axis ratio when viewed edge-on \citep{Fouque90}. We assume a universal $q_0=0.2$ \citep{Holmberg46, Ho11, Noordermeer07} although it has been shown to vary with galaxy type, mass and luminosity \citep{Heidmann72, Bottinelli83, Yuan04, Rodriguez13}. 

As explained in the next sections, we also consider stellar parameters derived from stellar populations that fit local and global photometry. Table~\ref{table-property} lists all the galaxy properties considered in this study.

\renewcommand{\arraystretch}{1.4}
\begin{table}
\begin{threeparttable}
\centering
\caption{Galaxy properties studied.}
\label{table-property}
\renewcommand{\arraystretch}{1.25}
\begin{tabular}{ccc}

Property & Description & Reference \\
\hline
\hline
Type & Galaxy classification & NED$^{\ast}$/Asiago \\
$\Delta\alpha(^{\circ})$ & SN angular offset  & Eq.~\ref{eq-offset} \\
$\overline{\Delta\alpha}$ & SN normalised offset & Eq.~\ref{eq-normoffset}\\  
$\Delta\alpha_{\mathrm{DLR}}$ & SN directional offset  & Eq.~\ref{eq-dlroffset}\\
$i(^{\circ})$ & Galaxy inclination & Eq.~\ref{eq-inclination} \\

\hline
$M_*^G/M_{\sun}$ & Global stellar mass & Sec.~\ref{sec-SEDfit} \\
$Z_*^G/Z_{\sun}$ & Global stellar metallicity & Sec.~\ref{sec-SEDfit} \\
$t_{\mathrm{age}}^G$(Gyr) & Global age & Sec.~\ref{sec-SEDfit} \\
$\tau^G$(Gyr) & Global e-folding time & Sec.~\ref{sec-SEDfit} \\
$A_V^G$ & Global attenuation & Sec.~\ref{sec-SEDfit} \\
$n^G$ & Global dust index & Sec.~\ref{sec-SEDfit} \\
SFR$_0^G (M_{\sun}$/yr) & Global SFR & Eq.~\ref{eq-sfh} \\
sSFR$_0^G$ (yr$^{-1}$) & Global specific SFR & Eq.~\ref{eq-ssfr} \\
\hline
$M_*^L/M_{\sun}$ & Local stellar mass & Sec.~\ref{sec-SEDfit} \\
$Z_*^L/Z_{\sun}$ & Local stellar metallicity & Sec.~\ref{sec-SEDfit} \\
$t_{\mathrm{age}}^L$(Gyr) & Local age & Sec.~\ref{sec-SEDfit} \\
$\tau^L$(Gyr) & Local e-folding time & Sec.~\ref{sec-SEDfit} \\
$A_V^L$ & Local attenuation & Sec.~\ref{sec-SEDfit} \\
$n^L$ & Local dust index & Sec.~\ref{sec-SEDfit} \\
SFR$_0^L (M_{\sun}$/yr) & Local SFR & Eq.~\ref{eq-sfh} \\
sSFR$_0^L$ (yr$^{-1}$)& Local specific SFR & Eq.~\ref{eq-ssfr} \\
\hline
\end{tabular}
\begin{tablenotes}
\small
\item $^{\ast}$ We use the NED homogenized classification scheme.
\end{tablenotes}
\end{threeparttable}
\end{table} 

\subsubsection{Galaxy photometry}
\label{sec:galphot}

The SNe in our sample occurred in nearby galaxies ($0.004<z<0.2$), for which there is extensive literature photometry. We use public optical images either from the Dark Energy Survey \citep[DES;][]{DES-DR1, DES-DR2, DES_pipeline, DECam}, Pan-STARRS1 \citep[PS1;][]{Waters20-PS1,Magnier20-PS1,Flewelling20-PS1} or from the Sloan Digital Sky Survey \citep[SDSS;][]{Abdurrouf22}, ultraviolet (UV) photometry from the NASA Galaxy Evolution Explorer \citep[GALEX;][]{GALEX, GALEX-DR7}, near-infrared (NIR) images either from the Two Micron All Sky Survey \citep[2MASS;][]{2MASS} or from the Visible and Infrared Survey Telescope for Astronomy \citep[VISTA;][]{VISTA, VISTA-pipeline, VISTA-pipeline2, VISTA-archive}, and infrared imaging from the unblurred co-adds of the Wide-field Infrared Survey Explorer \citep[unWISE;][]{unWISE,unWISE3yr,unWISE1yr}. We require one optical dataset, giving preference to DES over PS1 over SDSS, if available, and one NIR dataset, prioritising VISTA over 2MASS because of their higher resolution. The list of surveys and filters used is shown in Table~\ref{table-survey}.

\renewcommand{\arraystretch}{1.2}
\begin{table}
\centering
\begin{threeparttable}
\setlength{\tabcolsep}{15pt}
\caption{Surveys and filters for SN host photometry.}
\label{table-survey}
\begin{tabular}{ccc}
Survey & Filter $^{\ast}$ & $\lambda_{\mathrm{eff}}$ (\AA) $^{\ast}$ \\
\hline
\hline
\multicolumn{3}{c}{\textbf{Ultraviolet}}\\
\hdashline
\multirow{2}{*}{GALEX} & $FUV$ & 1 549\\
&$NUV$ & 2 303\\
\hline
\multicolumn{3}{c}{\textbf{Optical}}\\
 \hdashline
 \multirow{5}{*}{SDSS} & $u$ & 3 608 \\
& $g$ & 4 672 \\
& $r$ & 6 141 \\
& $i$ & 7 458 \\
& $z$ & 8 923 \\
\multirow{5}{*}{DES} & $g_{\mathrm{DECam}}$ & 4 770 \\
& $r_{\mathrm{DECam}}$ & 6 370 \\
& $i_{\mathrm{DECam}}$ & 7 774 \\
& $z_{\mathrm{DECam}}$ & 9 155 \\
& $Y_{\mathrm{DECam}}$ & 9 887 \\
\multirow{5}{*}{PS1} & $g_{\mathrm{PS1}}$ & 4 810 \\
& $r_{\mathrm{PS1}}$ & 6 156 \\
& $i_{\mathrm{PS1}}$ & 7 503 \\
& $z_{\mathrm{PS1}}$ & 8 668 \\
& $Y_{\mathrm{PS1}}$ & 9 614 \\
\hline
\multicolumn{3}{c}{\textbf{Near-infrared}}\\
\hdashline
\multirow{4}{*}{VISTA} & $Y_{\mathrm{VISTA}}$ & 10 196 \\
& $J_{\mathrm{VISTA}}$ & 12 481 \\
& $H_{\mathrm{VISTA}}$ & 16 347 \\
& $Ks_{\mathrm{VISTA}}$ & 21 376 \\
\multirow{3}{*}{2MASS} & $J$ & 12 350 \\
& $H$ & 16 620 \\
& $Ks$ & 21 590 \\
\hline
\multicolumn{3}{c}{\textbf{Infrared}}\\
\hdashline
\multirow{4}{*}{unWISE} & $W1$ & 33 526 \\
& $W2$ & 46 028 \\
& $W3$ & 115 608 \\
& $W4$ & 220 883 \\
\hline
\end{tabular}
\begin{tablenotes}
\small
\item $^{\ast}$ See the SVO Filter Profile Service: \url{http://svo2.cab.inta-csic.es/theory/fps/}
\end{tablenotes}
\end{threeparttable}
\end{table} 

\begin{figure*}
\centering
\includegraphics[width=0.49\textwidth]{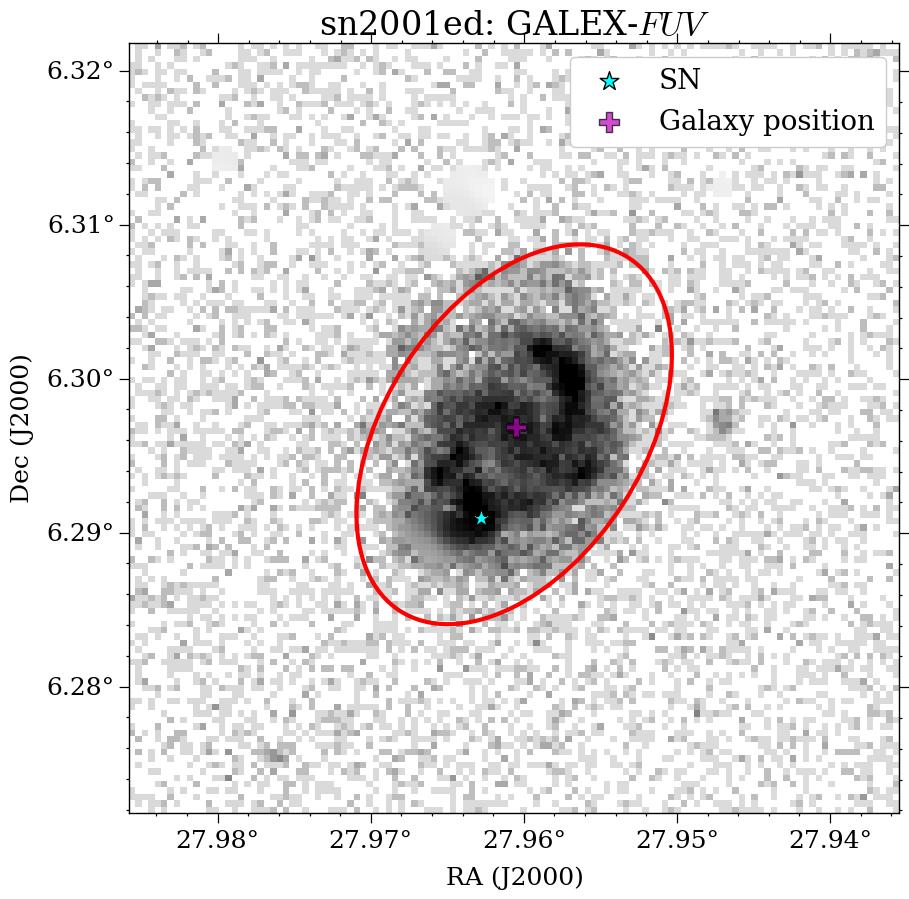}
\includegraphics[width=0.49\textwidth]{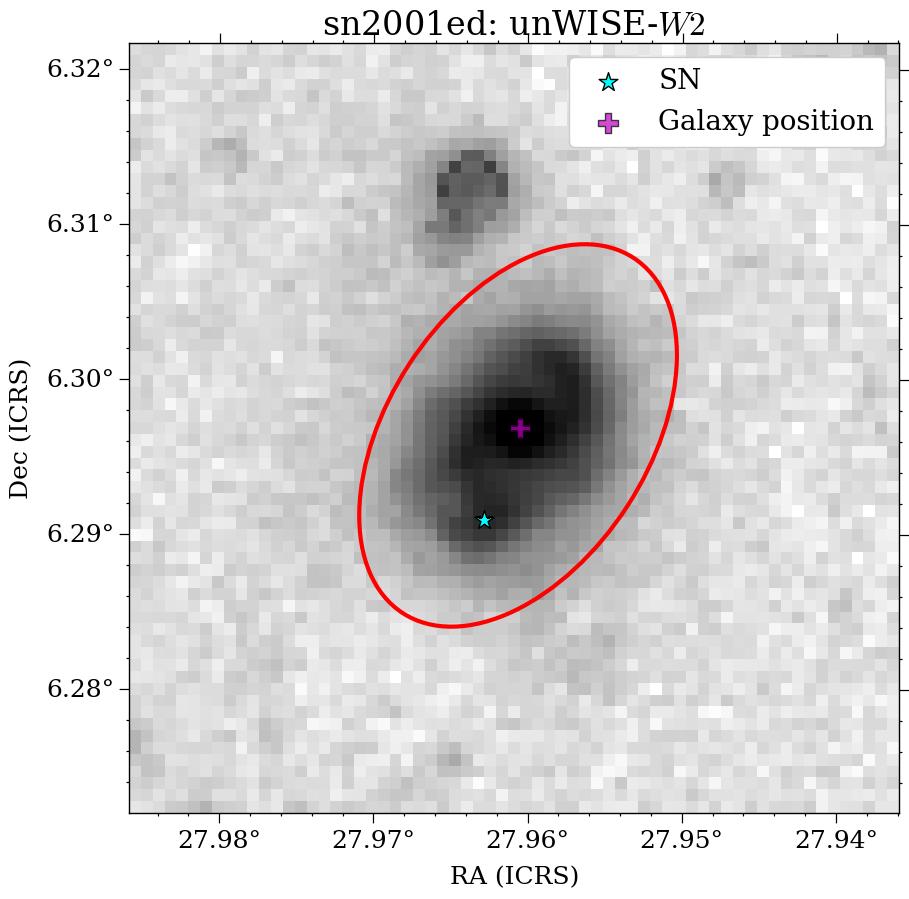}
\caption{Representative \textsc{hostphot} images of SN~2001ed  with the filters: $FUV$ (GALEX) and $W2$ (unWISE). Masked sources are shown in grey, and the global Kron apertures are shown as red ellipses. The galaxy's centre is shown as a purple cross, whereas the SN position is a cyan star.}
\label{fig:globalphot}
\end{figure*}

To download the images and perform photometry, we use the software \textsc{hostphot}\footnote{\url{https://hostphot.readthedocs.io/en/latest/}} \citep{Muller-Bravo22} tailored to SN host galaxies. The images are first masked for foreground stars cross-matched to the \textit{Gaia} catalogue \citep{Gaia-mission, Gaia-DR3}.  We then obtain global photometry in each filter image using Kron apertures \citep{Kron80} defined on the stack of the optical filters (see Figure~\ref{fig:globalphot}). 

\begin{figure*}
\centering
\includegraphics[width=0.51\textwidth]{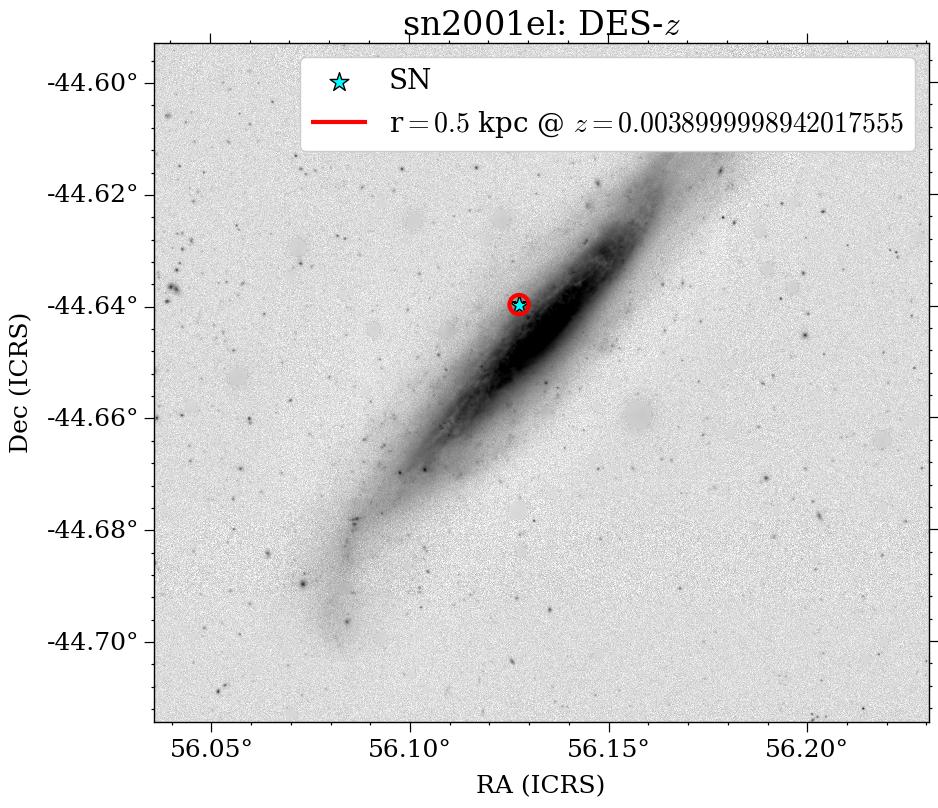}
\includegraphics[width=0.48\textwidth]{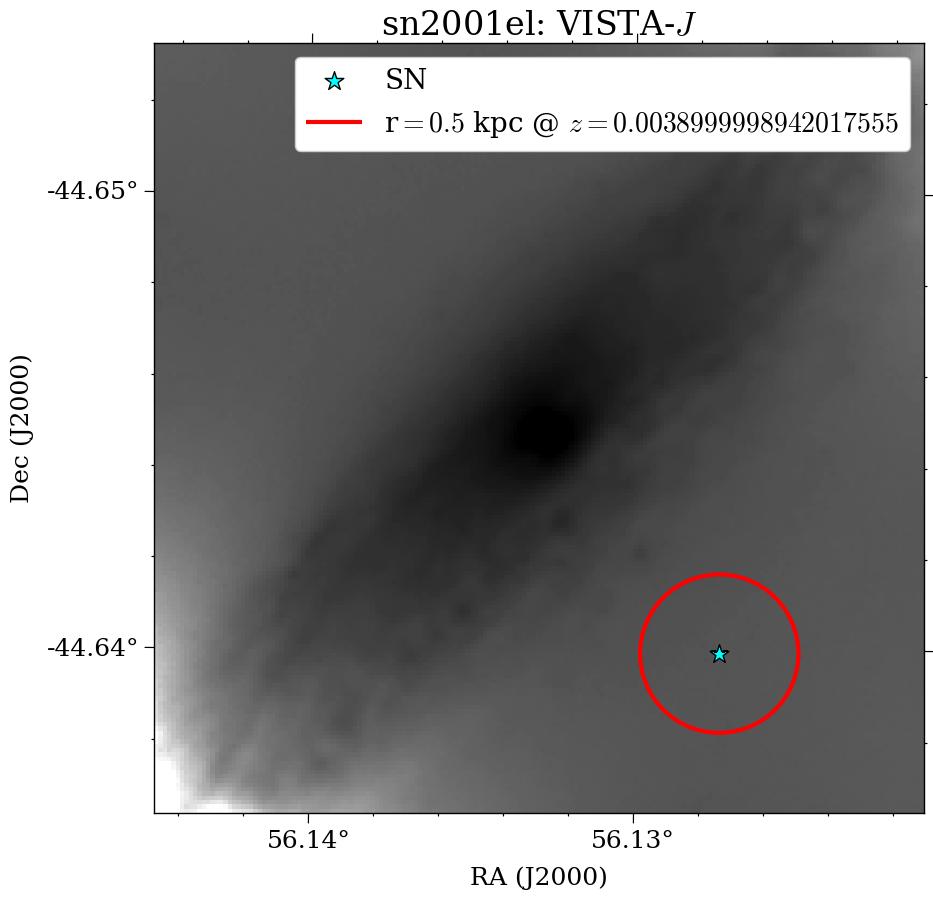}
\caption{Representative \textsc{hostphot} images of SN~2001el  with the filters: $z_{\mathrm{DECam}}$ (DES) and $J_{\mathrm{VISTA}}$ (VISTA). Masked sources are shown in grey, and the local circular apertures for $r=0.5$ kpc are shown as red lines. The cyan star denotes the SN position. Note that the images have different orientations and resolutions.}
\label{fig:localphot}
\end{figure*}

Local photometry is performed with circular apertures of radii of 0.5 kpc centred at the SN position (see Figure~\ref{fig:localphot}). The physical apertures are obtained with the redshift, assuming a standard cosmology ($H_0=70$ km/s/Mpc and $\Omega_m=0.3$). Since most of the optical images we use come from the SDSS, which has a lower resolution than DES, we set an upper redshift limit of 0.02, corresponding to a radial aperture of 0.5 kpc with the typical SDSS seeing of 1.2''. Higher redshifts have an aperture below the seeing limit and are not considered. The number of SNe with global and local photometry was reduced to 613 and 357, respectively (see Table~\ref{table:sne}).

\subsubsection{Spectral Energy Distribution fitting}
\label{sec-SEDfit}

We use \textsc{prospector}\footnote{\url{https://prospect.readthedocs.io/en/latest/}} \citep{Johnson21-prospector, Leja17} to fit stellar populations to the non-zero fluxes of the available filters of our local and global host galaxy photometry obtained with \textsc{hostphot}. The composite stellar populations are built with FSPS \citep{Conroy10, Conroy09} using the MIST isochrones \citep{Dotter16, Choi16, Paxton15, Paxton13, Paxton11} and the \textsc{miles} spectral library \citep{Falcon-Barroso11-miles}, with a Kroupa initial mass function \citep{Kroupa01} and a delayed-$\tau$ star formation history given by a star formation rate (SFR) as a function of lookback time, $t$:

\begin{equation}\label{eq-sfh}
\mathrm{SFR}(t) \propto (t_{\mathrm{age}}-t)e^{-(t_{\mathrm{age}}-t)/\tau}, \;\; 0 < t < t_{\mathrm{age}},
\end{equation}
where $t_{\mathrm{age}}$ and $\tau$ are the age and the characteristic e-folding time in Gyr of the star-forming episode, both free parameters of the fit along with the stellar metallicity (Z$_*$/\Zsun), and the stellar mass of existing stars and stellar remnants (M$_*$/\Msun). Although this treatment of the star formation history is simplistic and may fail to reproduce the extremes of galaxy populations \citep{Simha14}, the relative differences among populations --our main focus-- are overall maintained. We use a diffuse dust attenuation given by \citet{Noll09} with two free parameters, the total dust attenuation $\tau_V$ or "dust2", related to the optical depth and $A_V$, and the dust attenuation index $n$ which controls the wavelength dependence:

\begin{equation}
\label{eq-attenuation}
    \tau(\lambda) = \frac{\tau_V}{R_{V,0}}[k(\lambda)+D(\lambda)]\left(\frac{\lambda}{\lambda_V}\right)^n,
\end{equation}
where $k(\lambda)$ is the attenuation curve of \citet{Calzetti00}, with $R_{V,0}=4.05$, and $D(\lambda)$ accounts for the UV-bump Drude profile. 

We thus have six free parameters: $t_{\mathrm{age}}$, $\tau$, Z$_*$, M$_*$, $\tau_V$ and $n$. Additionally, the recent star formation rate (SFR$_0$), averaged over the last 100 Myr, can be directly obtained from Eq.~\ref{eq-sfh}. Thus, the recent specific star formation rate (sSFR$_0$) is defined as:

\begin{equation}\label{eq-ssfr}
    \mathrm{sSFR}_0 = \mathrm{SFR}_0 / M_*
\end{equation}

\begin{figure*}
\centering
\includegraphics[width=0.495\textwidth]{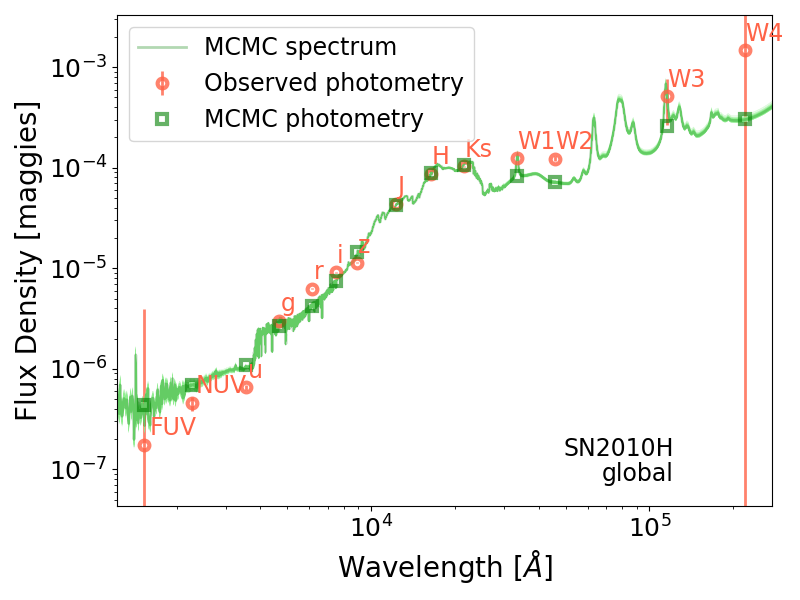}
\includegraphics[width=0.495\textwidth]{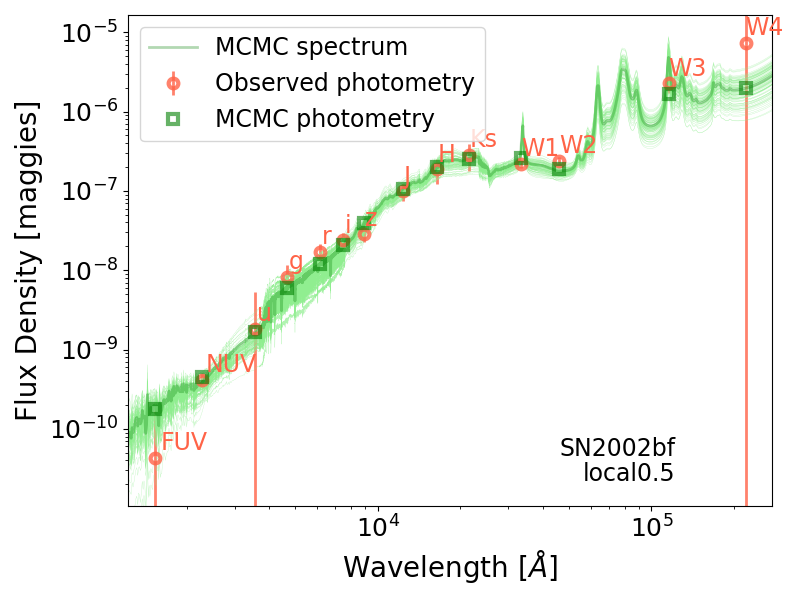}
\caption{\textsc{prospector} SED fits to the observed global photometry of the host galaxy of SN~2010H (left) and local photometry of the host galaxy of SN~2002bf (right). The observed photometry is shown with orange circles with corresponding filter names. The median MCMC fit is shown in dark green with synthetic photometry as green squares, and 100 MCMC realisations are shown in light green.}
\label{fig:SEDfits}
\end{figure*}

\begin{figure*}
\centering
\includegraphics[width=0.495\textwidth]{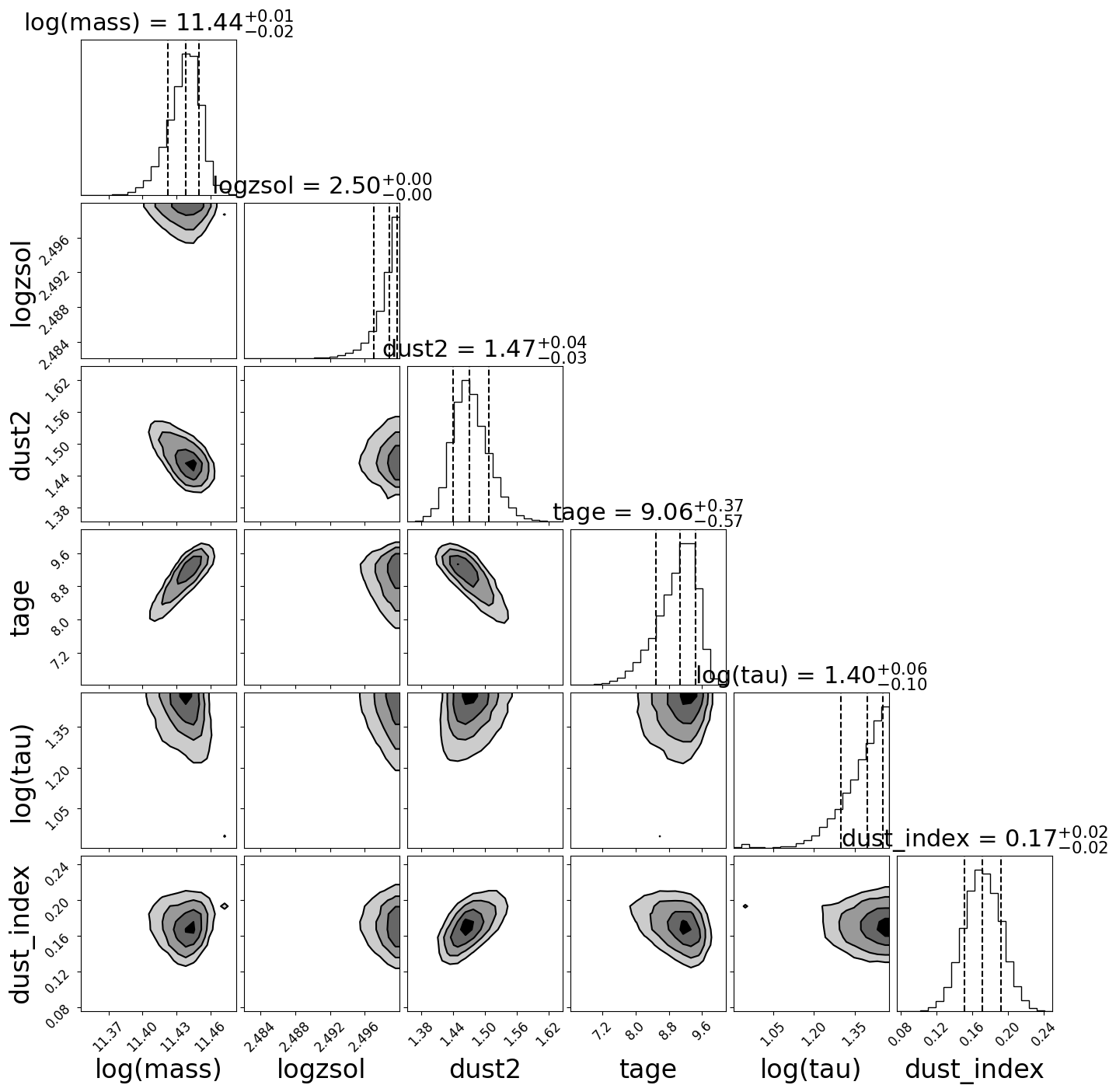}
\includegraphics[width=0.495\textwidth]{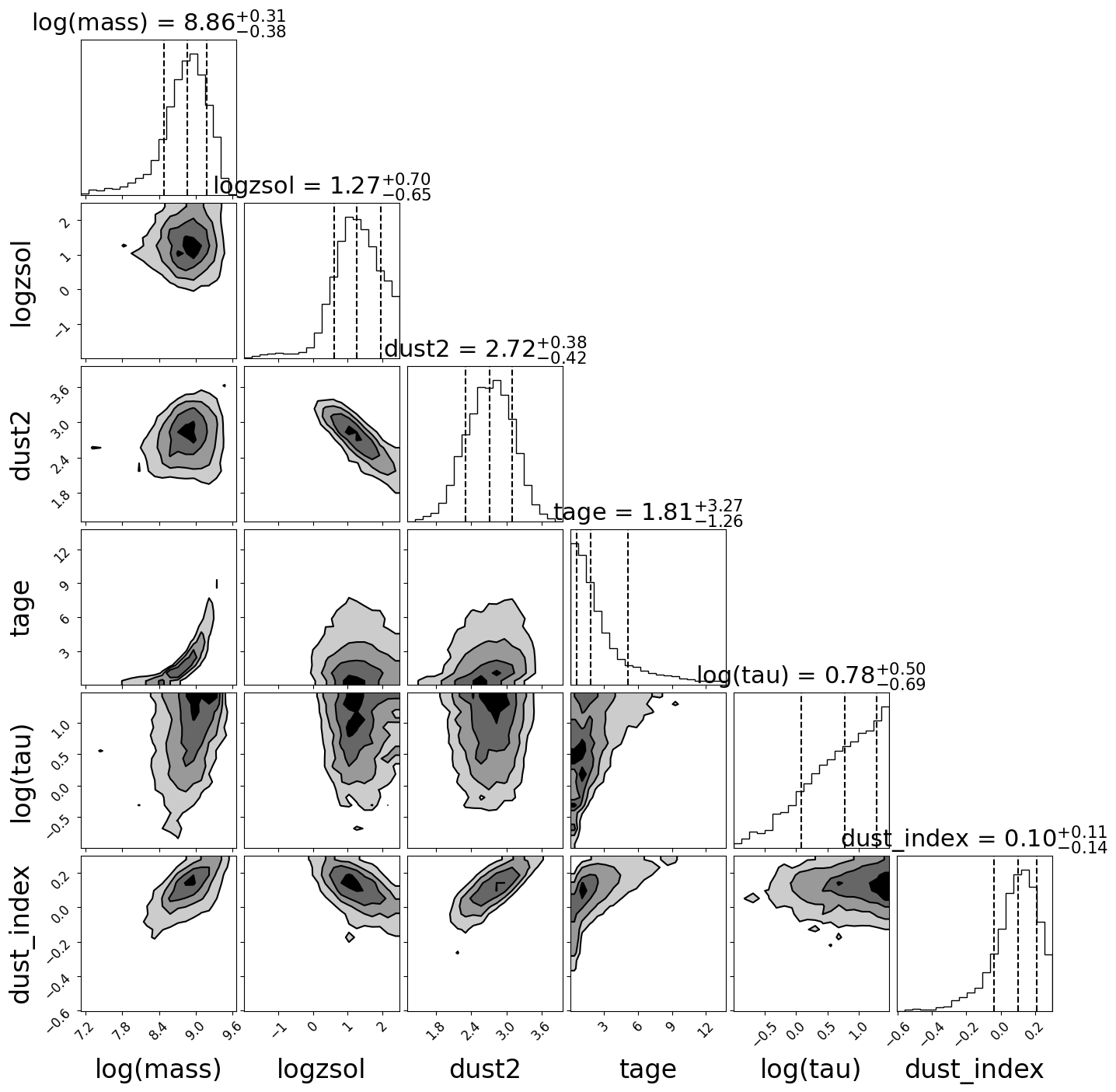}
\caption{Corner plots resulting from  \textsc{prospector} SED fits to the observed global/local photometry of the host galaxy of SN~2010H/SN~2002bf (left/right). The six free parameters ($\log$M$_*$, $\log$Z$_*$, $\tau_V$ or "dust2", $t_{\mathrm{age}}$, $\log\tau$, and dust index $n$) are shown in each panel. 
}
\label{fig:corner}
\end{figure*}

The parameter space is sampled with a Markov Chain Monte Carlo \citep{emcee} of 2512 iterations and 128 walkers. The starting point is given by a distribution around the best-fit parameters of 10 initial optimisations. A burn-in of 852 steps ensures the parameters have converged in most cases, but we do a visual check and extend the burn-in and iterations when necessary. If the fits are still visually unsatisfactory, these SNe are removed from the analysis (6\% of the sample). The best-fit spectral energy distribution (SED) for two representative galaxies and their corresponding corner plots are presented in Figures~\ref{fig:SEDfits} and~\ref{fig:corner}, respectively. As seen in Figure~\ref{fig:SEDfits},
the reddest WISE filters, particularly $W4$ (22$\mu$m), tend to show higher fluxes than the models, perhaps due to a contribution from dust emission that is not modelled here. Also, as seen in Figure~\ref{fig:corner}, given its large uncertainties, the e-folding time $\tau$ is not always well constrained. 

\subsubsection{Sample comparison}

\begin{figure*}
\centering
\includegraphics[width=\textwidth]{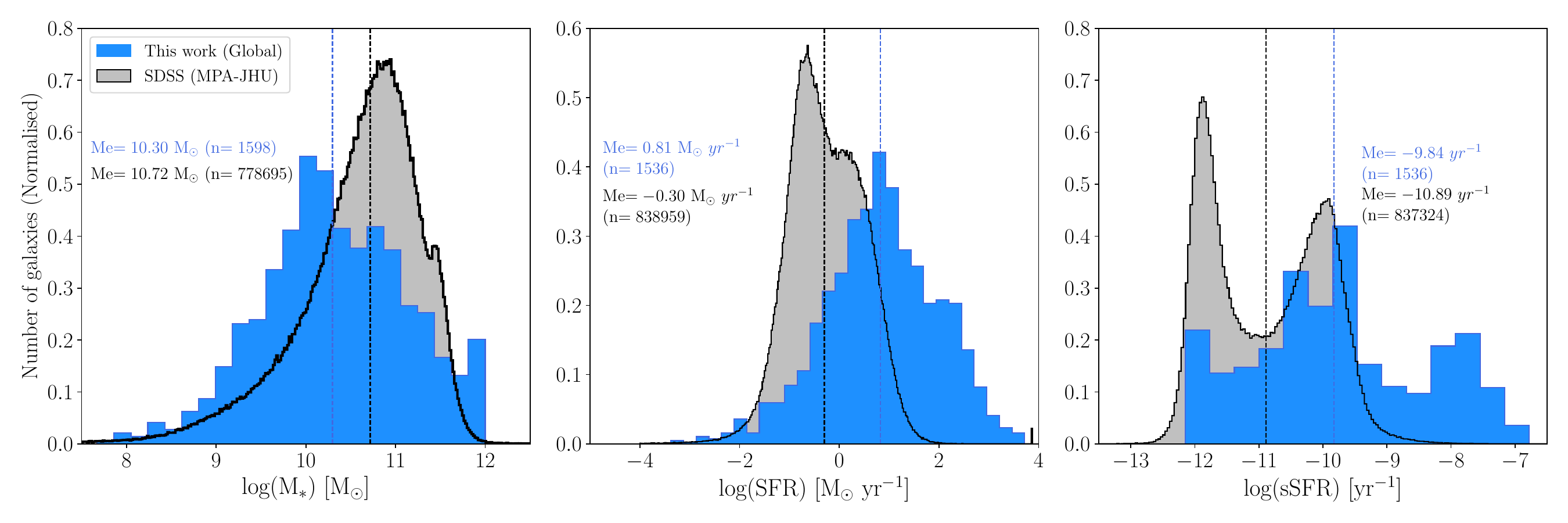}
\caption{Stellar mass (left), SFR (middle) and sSFR (right) distributions of our sample (blue) compared with the SDSS measurements (grey) from the MPA/JHU catalogue. Vertical dashed lines mark the median values for each distribution.  
}
\label{fig:galglobal}
\end{figure*}

The final stellar mass M$_*$, SFR and sSFR parameters of our sample obtained from global photometry are shown in Figure~\ref{fig:galglobal}. For comparison, we show the distributions of the SDSS galaxies from the MPA/JHU catalogue\footnote{\url{http://www.mpa-garching.mpg.de/SDSS/DR7}}. The SDSS stellar masses were obtained from photometric fits \citep[following][]{Kauffmann03, Salim07}\footnote{They used a Kroupa IMF as in the current work.}, and they covered a similar range to our sample (left panel). The SFR (middle panel) and sSFR (right panel) values, on the other hand, were obtained with spectral line diagnostics \citep{Brinchman04}; therefore, a direct comparison with our stellar fits is more difficult. With that in mind, we see clear differences towards more strongly star-forming galaxies in our low-$z$ sample. This could be due in part to known overestimates in the SFR integrated over short time periods of 100 Myr \citep{Boquien14}. On the other hand, our nearby targeted sample is likely comprised of larger, brighter star-forming galaxies. For the purpose of our study, absolute shifts are irrelevant since we care only about relative differences within the sample. 

\section{Analysis and Results}
\label{sec:res}

In this section, we compare the EW and VEL distributions of the narrow interstellar \naid\ line measured for the SN sample presented in Section~\ref{sec:EWmeas} and their corresponding galaxy properties obtained in Section~\ref{sec:galmeas}. 
To this end, we generate cumulative distributions to test whether two samples, divided based on a given galaxy property, originate from the same parent EW/VEL distribution using the Kolmogorov-Smirnov (K-S) statistic \citep{ks-test1,ks-test2}\footnote{We also tried the Anderson-Darling test \citep{Anderson-Darling52} with consistent results.}. If the K-S test has a p-value $\rho_{KS}<0.05$ ($\rho_{KS}<0.01$), then the null hypothesis that the two populations are drawn from the same distribution can be marginally (strongly) rejected. To test the robustness of these tests against outliers in the sample, we also do a bootstrap, generating 1000 random sub-samples from which we calculate the probability that the K-S test p-value is less than 0.05. Since our sample is incomplete, we also ensure that the redshift distributions of the two samples are consistent in each bootstrap iteration. This is done first by obtaining the optimal redshift bins \citep[via Bayesian Blocks, see][]{Scargle13-bayesianblocks} and drawing from each of these bins several SNe in one sample that are within the Poisson error of the other sample. We call this a "$z$-matched" bootstrap K-S test. Moreover, since the division of the two samples can be rather arbitrary, instead of using only the median of the galaxy property to obtain the two samples to be tested, we do a sweep of 10 different dividing values around the median (from 40\% to 60\% percentile of the distribution). For each division point, we do the $z$-matched bootstrap so that at the end, we have $10\times1000=10000$ realisations of the K-S test. The final value we quote is the median of all these iterations and the corresponding probability of the p-value being less than 0.05. Although using the minimum K-S test instead of the median (see \citealt{Forster13}) would provide the best division at which the samples are most different from each other, our approach is more robust to statistical fluctuations. In Appendix~\ref{ap:KSquart}, we also show the K-S tests based on the lower and upper quartiles of each distribution, as well as considering the look-elsewhere effect. Finally, to be certain that any differences between samples are not coming from measurement biases, we always check the K-S test obtained by measuring the same lines but at the wavelengths corresponding to the redshift of the Milky Way (MW), for which we expect the populations to be fully consistent. 

\begin{figure*}
\centering
\includegraphics[width=0.485\textwidth]{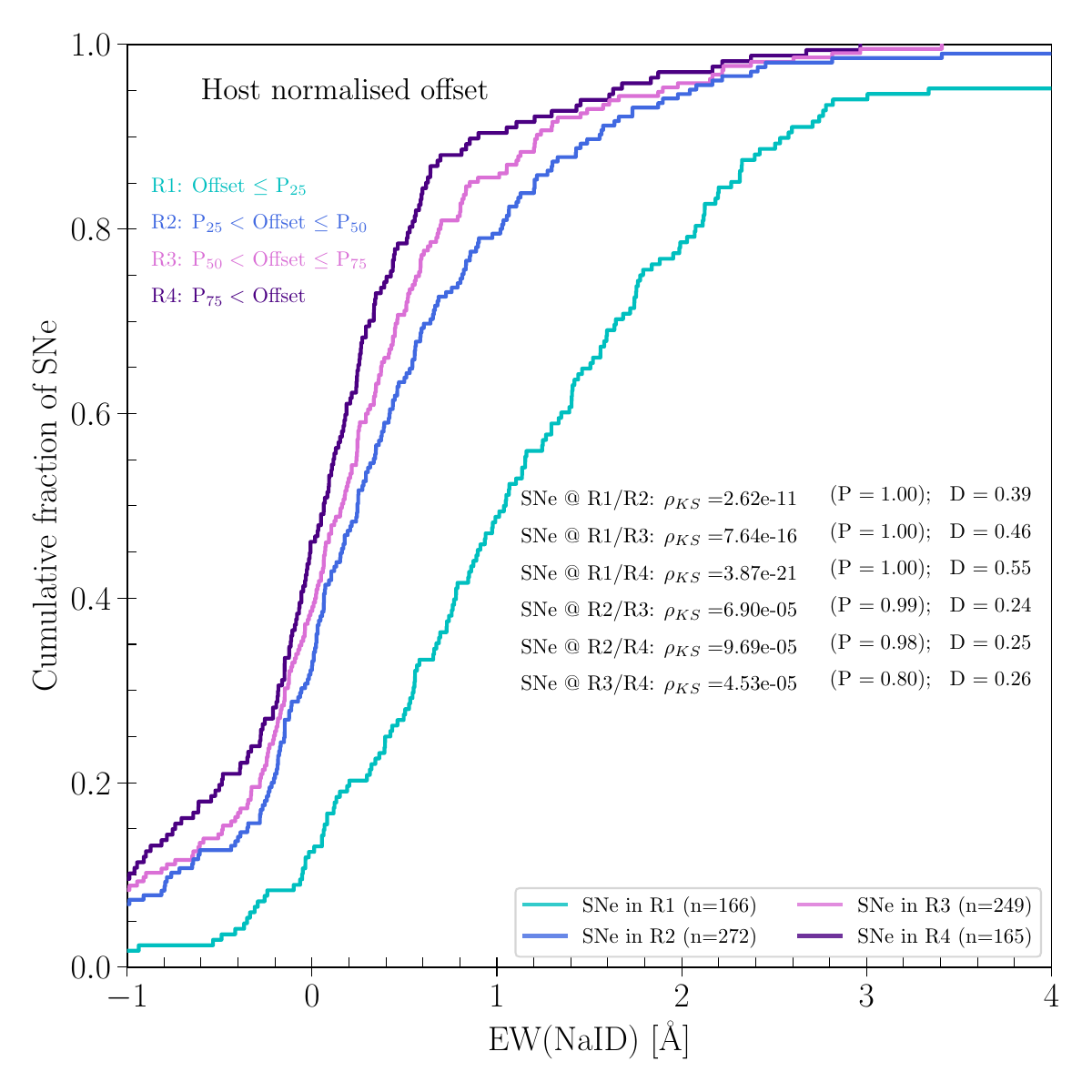}
\includegraphics[width=0.495\textwidth]{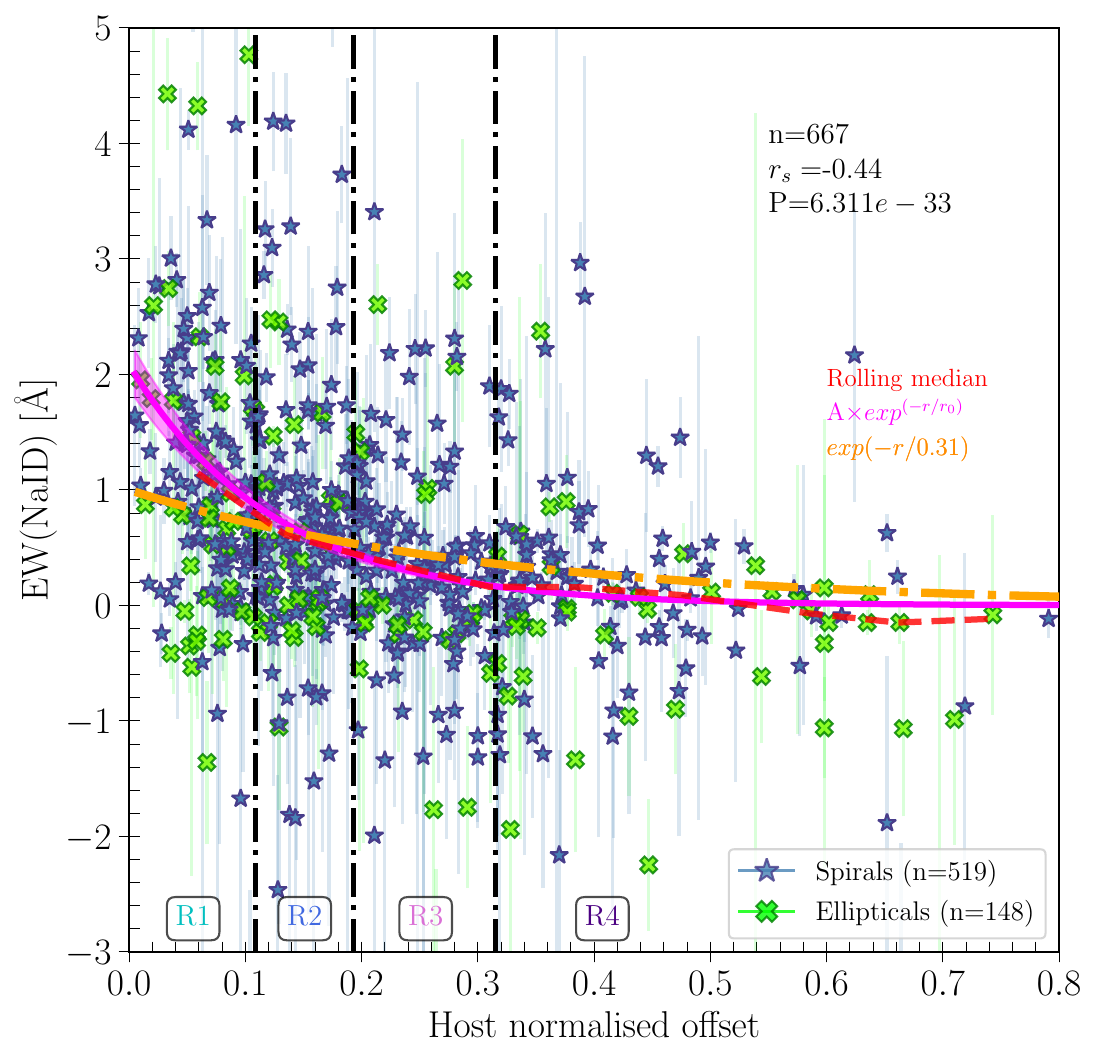}
\caption{\emph{Left:} Cumulative distributions of \naid\ EW divided into four bins according to normalised offset. K-S tests indicate that all bins are not drawn from the same distribution. \emph{Right:} \naid EW vs normalised offset. SNe located in spirals/ellipticals are highlighted in blue/green stars/crosses. On top of the figure, \textbf{$n=667$} is the number of events, \textbf{r$_S=-0.44$} is the Spearman correlation coefficient, and \textbf{P$=6.3\times10^{-33}$} is the probability of detecting a correlation by chance. The vertical dash-dotted lines represent the 25th, 50th and 75th percentiles of the sample from which the cumulative distributions are obtained. The red dashed line is the rolling median, the purple solid line is the exponential fit to the medians, with the shaded purple area indicating the 1$\sigma$ uncertainty, and the orange line an exponential function from a symbolic regression (see \S~\ref{sec:discussion-importance}).}
\label{fig:offnorm}
\end{figure*}

\begin{figure}
\centering
\includegraphics[width=\columnwidth]{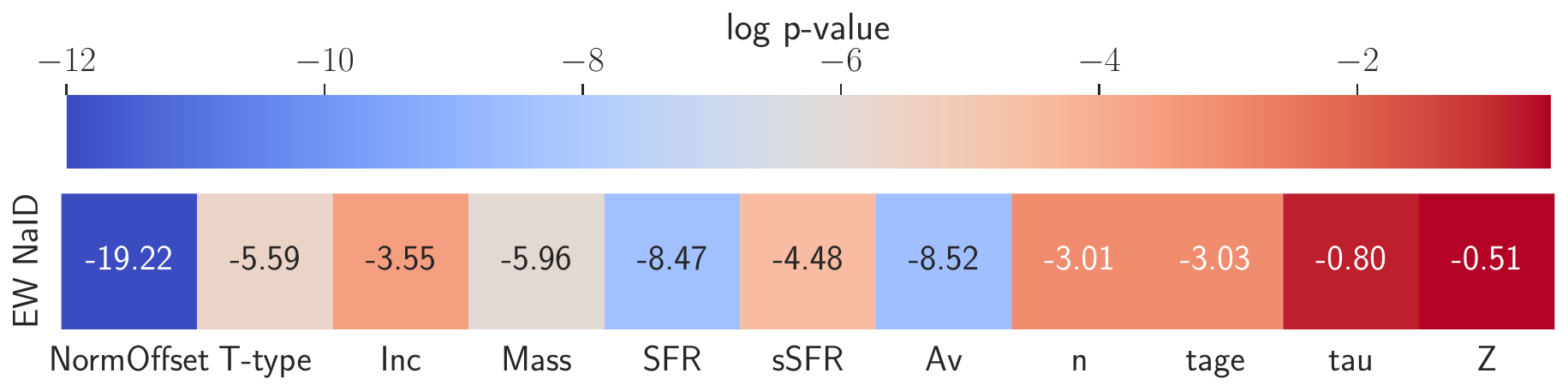}
\caption{
K-S test p-values (log) of \naid\ EW divided according to galaxy properties: normalised offset, T-type, inclination and local stellar and dust properties.}
\label{fig:EWcorr}
\end{figure}

We start by dividing each galaxy property into two samples based on its median and then comparing the two samples' EW (or VEL) distributions. We also analyse if the EW (or VEL) correlates with each property via the non-parametric Spearman's rank coefficient, $r_s$ \citep{Spearman1904}. The K-S statistics and correlations for \naid\ EW distributions divided according to galaxy properties are presented in Table~\ref{table:KS} and Figure~\ref{fig:EWcorr}. Many of the properties present strong K-S test statistics, as highlighted in bold, according to their $z$-matched bootstrap probability of being drawn from different parent populations, i.e. probability higher than 50\% of a K-S p-value less than 0.05: $\mathbf{P}(p_{MC}<0.05) > 50\%$. The strongest K-S rejections are found for the offset of the SN from the host galaxy centre, especially when normalised by the semi-major axis of the galaxy. In this case, the median bootstrap K-S p-value is $\sim10^{-20}$ or a 100\% MC probability that the two \naid EW distributions of low and high normalised offset are not drawn from the same distribution. Since the relation is so strong, we further divide the sample into four equal offset bins as shown in the left Figure~\ref{fig:offnorm}: all four samples are statistically different according to the K-S tests. This can be further seen in the right scatter plot, where clearly the EW values inside 25\% of the galaxy's major axis cover a wider range, extending to values as high as 4-5 \AA. At high separations from the centre, the lines are, on average, weaker. We caution that the offset range seen in Figure~\ref{fig:offnorm} may seem shifted to low values (the majority of values are below the semi-major axis, i.e. $\overline{\Delta\alpha}<1$) compared with separations in the literature. This is due to the large semi-major axes from NED as compared to those from e.g. Kron apertures used to normalise the angular offsets (see Appendix~\ref{ap:NED-HP}). Although the K-S tests still show significant differences between the samples for the angular offset ($\sim10^{-7}$) and the DLR offset ($\sim10^{-14}$), they are weaker than the normalised offset. The normalised offset is normalised by the semi-major axis and washes out any distance uncertainties, as well as different galactic sizes. The DLR also normalises by the ellipse axis into the SN direction and has shown to be the most robust parameter for SN host galaxy association \citep{Gupta16}; however, all our tests with narrow lines are stronger for the normalised offset. This is likely because the DLR partly washes out the galactic inclination, as explained in the following paragraph. For this reason, we focus only on the normalised offset in the remainder of the paper. 

\renewcommand{\arraystretch}{1.3}
\begin{table*}
\begin{threeparttable}
\small
\centering
\caption{K-S statistics and correlations for \naid EW divided according to galaxy properties.}
\label{table:KS}
\renewcommand{\arraystretch}{1.6}
\begin{tabular}{c|c|cccc|cccc}
& & \multicolumn{4} {c|} {\bfseries HOST} & \multicolumn{4} {c} {\bfseries MW} \\
\hline
Property & Nr & $<D^{\mathrm{EW}}_{\mathrm{MC}}>$ & $<p_{MC}^{\mathrm{EW}}>^{\ast}$ & $\mathbf{P}\left(p^{\mathrm{EW}}_{MC}<0.05\right)^{\ast}$ & $r_s^{\mathrm{EW}}$ & $<D^{\mathrm{EW}}_{\mathrm{MC}}>$ &$<p_{MC}^{\mathrm{EW}}>$ & $\mathbf{P}\left(p^{\mathrm{EW}}_{MC}<0.05\right)^{\ast}$ & $r_s^{\mathrm{EW}}$ \\
\hline                                         
\hline
\multicolumn{10}{c}{\textbf{General properties}}\\
\hdashline 
\boldsymbol{$\Delta\alpha(^{\circ})$}        & 697            & 0.25 & $2.3\times10^{-7}$ & {\bf 100\%} & $-0.22$  & 0.13 & 0.10 & 24\% & $0.19$ \\
\boldsymbol{$\overline{\Delta\alpha}$}       & 667            & 0.39 & $6.0\times10^{-20}$ & {\bf 100\%} & $-0.44$ & $0.09$ & $0.36$ & 6\%  & $0.00$ \\ 
\boldsymbol{$\Delta\alpha_{\mathrm{DLR}}$}            & 662            & 0.33  & $2.5\times10^{-14}$  & {\bf 100\%} & $-0.36$ & $0.08$ & $0.45$ & 3\% & $-0.01$  \\
{\bf T-type$^{\dag}$}                        & 660            & 0.34 & 
$2.6\times10^{-6}$  & {\bf 100\%} & 0.07  & $0.13$ & $0.39$ & 6\%  & $-0.13$ \\
\boldsymbol{$i(^{\circ})$}                   & 666            &  0.18               & $2.8\times10^{-4}$  & {\bf 98\%}  & 0.18  & 0.09 & 0.29 & 7\%  & $-0.09$ \\
\hline                                         \multicolumn{10}{c}{\textbf{Local properties}}\\
\hdashline     
\bf{M$_*^L$/ \Msun}                          & $357^{\oplus}$            & 0.31  & $1.2\times10^{-6}$  & \bf{100\%}  & 0.32  & 0.13 & 0.25 & 11\%  & $-0.02$  \\
\bf{SFR\boldsymbol{$_0^L$} (\Msun /yr)}      & $357$ & 0.36  & $3.7\times10^{-9}$  & {\bf 100\%} & 0.39  & $0.11$ & $0.46$ & 3\%  & $-0.09$ \\
\bf{sSFR\boldsymbol{$_0^L$} (yr$^{-1}$)}     & 357            & 0.30  & $3.3\times10^{-5}$  & \bf{99\%}  & 0.25  & 0.13 & 0.22 & 13\% & $-0.16$ \\
\boldsymbol{$A_V^L$}                         & 357            & 0.36               & $3.0\times10^{-9}$               & \bf{100\%}   & 0.35  & 0.12 & 0.32 & 7\% & $-0.02$  \\
\boldsymbol{$n^L$}                                        & 357            & 0.22                & $9.8\times10^{-4}$                & {\bf 84\%}        & 0.18  & 0.14 & $0.15$ & 17\%  & $-0.16$  \\
\boldsymbol{$t_{\mathrm{age}}^L$}\bf{(Gyr)}  & 357            & 0.22  & $9.4\times10^{-4}$  & \bf{94\%}  & $-0.21$ & 0.11 & 0.44 & 5\%  & 0.08  \\
$\tau^L$(Gyr)                                & 357            & $0.13$                & $0.16$                & 17\%    & $-0.09$    & $0.13$  & $0.24$  & 9\%  & $0.07$ \\
$Z_*^L/Z_{\sun}$                             & 357            & 0.11                & $0.31$                & 7\%        & $0.05$ & 0.11 & 0.45 & 4\% & $-0.01$  \\
\hline                                         \multicolumn{10}{c}{\textbf{Global properties}}\\
\hdashline              
M$_*^G$/\Msun                                & 613            & 0.12                & 0.045                & 40\%       & 0.07  & 0.11 & 0.21 & 15\% & 0.14  \\
SFR$_0^G$ (\Msun /yr)                         & 613            & 0.07                & 0.52                & 2\%        & 0.03  & 0.09 & 0.41 & 6\%  & 0.07  \\
sSFR$_0^G$ (yr$^{-1}$)                       & 613            & 0.08                & 0.41                & 5\%         & 0.00  & 0.09 & 0.43 & 3\%  & $-0.01$ \\
$A_V^G$                         & 613            & 0.11                & 0.12                & 21\%   & 0.07  & 0.09 & 0.38 & 5\%  & $-0.02$  \\
$n^G$                                        & 613            & 0.08                & 0.34                & 5\%         & 0.01  & 0.10 & 0.29 & 9\%  & $0.03$ \\
$t_{\mathrm{age}}^G$(Gyr)                    & 613            & 0.08                & 0.34                & 7\%         & $-0.01$  & 0.08 & 0.46 & 5\% & 0.09  \\
$\tau^G$(Gyr)                                & 613            & 0.09                & 0.24                & 12\%         & 0.06  & 0.09 & 0.42 & 4\%  & $0.00$  \\
$Z_*^G$/\Zsun                                & 613            & 0.09              & 0.23                & 13\%        & $-0.07$ & 0.14 & 0.19 & 12\% & 0.11  \\
\hline
\end{tabular}
\tablefoot{The null hypothesis of the K-S test is that the \naid EW of two samples divided according to a value between the 40\% and 60\% percentile of the galaxy property indicated in the leftmost column comes from the same parent population. The K-S statistic, $D$, the $p$-value, the probability $\mathbf{P}$ of the $p$-value being lower than 0.05 according to a bootstrap (see~\S~\ref{sec:res}), and the correlation $r_s$ are shown for the host and MW. Significant rejections of the hypothesis ($\mathbf{P}>50\%$) are highlighted in bold. 
}
\begin{tablenotes}
\small
\item $^{\ast}$ This probability is obtained from the median of 1000 bootstrap "$z$-matched" simulations on the two samples recalculating the K-S statistic at each iteration and additionally dividing the sample in two at 10 different positions around the median (40-60\% percentile).
\item $^{\dag}$ The T-type samples are divided between spirals and ellipticals (at a fixed T-type $=0$) instead of the range around the median ($<$T-type$>\sim 2$).
\item $^{\oplus}$ The number of SN hosts with local properties is smaller than for global because of the upper redshift limit of 0.02 (see \S~\ref{sec:galphot}).
\end{tablenotes}
\end{threeparttable}
\end{table*} 

Besides the offset location from the galaxy, we find strong evidence for a difference between the EW of the \naid doublet among morphological galaxy types, with a median K-S p-value of $10^{-6}$ (or 100\% probability of being different), passive galaxies host SNe with less sodium absorption. In comparison, younger spirals tend to have stronger absorption. The inclination of the galaxy also plays a substantial role in differentiating the column density of sodium gas: as expected in highly inclined galaxies (viewed edge-on), the column density is significantly higher than in less inclined hosts (viewed face-on) as given by a K-S p-value of $\sim10^{-4}$ ($\mathbf{P}_{MC} = 98\%$). As the DLR considers both semi-major and semi-minor axes to get the vector distance to the SN position, it includes the inclination of the host galaxy partly. By normalising through the DLR, we subtract part of this influence, and that is why the normalised offset is indeed a better tracer of EW. In fact, if we do a K-S test for only face-on galaxies with inclinations larger than 70\textdegree, we find that the DLR is indeed stronger ($D=0.37$, p-value$=4.1\times10^{-4}$ and $\mathbf{P}_{MC}=95\%$) than the normalised offset ($D=0.34$, p-value$=1.8\times10^{-3}$ and $\mathbf{P}_{MC}=89\%$), showing that the inclination plays an important role in the EW of SNe.

\begin{figure*}
\centering
\includegraphics[width=0.493\textwidth]{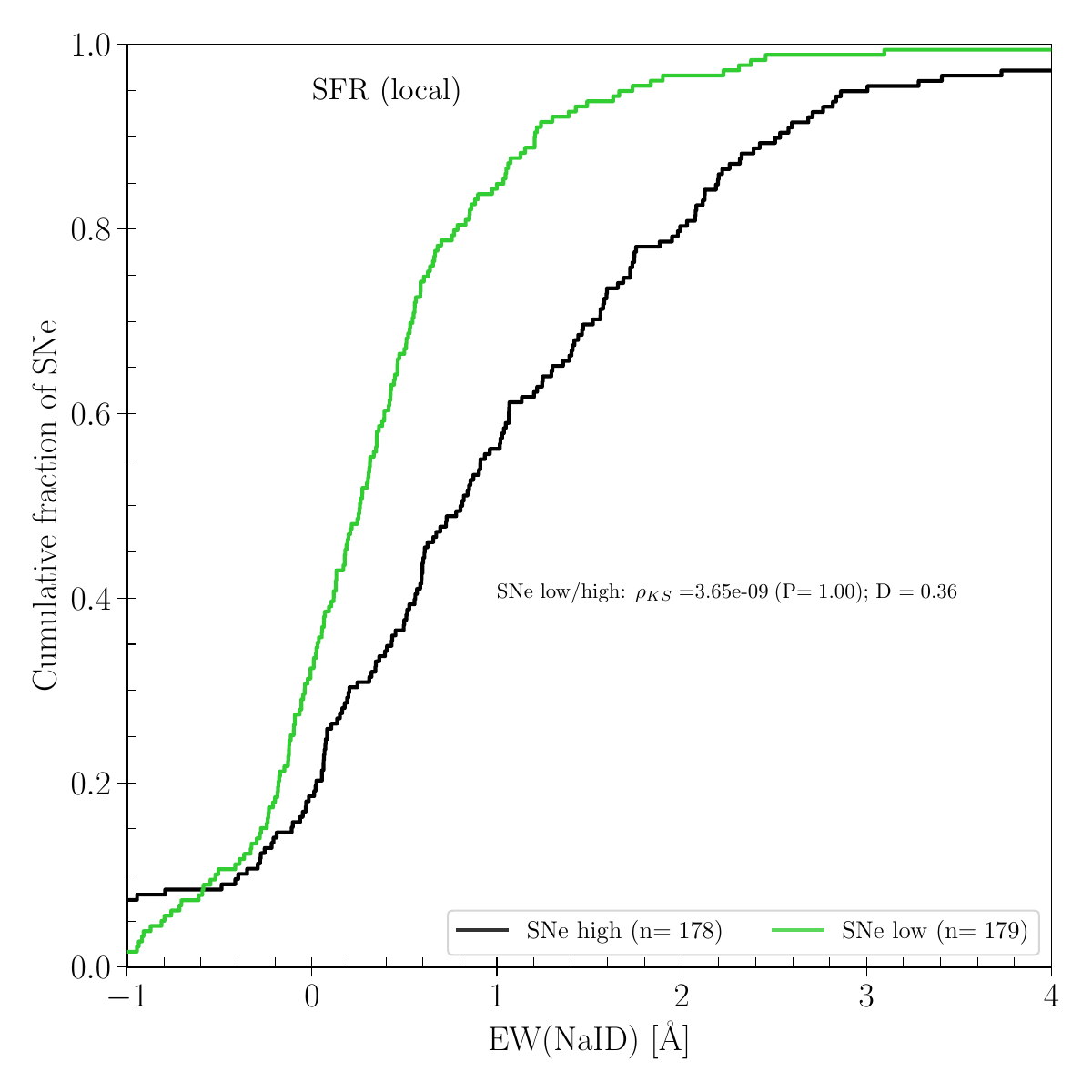}
\includegraphics[width=0.49\textwidth]{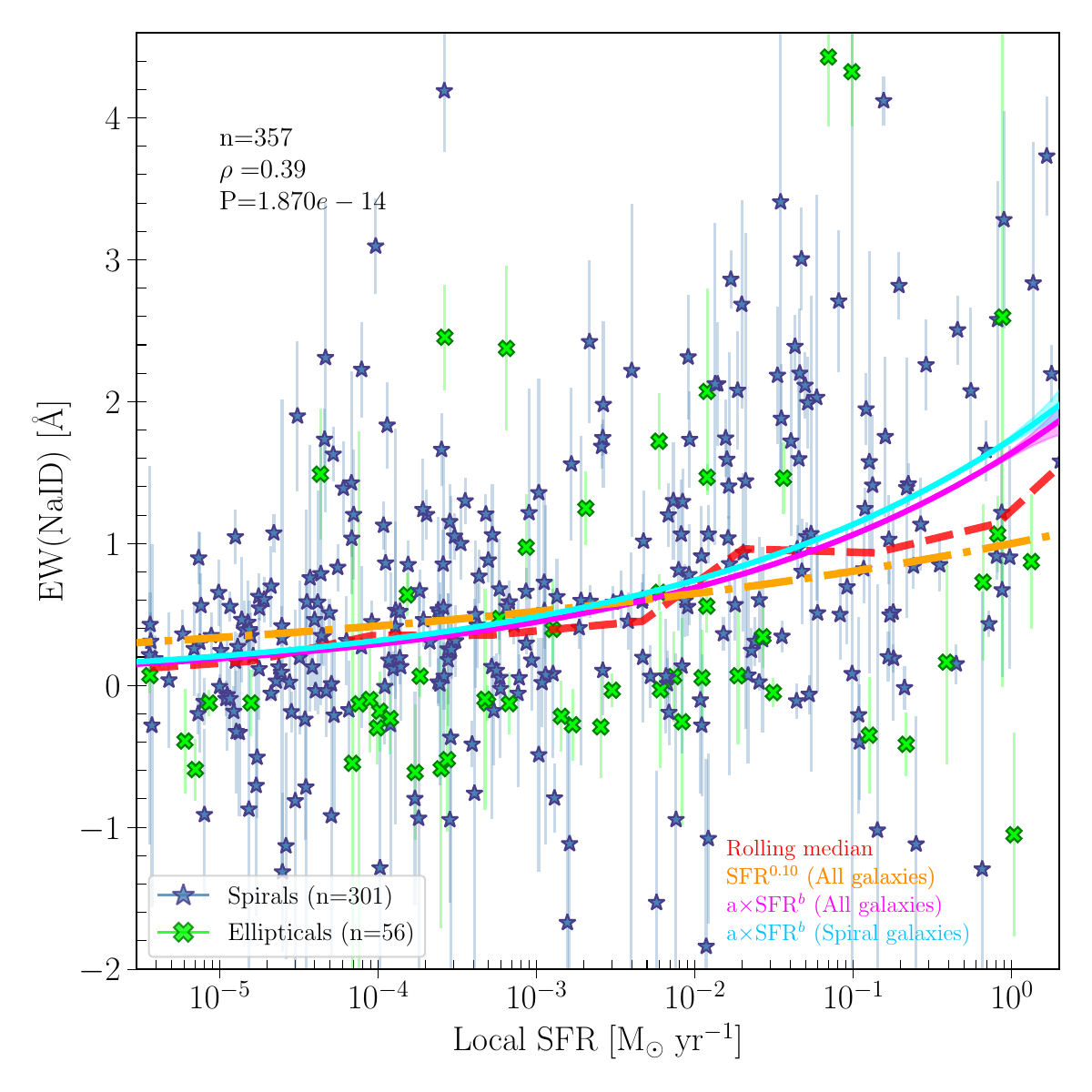}
\caption{\emph{Left:} Cumulative distributions of \naid\ EW divided into two bins according to local SFR: SNe in high and low SFR regions are shown in black and green, respectively. \emph{Right:} \naid EW vs local SFR ($\log$ scale). Local environments of spiral/elliptical galaxies are shown in blue/green stars/crosses. The red dashed line is the median, the purple is a power-law fit, and the orange is a power-law from a symbolic regression (see \S~\ref{sec:discussion-importance}). 
}
\label{fig:locsfr}
\end{figure*}

\begin{figure*}
\centering
\includegraphics[width=0.485\textwidth]{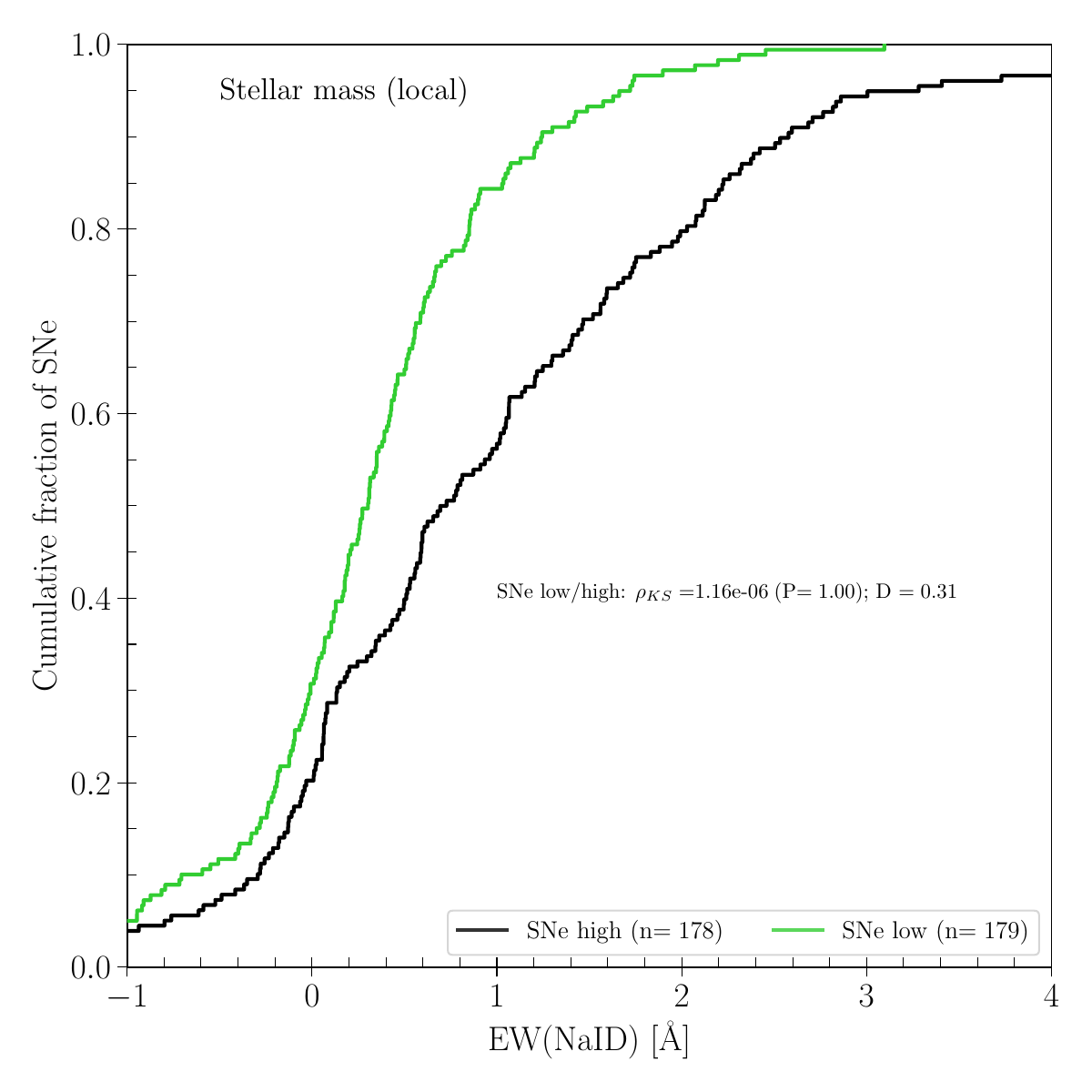}
\includegraphics[width=0.49\textwidth]{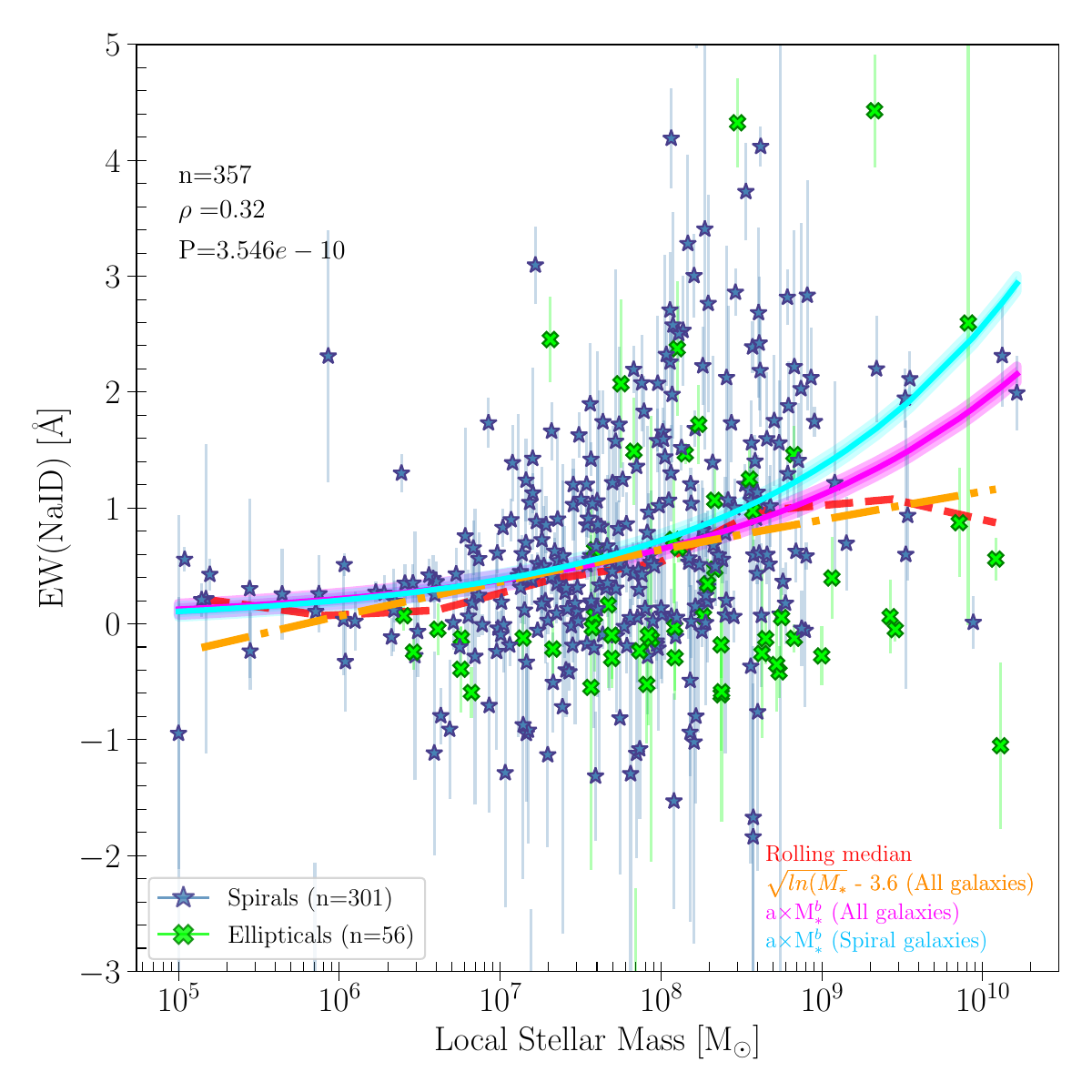}
\caption{\emph{Left:} Cumulative distributions of \naid\ EW divided into two bins according to local stellar mass: SNe in high and low stellar mass regions are shown in black and green, respectively. \emph{Right:} \naid EW vs local stellar mass ($\log$ scale). Local environments of spiral/elliptical galaxies are shown in blue/green stars/crosses. The red dashed line is the rolling median, the purple is a power-law fit, and the orange is a power-law from a symbolic regression (see \S~\ref{sec:discussion-importance}).}
\label{fig:locmass}
\end{figure*}

Regarding the stellar and dust parameters obtained from the local ($r=0.5$ kpc) SED fits, we find strong differences ($\mathbf{P}_{MC} > 80\%$) in EW distribution for the recent ($<100$ Myr) star formation rate SFR$_0^L$, the stellar mass $M_*^L$, the dust attenuation $A_V^L$ and dust slope $n$, the stellar age $t_\mathrm{age}^L$ and the specific star formation rate sSFR$_0^L$. Higher EWs of \naid are found in star-forming, more massive, younger and dustier (more attenuated) environments. An example for the local SFR is shown in Figure~\ref{fig:locsfr}, and local stellar mass is shown in Figure~\ref{fig:locmass}. 

For the global stellar and dust properties, none of the parameters show strong significance. 
This clearly suggests that the SN narrow line characteristics are indicators of the \emph{local} galactic environment rather than the global properties of the hosts. Interestingly, of the fitted global properties, the stellar mass seems to be the best indicator of the ISM ($\mathbf{P}_{MC}=40\%$), even more than the dust attenuation ($\mathbf{P}_{MC}=21\%$).    

\begin{table}
\centering
\caption{Significant K-S statistics and correlations for \naid VEL divided according to galaxy properties.}
\label{table:velKS}
\renewcommand{\arraystretch}{1.4}
\begin{tabular}{c|c|cccc}
\hline
Property & Nr & $<D^{\mathrm{VEL}}_{\mathrm{MC}}>$ & $<p_{\mathrm{MC}}^{\mathrm{VEL}}>$ & $\mathbf{P}$ & $r_s^{\mathrm{VEL}}$  \\
\hline
\hline
\boldsymbol{$\overline{\Delta\alpha}$} & 462 & 0.16 & 0.011 & {\bf 71\%} & $-0.08$ \\ 
\hline
\end{tabular}
\tablefoot{Similar to Table~\ref{table:KS} for the \naid VEL (for SNe with EW$>|0.3|$\AA\,) only for galaxy properties with low K-S p-values ($\mathbf{P} > 35$\%) and significantly low p-values highlighted in bold ($\mathbf{P} > 50$\%). Values for the MW lines are not shown, but are consistent with being drawn from the same parent population. 
}
\end{table}

\begin{figure*}
\centering
\includegraphics[width=0.48\textwidth]{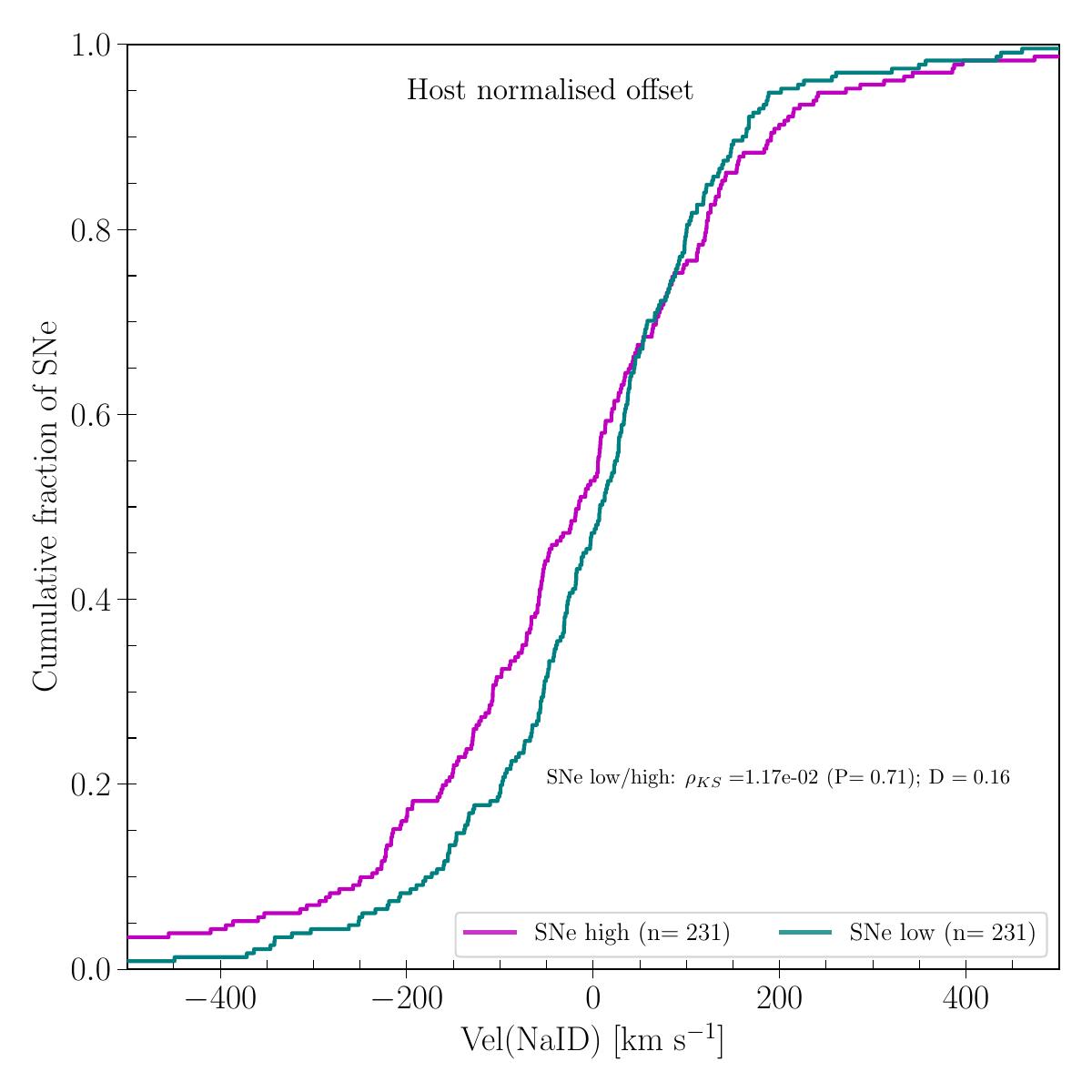}
\includegraphics[width=0.48\textwidth]{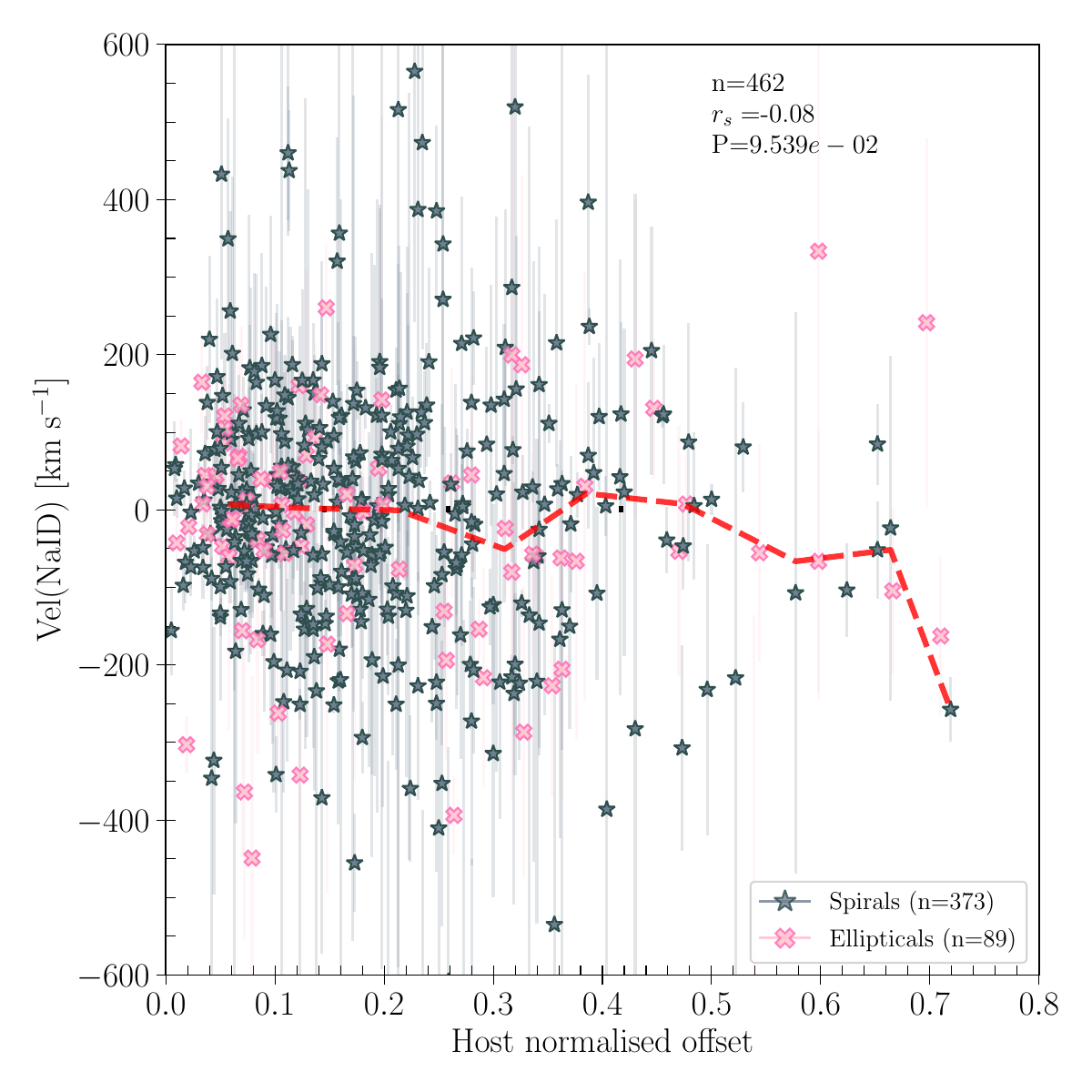}
\caption{\emph{Left:} Cumulative distributions of \naid\ VEL divided into two bins according to normalised offset from the galaxy centre: SNe with high and low offset are shown in purple and blue, respectively. \emph{Right:} \naid VEL vs normalised offset. The red dashed line is the median. 
}
\label{fig:vel}
\end{figure*}

In Table~\ref{table:velKS}, we also show the results for the K-S tests and correlations of the most significant \naid\ VEL distributions. We note that the velocities of very weak lines (i.e. EW$<|0.3|$\AA) are very hard to measure and thus not considered in these statistics (see Appendix~\ref{ap:EW-VEL}). We see that the normalised offset again has the lowest p-value of the K-S test, which can also be seen in Figure~\ref{fig:vel}: high/low-offset SNe are generally more blue/red-shifted with respect to the SN systemic velocity. However, there is a large dispersion, and the correlation between both parameters is weak. In fact, the blueshift is driven mainly by fewer SNe at high offsets. The VEL distributions, divided according to other galaxy properties as well as to the EW, are not significantly different from each other.

Lastly, we confirm some of the trends found with \naid\ EW with absorption lines from other species, in particular for \caii H\&K, the DIB-5780 and \ki\ divided according to the normalised offset, local SFR and age, and a few other properties shown in Table~\ref{table:KS-DIB5780}.  Interestingly, we find that the distributions of the EW of DIB-5780, divided according to global stellar mass and age, are significantly different. We note that the more significant relations are seen for the lines that are generally stronger (see Table~\ref{table:medlines}). However, they  are generally weaker than \naid\ and, in low-resolution spectra, harder to measure.  

\begin{table}
\centering
\caption{Significant K-S statistics and correlations for the EW and VEL of narrow lines (other than \naid) divided according to galaxy properties.}
\label{table:KS-DIB5780}
\renewcommand{\arraystretch}{1.4}
\begin{tabular}{c|c|cccc}
Property &  Nr & $<D_{\mathrm{MC}}>$ & $<p_{\mathrm{MC}}>$ & $\mathbf{P}$ & $r_s$  \\
\hline
\hline
\multicolumn{6}{c}{\textbf{Host \caii H EW}}\\
\hdashline    
\boldsymbol{$\overline{\Delta\alpha}$} & 226 & 0.29 & $6.6\times10^{-4}$ & {\bf 92\%} & -0.27 \\ 
{\bf M$_*^L$} & 137 & 0.29 & 0.014 & \bf{62\%} & 0.19 \\ 
\bf{SFR$_0^L$} & 137 & 0.30 & $9.7\times10^{-3}$ & \bf{70\%} & 0.28 \\
\bf{sSFR$_0^L$} & 137 & 0.28 & 0.022 & \bf{54\%} & 0.27 \\
\boldsymbol{$t_{\mathrm{age}}^L$} & 137 & 0.29 & 0.013 & \bf{60\%} & $-0.21$ \\
\hline
\multicolumn{6}{c}{\textbf{Host \caii K EW}}\\
\hdashline    
\boldsymbol{$\overline{\Delta\alpha}$} & 170 & 0.25 & $0.024$ & {\bf 58\%} & -0.26 \\ 
\bf{SFR$_0^L$} & 106 & 0.36 & $7.0\times10^{-3}$ & \bf{77\%} & 0.16 \\
sSFR$_0^L$ & 106 & 0.32 & 0.031 & 49\% & 0.21 \\
\boldsymbol{$t_{\mathrm{age}}^L$} & 106 & 0.34 & 0.010 & \bf{69\%} & $-0.24$ \\
\hline
\multicolumn{6}{c}{\textbf{Host \caii K VEL}}\\
\hdashline  
$i(^{\circ})$ & 113 & 0.30 & 0.038 & 43\% & 0.20 \\
\hline     
\multicolumn{6}{c}{\textbf{Host \ki 1 EW}}\\
\hdashline
\boldsymbol{$\overline{\Delta\alpha}$} & 437 & 0.21 & $3.9\times10^{-4}$ & {\bf 96\%} & $-0.22$ \\
{\bf M$_*^L$} & 281 & 0.20 & 0.019 & \bf{58\%} & 0.16 \\   
\bf{SFR$_0^L$} & 281 & 0.24 & $2.0\times10^{-3}$ & \bf{85\%} & 0.18 \\
\hline
\multicolumn{6}{c}{\textbf{Host DIB 5780 EW}}\\
\hdashline                               
\boldsymbol{$\overline{\Delta\alpha}$} & 488 & 0.20 & $6.2\times10^{-4}$ & {\bf 92\%} & -0.12 \\ 
M$_*^L$                        & 303            & 0.18  & $0.032$  & 47\%  & 0.16  \\
$A_V^L$ & 303 & 0.17 & 0.046 & 40\% & 0.17 \\
\boldsymbol{$M_*^G$}                        & 466            & 0.16  & $0.015$  & \bf{62\%}  & 0.16  \\
\boldsymbol{$t_{\mathrm{age}}^G$} & 466 & 0.16 & 0.012 & \bf{58\%} & 0.08 \\
\hline
\multicolumn{6}{c}{\textbf{Host DIB 5780 VEL}}\\
\hdashline
M$_*^L$                        & 113            & 0.29  & $0.039$  & 40\%  & 0.25  \\
$A_V^G$ & 208 & 0.22 & 0.034 & 47\% & $-0.09$ \\

\hline
\multicolumn{6}{c}{\textbf{Host DIB 6283 EW}}\\
\hdashline
T-type & 560 & 0.21 & 0.040 & 49\% & 0.11 \\
$t_{\mathrm{age}}^L$  & 342            & 0.17  & 0.041  & 41\%  & $-0.12$ \\
\boldsymbol{$n^L$} & 342 & 0.18 & 0.020 & {\bf 58\%} & 0.14 \\
\end{tabular}
\tablefoot{Similar to Table~\ref{table:KS} for \caii H \& K, \ki 1 \& 2 and the DIB 5780, 4428 and 6283 lines only for galaxy properties with low K-S p-values ($\mathbf{P} >$ 40\%) and significantly low p-values ($\mathbf{P} >$ 50\%) highlighted in bold. Values for the MW lines are not shown, but are consistent with being drawn from the same parent population.}
\end{table}

\section{Discussion}
\label{sec:discussion}

We now move to the discussion of the results found in the previous section. We analyse the results of all SNe together as a function of galaxy properties in Section~\ref{sec:discussion-all}, trying to find the most relevant relations in Section~\ref{sec:discussion-importance}. In Section~\ref{sec:discussion-nonISM}, we comment on the influence of non-ISM, SN-related physics in the obtained EW and VEL distributions. 

\subsection{Comparison to previous galaxy studies}\label{sec:discussion-all}

This work confirms the findings of multiple ISM characteristics and galaxy properties. Firstly, we find that the column density of ISM lines decreases sharply with galactic radius. Despite the heterogeneous distribution of the ISM structured in gas clouds, \ion{H}{ii} regions, voids and filaments, the ISM in spiral galaxies is, on average, distributed in a disk with an exponential profile decreasing with a radius similar to the distribution of stars \citep[e.g.][]{Bianchi07, Munoz-Mateos11, Casasola17}. We also see a qualitative decrease of \naid EW with radius, as seen in Figure~\ref{fig:offnorm}, despite the numerous different galaxy types and viewing angles that are showcased. Although the dispersion is too large to allow for a quantitative fit to all SNe, we can perform an exponential fit to the median values, finding a relation: EW$(r) = A\exp(-r/r_0)$, with $r$ being the normalised offset, $r=\overline{\Delta\alpha}$. We find $A=2.1\pm0.2$\,\AA\ and a scale length of $r_0=0.13\pm0.02$. Taking the average size of $\sim$28.3 kpc for disc galaxies \citep{Goodwin98}, we obtain an average ISM scale length of 3.7 kpc, which is of the order of the average stellar disc scale length of 3.8 kpc \citep{Fathi10}. It is worth mentioning that the \ki 1 line and DIB-5780 also show significant K-S tests when divided by the offset. Moreover, a relation of sodium EW strength and offset was also previously reported for SNe~Ia \citep{Clark21}.

A second set of galaxy properties that show strong differences in ISM abundance are related to the stars: the star formation rate and stellar age. It is well documented that star-forming regions have a higher gas content from which stars form \citep[e.g.][]{Frerking82, Orellana17}, thus explaining the higher EW of \naid in star-forming regions (see Figure~\ref{fig:locsfr}). A power-law behaviour also seems to represent the median of the data well: EW$=c\times\mathrm{SFR}^{\gamma}$ with $c=0.19\pm0.03$\,\AA\ and $\gamma=1.64\pm0.14$. Interestingly, \citet{Feldmann20} finds a power-law relation for the H$_2$ gas mass following a power-law relation with SFR of $\gamma=0.76$ for star-forming galaxies. If we remove elliptical galaxies from the fit, we obtain a consistent value with the full sample of $\gamma=1.74\pm0.15$. Likewise, stellar age also relates to column density \citep[e.g.][]{Lin20} because younger stars have had less time to move away from their birthplace in gas- and dust-abundant star-forming regions, which in turn disperse their surroundings after a short timescale of $<$30 Myr \citep{Chevance20}. This could partly explain why, according to our K-S tests, the age is a poorer tracer of the ISM than the recent SFR.

Galaxy morphology is also known to relate to gas fraction \citep{Roberts94, Namiki21}, and it is interesting to see that the galaxy T-type is a much cleaner tracer of its abundance than are global characteristics such as star formation or the age of the galaxy. The T-type is, in fact, the strongest global property found in this study, perhaps indicating a hint that morphological type is more representative of the ISM properties \citep{Davies19}. However, stellar population parameters from SEDs can be highly degenerate \citep[e.g.][]{Bell01, Walcher11}, whereas the morphology of nearby galaxies is quite robust.

The stellar mass of the galaxy is known to strongly correlate with the dust mass \citep{Garn10} and gas fraction \citep{Morokuma-Matsui15}, an observation that we confirm here locally with the EW of the \naid absorption lines. The local dust attenuation is not a stronger tracer than stellar mass, possibly because of systematic effects in dust attenuation estimates \citep{Qin22} arising partly from scattering effects in integrated light \citep{Duarte25}. We note that \citet{Feldmann20} finds that the gas mass also relates to stellar mass with a power-law relation of $\gamma_M=0.28$. We fit a power-law to the EW with stellar mass, EW$=c_M\times$M$_{*}^{\gamma_M}$, obtaining: $c_M=0.25\pm0.04$\,\AA\ and $\gamma_M=0.004\pm0.003$. When restricting to SF galaxies, we obtain $\gamma_M=0.006\pm0.005$ (see Figure~\ref{fig:locmass}). 
We note that the local dust slope index also shows a trend with EW, which is expected as it has a strong correlation with the attenuation, $A_V$ \citep{Qin22, Duarte23}.


The inclination of the galaxy with respect to the observer also plays an important role: galaxies viewed edge-on have a higher column density of dust and gas than face-on galaxies \citep[e.g.][]{Holmberg75, Tuffs04, Yuan21}. Although this effect is quite noticeable in our dataset with very low p-values in their K-S tests, it is less dominant than the T-type and the local SFR, stellar mass and dust attenuation.

It is noteworthy that the inferred stellar metallicity does not show signs of relation with the \naid\ abundance. Gas-phase and stellar metallicities are closely related \citep{Gallazzi05,Fraser-McKelvie22}, and previous studies have shown that their distribution in galaxies also follows an exponential decline with some deviations \citep{Sanchez-Menguiano18,Bresolin19,Easeman22}. We attribute this lack of correlation to the poor stellar metallicity estimates from {\sc prospector}, as shown also in \citet[][see their appendix B]{Duarte23}\footnote{In fact, we do not recover the mass-metallicity relation neither for our local nor global stellar fits.}.

The fact that the integrated stellar mass is a stronger tracer of the EW than the global dust attenuation reflects the difficulty in obtaining dust properties from stellar population fits \citep{Duarte25}, and perhaps indicates how mass carries out important dust information, as also found with the "mass-step" in SN~Ia cosmology standardisation \citep[e.g.][]{Brout21,Gonzalez21}.

Lastly, galactic gas outflows originate from ISM gas clouds that are accelerated outwards to speeds of $10^2-10^3$ km/s by initially tenuous, fast winds (from stellar winds and SNe); their velocities have been shown to strongly correlate with SFR, with more star-forming galaxies showing more blueshifted absorption \citep{Rupke05b, Martin05}. We find no evidence for such a relation in our data. 
On the other hand, we find that \naid\ outflow velocities become slightly more significant (more blueshifted) at large distances from the galaxy centre, a trend that has also been both observed \citep{Xu23} and theoretically predicted \citep{Fielding22}. However, as previously stated, in our case this is mainly driven by few SNe at the extremes of the population.

The trends in the literature mentioned above that we confirm with our SN line indicators were obtained independently by the respective authors using other ISM tracers different from ours. Only a few works have also used \naid absorption lines consistent with our study. \citet{Chen10} analysed a large sample of stacked star-forming galaxy spectra from SDSS to show how the strength of the narrow \naid\ interstellar line from the galaxies themselves depends on galaxy physical properties and to look for evidence of galactic winds. They measured the velocity and EW of the outflow component and the EW of the galaxy systemic velocity. They found that the EW of the outflow component is smaller at higher inclinations, whereas the opposite behaviour was observed for the systemic EW. For inclinations higher than 60\textdegree, they found that the velocity drops abruptly. Regarding the physical properties of the galaxy, they found that \naid\ EW of the systemic and outflow components increase strongly and nearly linearly with SFR, M$_*$ and A$_V$.

In contrast to \citet{Chen10}, we cannot divide the lines into different components due to the low spectral resolution. Nonetheless, although our sample shows a trend between the EW and the galaxy inclination (but with a much weaker correlation), it does not show a trend with VEL. For the galaxy properties, we find very low p-values but weak correlations ($p\sim10^{-8}$, $r_s=0.35$) between \naid\ EW and M$_*$ as well as with A$_V$ ($p\sim10^{-7}$, $r_s=0.26$) and a very significant low p-value and moderate correlation with SFR ($p\sim10^{-11}$, $r_s=0.45$)\footnote{For the significance of the correlation, we use the same ranges as those presented in \citet{Gutierrez17b}.}. Although our correlations are not as strong as those found by \citet{Chen10}, they follow the same trend (low \naid\ EWs at low SFR, M$_*$ and A$_V$). Surprisingly, \citet{Chen10} did not find a correlation between \naid\ EW and the sSFR, while we find that these two parameters weakly correlate ($r_s=0.27$). If we consider only star-forming galaxies (T-type $> 0$), we find stronger correlations between \naid\ EWs and A$_V$ ($r_s=0.38)$ and M$_*$ ($r_s=0.42$). For SFR, the correlation is weaker ($r_s=0.39$). Our correlations are weaker, probably because we have data for individual sight lines, whereas \citet{Chen10} use the integrated light of entire galaxies, which averages out local variations. 




\subsection{Importance of each galaxy parameter}
\label{sec:discussion-importance}

As shown in the K-S tests of Table~\ref{table:KS}, several galactic parameters strongly influence the \naid EW. The causality of each of these variables in the column density of the ISM gas is difficult to trace as many galaxy properties are actually correlated with each other, e.g. star-forming environments that are more massive are simultaneously more metal-rich, have more dust content and are forming stars more actively than less massive ones. We perform several statistical and machine learning tests on our dataset to infer the driving galactic properties and to what extent they can predict the EW.

\begin{itemize}
\item \textbf{Key driver analysis}: We use a key driver analysis \citep[KDA;][]{kda}\footnote{Key driver analysis is also referred to as Importance Analysis or Relative Importance Analysis and it uses Johnson's Relative Weights, a combination of principal component analysis on the predictor variables and least square regression to predict the target quantity.} which quantifies the relative importance of a set of variables, here the environmental properties, in predicting a target quantity, in this case the EW of \naid. The results are shown in Table~\ref{table:kda} for three cases: all galactic properties, only local properties, and only global properties. We show the five most relevant properties in each case. As expected, local properties, particularly the normalised offset and the local star formation rate, are generally more important, although the galaxy T-type plays a substantial role in the EW. It is interesting to note that the absolute score of the KDA given by $R^2$, i.e., the proportion of variance explained, using both local \emph{and} global information, is still substantially better than using only local information. This means there is some information in the galaxy T-type (third most important) and the galaxy inclination (sixth most important) that is not present uniquely in the local properties inferred in this study. Additionally, it is worth mentioning that even the highest $R^2$ score (0.36) is far from the optimal fully predictable scenario, meaning that it is difficult to recover the EW solely from the properties considered here. 
To compare, the following six most relevant features according to a gradient-boosting tree algorithm confirm our KDA findings: $\overline{\Delta\alpha}$, M$_*^L$, SFR$_0^L$, T-type, $A_V^L$ and  $t_{\mathrm{age}}^L$.

As for the velocity of \naid, we find a low $R^2$ of 0.073 driven mainly by inclination (26\%), normalised offset (23\%) and global dust (13\%). According to the KDA, the EW contributes less than 10\% in importance to the VEL.
\begin{table}
\centering
\caption{Key driver analysis of the relative importance of galaxy properties for the \naid EW.}
\label{table:kda}
\renewcommand{\arraystretch}{1.5}
\begin{tabular}{ccc}
Feature & \makecell{Relative \\ importance}& \makecell{Normalized \\ relative \\ importance (\%)} \\
\hline
\hline
\multicolumn{3}{c}{\textbf{All properties - $R^2$ score = 0.36}}\\
        $\overline{\Delta\alpha}$ &   0.079  &   30.7 \\
        SFR$_0^L$  &   0.043  &   16.8 \\
          T-type  &   0.034  &   13.2 \\
       $A_V^L$ &   0.032   &  12.5 \\
    $t_{\mathrm{age}}^L$  &   0.018   &   6.9 \\
 \hline
\multicolumn{3}{c}{\textbf{Local properties - $R^2$ score = 0.24}}\\
    $\overline{\Delta\alpha}$ &   0.081  &   40.4 \\
           $A_V^L$ &   0.043   &  21.4 \\
    SFR$_0^L$  &   0.039  &   19.4 \\
 $t_{\mathrm{age}}^L$ & 0.020 & 9.9 \\
$n^L$ & 0.009 & 4.2 \\
\hline
\multicolumn{3}{c}{\textbf{Global properties - $R^2$ score = 0.11}}\\
T-type  &   0.029  &   45.0 \\
$i(^{\circ})$ & 0.018 & 27.3 \\
$A_V^G$  &   0.004   &   6.8 \\
SFR$_0^G$ &   0.004   &  6.2 \\
$Z_*^G$ & 0.003 & 5.4 \\
\hline                                         
\end{tabular}
\tablefoot{We show only the five most relevant environmental properties in predicting the \naid EW for three cases: a) all properties, b) local properties and c) global properties. Of the offset parameters, we only used the normalized offset as input.}
\end{table}

\item \textbf{Symbolic regression}: Symbolic Regression \citep[see, e.g.][]{Angelis23} is a machine learning tool to find a mathematical expression relating input features, here the galaxy properties that optimize the prediction of a target, here the EW of \naid, with the least possible mathematical complexity. The algorithm\footnote{This is a multi-population evolutionary algorithm which applies behaviours of biological organisms such as reproduction, mutation, and recombination and follows a natural selection process to solve an optimisation problem.} searches for combinations of mathematical operators (e.g. addition, $+$, or power, \verb!^!), analytical functions (e.g. $\cos$ or $\log$) and constants relating the input galaxy properties to minimize a loss function: LOSS$=\mathrm{w}(\mathrm{EW}_{\mathrm{pred}}-\mathrm{EW}_{\mathrm{obs}})^2$, with the weights given by: $\mathrm{w}=1/\sigma_{\mathrm{EW}}^2$. We find that all the recovered expressions have an RMS larger than 0.7\,\AA, showing that it is difficult to accurately predict the full diversity of \naid EW from our galactic properties. On the other hand, the equation with the best score is only dependent on the normalised offset and is an exponential decreasing relation as expected for a typical ISM radial distribution (see previous section): EW$(r)=\exp{(-r/0.31)}$ (see orange line in Figure~\ref{fig:offnorm}) with a larger ISM scale length than when doing a fit, and larger than the disc scale length, as found in \citet{Casasola17}. If we remove the offset from the input properties, we recover a power-law relation with SFR, i.e. EW$=$SFR$^{\gamma}$ with $\gamma=0.10$, shown in Figure~\ref{fig:locsfr}. Similarly, after removing both offset and SFR, we find a power-law relation with log stellar mass, EW$=(\log M_*)^{0.5}-3.6$ shown in orange in Figure~\ref{fig:locmass}. Thus, we confirm with symbolic regression all trends found in the literature and that we previously fitted, although the relations are shallower, as can be seen in the orange lines of the Figures. Regressions relating the EW to a combination of more of these parameters are not preferred by the algorithm.

Regarding the VEL prediction with Symbolic Regression, we find that it is even more difficult to obtain than the EW. The heterogeneous equations involve different variables such as the normalised offset, the local SFR or age, but the RMS is larger than 150 km/s, making any prediction very unreliable.

\end{itemize}

\subsection{Non-ISM contamination}
\label{sec:discussion-nonISM}

It has been previously reported that the narrow absorption lines of SNe can be related to the intrinsic properties of the explosions and their progenitors. Evolving \naid\ lines in SN~Ia spectra are associated with CSM \citep[e.g.][]{Ferretti16} or to the interaction of the SN radiation with very nearby ISM \citep[e.g.][]{Hoang19}. In either case, the EW measurement will be dependent on the SN and thus not a trustworthy tracer of the ISM. Other lines, such as \ki\ or even DIBs, have also shown evolution during SN lifetimes \citep{Graham15, Milisavljevic14}. Although statistically, the evolution of intervening SN lines in our sample seems rather absent (see Paper I), the strength of the EW could be affected by SN-interacted nearby material. Indeed, \citet{Phillips13} and \citet{Maguire13} find that there is an excess absorption for possible candidates of SNe with CSM. 
Regarding the velocity, blueshifted emission lines have been used as indicators of expelled material from the progenitor in various SNe, notably in interacting SNe \citep[e.g.][]{Fransson02} but also in narrow absorption lines of SNe~Ia \citep[e.g.][]{Sternberg13}. This material could potentially influence conclusions on galactic gas outflows if it is present.

Moreover, most of the relations found here are heavily driven by SNe~Ia, which appear in larger numbers than core-collapse (CC) SNe but, more importantly, occur in a variety of environments and are thus more representative of the ISM extent. To test for possible biases from the SN type, we divide the sample into two different main groups, SNe~Ia and CC~SNe, and repeat the study carried out for the full sample. We recover the trends with galaxy properties (although with fewer statistics and less significant p-values), as shown in Table~\ref{table:iacc}. However, we also find variations among the different types, notably for the age: SNe~Ia show strong differences in EW when dividing the sample into age populations, and CC~SNe do not. This is most likely due to the fact that SNe~Ia span a much wider range of ages ($\sim$Gyr) than CC SNe ($\sim$Myr), and in the simple SFH treatment of Equation~\ref{eq-sfh}, subtle age differences are less evident. Other smaller variations, as for the inclination, can be explained as SNe~Ia include elliptical galaxies where orientation plays a weaker role. 
The findings divided by each SN type, and the implications for progenitors will be investigated further in a forthcoming study. The unequivocal galaxy trends found in this work clearly show that SN narrow lines are \emph{statistically} robust tracers of the local ISM, but care should be taken with small samples.

\begin{table*}
\centering
\caption{K-S statistics and correlations for the EW of \naid divided according to representative galaxy properties for only SNe~Ia and only CC SNe.}
\label{table:iacc}
\renewcommand{\arraystretch}{1.4}
\begin{tabular}{c|ccccc|ccccc}
& \multicolumn{5} {c|} {\bf Ia} & \multicolumn{5} {c} {\bf CC} \\
\hline
Property & Nr & $<D^{\mathrm{EW}}_{\mathrm{MC}}>$ & $<p_{\mathrm{MC}}^{\mathrm{EW}}>$ & $\mathbf{P}$ & $r_s^{\mathrm{EW}}$  &
Nr & $<D^{\mathrm{EW}}_{\mathrm{MC}}>$ & $<p_{\mathrm{MC}}^{\mathrm{EW}}>$ & $\mathbf{P}$ & $r_s^{\mathrm{EW}}$  \\
\hline
\hline
$\overline{\Delta\alpha}$ & 349 & 0.45 & $1.8\times10^{-13}$ & {\bf 100\%} & $-0.46$ & 
296 & 0.34 & $1.4\times10^{-6}$ & {\bf 100\%} & $-0.37$ \\
$i(^{\circ})$ & 349 & 0.18 & 0.017 & {\bf 54\%} & 0.15 & 295 & 0.20 & 0.015 & {\bf 65\%} & 0.22 \\
M$_*^L$/ \Msun & 167 & 0.34 & $8.2\times10^{-4}$ & {\bf 89\%} & 0.8 & 190 & 0.29 & $1.9\times10^{-3}$ & {\bf 94\%} & 0.36\\
SFR$_0^L$ (\Msun /yr) & 167 & 0.42 & $1.1\times10^{-5}$ & {\bf 100\%} & 0.45 & 190 & 0.36 & $5.8\times10^{-5}$ & {\bf 96\%} & 0.32 \\
sSFR$_0^L$ (yr$^{-1}$) & 167 & 0.34 & $9.7\times10^{-4}$ & {\bf 92\%} & 0.36 & 190 & 0.28 & $3.5\times10^{-3}$ & {\bf 88\%} & 0.16 \\ 
$A_V^L$ & 167 & 0.30 & $3.0\times10^{-3}$ & {\bf 82\%} & 0.29 & 190 & 0.38 & $1.6\times10^{-5}$ & {\bf 100\%} & 0.40 \\
$n^L$ & 167 & 0.25 & 0.031 & 46\% & 0.21 & 190 & 0.22 & 0.045 & 37\% & 0.16 \\
$t_{\mathrm{age}}^L$(Gyr) & 167 & 0.35 & $5.5\times10^{-4}$ & {\bf 96\%} & $-0.35$ & 190 & 0.20 & 0.077 & 30\% & $-0.10$\\
\hline
\end{tabular}
\tablefoot{The split between samples is done according to the 40-60\% of each sub-sample.}
\end{table*}


\section{Conclusions}
\label{sec:conc}

We conducted the first statistical study of ISM properties using the narrow absorption lines in SN spectra. With the advantage of being bright and occurring in many environments, SNe are probes of individual sight lines and, consequently, of very local ISM properties. Most notably, with the strength of the EW of \naid\, we confirm literature findings of relations between gas abundance and the distance from the galaxy centre, as well as with local (0.5 kpc) properties such as the star formation rate, the stellar mass, the dust attenuation and the stellar age. We also recover the dependence of the gas abundance with global properties like the galaxy T-type and the inclination of the galaxy.

According to our statistical K-S tests and machine learning approaches, the most important parameters determining the \naid\ EW are the normalised offset from the galaxy centre, the local star formation rate and the local stellar mass. However, the other parameters also contribute to the diversity of values. We find, both by fitting and through a symbolic regression, that the EW a) decreases exponentially with the distance from the galactic centre and b) increases as a power-law with the star formation rate and the stellar mass of the local environment. These findings agree well with previous studies based on other ISM tracers.

We conclude that narrow absorption lines within SN spectra are appropriate for statistically investigating the ISM properties of galaxies. At the same time, in a companion paper \citep{Gutierrez24}, we show that they also trace very nearby material in the immediate vicinity of the explosion, contaminating the results of individual events or of smaller samples. In the future, with more statistics, it will be possible to increase the S/N by stacking spectra of different SNe per environmental bin, similarly to galaxy spectra, to obtain more accurate results.

\begin{acknowledgements}

We thank the anonymous referee for the comments and suggestions that have helped us to improve the paper.
S.G.G thanks FCT for financial support through Project~No.~UIDB/00099/2020 and for support from the ESO Scientific Visitor Programme.
C.P.G. acknowledges financial support from the Secretary of Universities and Research (Government of Catalonia) and by the Horizon 2020 Research and Innovation Programme of the European Union under the Marie Sk\l{}odowska-Curie and the Beatriu de Pin\'os 2021 BP 00168 programme, C.P.G. and L.G. recognise the support from the Spanish Ministerio de Ciencia e Innovaci\'on (MCIN) and the Agencia Estatal de Investigaci\'on (AEI) 10.13039/501100011033 under the PID2023-151307NB-I00 SNNEXT project, from Centro Superior de Investigaciones Cient\'ificas (CSIC) under the PIE project 20215AT016 and the program Unidad de Excelencia Mar\'ia de Maeztu CEX2020-001058-M, and from the Departament de Recerca i Universitats de la Generalitat de Catalunya through the 2021-SGR-01270 grant. A.M.G. acknowledges financial support from grant PID2023-152609OA-I00, funded by the Spanish Ministerio de Ciencia, Innovación y Universidades (MICIU), the Agencia Estatal de Investigación (AEI, 10.13039/501100011033), and the European Union's European Regional Development Fund (ERDF). S.M. acknowledges support from the Research Council of Finland project 350458.



Computations were performed at the cluster COIN, the CosmoStatistics Initiative, whose purchase was made possible due to a CNRS MOMENTUM 2018-2020 under the project "Active Learning for large scale sky surveys”.\\

This research has made use of the NASA/IPAC Extragalactic Database (NED), which is funded by the National Aeronautics and Space Administration and operated by the California Institute of Technology.\\

This project used public archival data from the Dark Energy Survey (DES), the Sloan Digital Sky Survey (SDSS), the NASA Galaxy Evolution Explorer (GALEX), the Two Micron All Sky Survey (2MASS), the Visible and Infrared Survey Telescope for Astronomy (VISTA).\\

Funding for the SDSS-V has been provided by the Alfred P. Sloan Foundation, the Heising-Simons Foundation, the National Science Foundation, and the Participating Institutions. SDSS acknowledges support and resources from the Center for High-Performance Computing at the University of Utah. SDSS telescopes are located at Apache Point Observatory, funded by the Astrophysical Research Consortium and operated by New Mexico State University, and at Las Campanas Observatory, operated by the Carnegie Institution for Science. The SDSS web site is \url{www.sdss.org}.

SDSS is managed by the Astrophysical Research Consortium for the Participating Institutions of the SDSS Collaboration, including Caltech, The Carnegie Institution for Science, Chilean National Time Allocation Committee (CNTAC) ratified researchers, The Flatiron Institute, the Gotham Participation Group, Harvard University, Heidelberg University, The Johns Hopkins University, L’Ecole polytechnique f\'{e}d\'{e}rale de Lausanne (EPFL), Leibniz-Institut f\"{u}r Astrophysik Potsdam (AIP), Max-Planck-Institut f\"{u}r Astronomie (MPIA Heidelberg), Max-Planck-Institut f\"{u}r Extraterrestrische Physik (MPE), Nanjing University, National Astronomical Observatories of China (NAOC), New Mexico State University, The Ohio State University, Pennsylvania State University, Smithsonian Astrophysical Observatory, Space Telescope Science Institute (STScI), the Stellar Astrophysics Participation Group, Universidad Nacional Aut\'{o}noma de M\'{e}xico, University of Arizona, University of Colorado Boulder, University of Illinois at Urbana-Champaign, University of Toronto, University of Utah, University of Virginia, Yale University, and Yunnan University.\\

Funding for the DES Projects has been provided by the U.S. Department of Energy, the U.S. National Science Foundation, the Ministry of Science and Education of Spain, the Science and Technology Facilities Council of the United Kingdom, the Higher Education Funding Council for England, the National Center for Supercomputing Applications at the University of Illinois at Urbana-Champaign, the Kavli Institute of Cosmological Physics at the University of Chicago, the Center for Cosmology and Astro-Particle Physics at the Ohio State University, the Mitchell Institute for Fundamental Physics and Astronomy at Texas A\&M University, Financiadora de Estudos e Projetos, Funda{\c c}{\~a}o Carlos Chagas Filho de Amparo {\`a} Pesquisa do Estado do Rio de Janeiro, Conselho Nacional de Desenvolvimento Cient{\'i}fico e Tecnol{\'o}gico and the Minist{\'e}rio da Ci{\^e}ncia, Tecnologia e Inova{\c c}{\~a}o, the Deutsche Forschungsgemeinschaft, and the Collaborating Institutions in the Dark Energy Survey.
The Collaborating Institutions are Argonne National Laboratory, the University of California at Santa Cruz, the University of Cambridge, Centro de Investigaciones Energ{\'e}ticas, Medioambientales y Tecnol{\'o}gicas-Madrid, the University of Chicago, University College London, the DES-Brazil Consortium, the University of Edinburgh, the Eidgen{\"o}ssische Technische Hochschule (ETH) Z{\"u}rich,  Fermi National Accelerator Laboratory, the University of Illinois at Urbana-Champaign, the Institut de Ci{\`e}ncies de l'Espai (IEEC/CSIC), the Institut de F{\'i}sica d'Altes Energies, Lawrence Berkeley National Laboratory, the Ludwig-Maximilians Universit{\"a}t M{\"u}nchen and the associated Excellence Cluster Universe, the University of Michigan, the National Optical Astronomy Observatory, the University of Nottingham, The Ohio State University, the OzDES Membership Consortium, the University of Pennsylvania, the University of Portsmouth, SLAC National Accelerator Laboratory, Stanford University, the University of Sussex, and Texas A\&M University.
Based in part on observations at Cerro Tololo Inter-American Observatory, National Optical Astronomy Observatory, which is operated by the Association of Universities for Research in Astronomy (AURA) under a cooperative agreement with the National Science Foundation.\\

The Pan-STARRS1 Surveys (PS1) and the PS1 public science archive have been made possible through contributions by the Institute for Astronomy, the University of Hawaii, the Pan-STARRS Project Office, the Max-Planck Society and its participating institutes, the Max Planck Institute for Astronomy, Heidelberg and the Max Planck Institute for Extraterrestrial Physics, Garching, The Johns Hopkins University, Durham University, the University of Edinburgh, the Queen's University Belfast, the Harvard-Smithsonian Center for Astrophysics, the Las Cumbres Observatory Global Telescope Network Incorporated, the National Central University of Taiwan, the Space Telescope Science Institute, the National Aeronautics and Space Administration under Grant No. NNX08AR22G issued through the Planetary Science Division of the NASA Science Mission Directorate, the National Science Foundation Grant No. AST-1238877, the University of Maryland, Eotvos Lorand University (ELTE), the Los Alamos National Laboratory, and the Gordon and Betty Moore Foundation.

GALEX is operated for NASA by the California Institute of Technology under NASA contract NAS5-98034.

2MASS, which is a joint project of the University of Massachusetts and the Infrared Processing and Analysis Center/California Institute of Technology, funded by the National Aeronautics and Space Administration and the National Science Foundation.\\

Based on data products created from observations collected at the European Organisation for Astronomical Research in the Southern Hemisphere under ESO programme 179.A-2010 and made use of data from the VISTA Hemisphere survey \citep{VISTA} with data pipeline processing
with the VISTA Data Flow System \citep{VISTA-pipeline,VISTA-pipeline2,VISTA-archive}.\\

This work has made use of data from the European Space Agency (ESA) mission {\it Gaia} (\url{https://www.cosmos.esa.int/gaia}), processed by the {\it Gaia} Data Processing and Analysis Consortium (DPAC,
\url{https://www.cosmos.esa.int/web/gaia/dpac/consortium}). Funding for the DPAC has been provided by national institutions, in particular, the institutions participating in the {\it Gaia} Multilateral Agreement.

\end{acknowledgements}

\section*{Software}
\small
This research has made use of the \textsc{python} packages  \textsc{hostphot} \citep{Muller-Bravo22} for galaxy photometry and \textsc{prospector} \citep{Johnson21-prospector,Leja17} for SED fitting. \textsc{hostphot} uses \textsc{astropy} \citep{astropy:2013,astropy:2018,astropy:2022}, \textsc{photutils} \citep{bradley23-photutils}, \textsc{sep} \citep{Barbary16-sep,Bertin96-sex}, \textsc{astroquery} \citep{Ginsburg19-astroquery}, \textsc{reproject}, \textsc{extinction} \citep{Barbary16-ext},
 \textsc{sfdmap}, \textsc{pyvo} \citep{Graham14-pyvo}, \textsc{ipywidgets}
and \textsc{ipympl6}. \textsc{prospector} also requires FSPS \citep{Conroy09,Conroy10}, \textsc{python}-FSPS \citep{johnson23-python-fsps} and \textsc{emcee} \citep{emcee}. We also made use of: \textsc{numpy} \citep{numpy}, \textsc{matplotlib} \citep{matplotlib}, \textsc{scipy} \citep{scipy} and \textsc{pandas} \citep{pandas,mckinney-pandas}. We use the python implementations of KDA: \textsc{key-driver-analysis}\footnote{\url{https://github.com/bnriiitb/key-driver-analysis}} and symbolic regression: \textsc{PySR} \citep{PySR}\footnote{\url{https://github.com/MilesCranmer/PySR}}.

%
\bibliographystyle{aa} 
\bibliography{biblio_short} 

\begin{thebibliography}{156}
\expandafter\ifx\csname natexlab\endcsname\relax\def\natexlab#1{#1}\fi

\bibitem[{{Abbott} {et~al.}(2018){Abbott}, {Abdalla}, {Allam}, {Amara},
  {Annis}, {Asorey}, {Avila}, {Ballester}, {Banerji}, {Barkhouse}, {Baruah},
  {Baumer}, {Bechtol}, {Becker}, {Benoit-L{\'e}vy}, {Bernstein}, {Bertin},
  {Blazek}, {Bocquet}, {Brooks}, {Brout}, {Buckley-Geer}, {Burke}, {Busti},
  {Campisano}, {Cardiel-Sas}, {Carnero Rosell}, {Carrasco Kind}, {Carretero},
  {Castander}, {Cawthon}, {Chang}, {Chen}, {Conselice}, {Costa}, {Crocce},
  {Cunha}, {D'Andrea}, {da Costa}, {Das}, {Daues}, {Davis}, {Davis}, {De
  Vicente}, {DePoy}, {DeRose}, {Desai}, {Diehl}, {Dietrich}, {Dodelson},
  {Doel}, {Drlica-Wagner}, {Eifler}, {Elliott}, {Evrard}, {Farahi}, {Fausti
  Neto}, {Fernandez}, {Finley}, {Flaugher}, {Foley}, {Fosalba}, {Friedel},
  {Frieman}, {Garc{\'\i}a-Bellido}, {Gaztanaga}, {Gerdes}, {Giannantonio},
  {Gill}, {Glazebrook}, {Goldstein}, {Gower}, {Gruen}, {Gruendl}, {Gschwend},
  {Gupta}, {Gutierrez}, {Hamilton}, {Hartley}, {Hinton}, {Hislop}, {Hollowood},
  {Honscheid}, {Hoyle}, {Huterer}, {Jain}, {James}, {Jeltema}, {Johnson},
  {Johnson}, {Kacprzak}, {Kent}, {Khullar}, {Klein}, {Kovacs}, {Koziol},
  {Krause}, {Kremin}, {Kron}, {Kuehn}, {Kuhlmann}, {Kuropatkin}, {Lahav},
  {Lasker}, {Li}, {Li}, {Liddle}, {Lima}, {Lin}, {L{\'o}pez-Reyes}, {MacCrann},
  {Maia}, {Maloney}, {Manera}, {March}, {Marriner}, {Marshall}, {Martini},
  {McClintock}, {McKay}, {McMahon}, {Melchior}, {Menanteau}, {Miller},
  {Miquel}, {Mohr}, {Morganson}, {Mould}, {Neilsen}, {Nichol}, {Nogueira},
  {Nord}, {Nugent}, {Nunes}, {Ogando}, {Old}, {Pace}, {Palmese},
  {Paz-Chinch{\'o}n}, {Peiris}, {Percival}, {Petravick}, {Plazas}, {Poh},
  {Pond}, {Porredon}, {Pujol}, {Refregier}, {Reil}, {Ricker}, {Rollins},
  {Romer}, {Roodman}, {Rooney}, {Ross}, {Rykoff}, {Sako}, {Sanchez}, {Sanchez},
  {Santiago}, {Saro}, {Scarpine}, {Scolnic}, {Serrano}, {Sevilla-Noarbe},
  {Sheldon}, {Shipp}, {Silveira}, {Smith}, {Smith}, {Smith}, {Soares-Santos},
  {Sobreira}, {Song}, {Stebbins}, {Suchyta}, {Sullivan}, {Swanson}, {Tarle},
  {Thaler}, {Thomas}, {Thomas}, {Troxel}, {Tucker}, {Vikram}, {Vivas},
  {Walker}, {Wechsler}, {Weller}, {Wester}, {Wolf}, {Wu}, {Yanny}, {Zenteno},
  {Zhang}, {Zuntz}, {DES Collaboration}, {Juneau}, {Fitzpatrick}, {Nikutta},
  {Nidever}, {Olsen}, {Scott}, \& {NOAO Data Lab}}]{DES-DR1}
{Abbott}, T.~M.~C., {Abdalla}, F.~B., {Allam}, S., {et~al.} 2018, \apjs, 239,
  18

\bibitem[{Abbott {et~al.}(2021)Abbott, Adamów, Aguena, Allam, Amon, Annis,
  Avila, Bacon, Banerji, Bechtol, Becker, Bernstein, Bertin, Bhargava, Bridle,
  Brooks, Burke, Carnero~Rosell, Carrasco~Kind, Carretero, Castander, Cawthon,
  Chang, Choi, Conselice, Costanzi, Crocce, da~Costa, Davis, De~Vicente,
  DeRose, Desai, Diehl, Dietrich, Drlica-Wagner, Eckert, Elvin-Poole, Everett,
  Evrard, Ferrero, Ferté, Flaugher, Fosalba, Friedel, Frieman,
  García-Bellido, Gaztanaga, Gelman, Gerdes, Giannantonio, Gill, Gruen,
  Gruendl, Gschwend, Gutierrez, Hartley, Hinton, Hollowood, Honscheid, Huterer,
  James, Jeltema, Johnson, Kent, Kron, Kuehn, Kuropatkin, Lahav, Li, Lidman,
  Lin, MacCrann, Maia, Manning, Maloney, March, Marshall, Martini, Melchior,
  Menanteau, Miquel, Morgan, Myles, Neilsen, Ogando, Palmese, Paz-Chinchón,
  Petravick, Pieres, Plazas, Pond, Rodriguez-Monroy, Romer, Roodman, Rykoff,
  Sako, Sanchez, Santiago, Scarpine, Serrano, Sevilla-Noarbe, Smith, Smith,
  Soares-Santos, Suchyta, Swanson, Tarle, Thomas, To, Tremblay, Troxel, Tucker,
  Turner, Varga, Walker, Wechsler, Weller, Wester, Wilkinson, Yanny, Zhang,
  Nikutta, Fitzpatrick, Jacques, Scott, Olsen, Huang, Herrera, Juneau, Nidever,
  Weaver, Adean, Correia, de~Freitas, Freitas, Singulani, \&
  Vila-Verde}]{DES-DR2}
Abbott, T. M.~C., Adamów, M., Aguena, M., {et~al.} 2021, The Astrophysical
  Journal Supplement Series, 255, 20

\bibitem[{{Abdurro'uf} {et~al.}(2022){Abdurro'uf}, {Accetta}, {Aerts}, {Silva
  Aguirre}, {Ahumada}, {Ajgaonkar}, {Filiz Ak}, {Alam}, {Allende Prieto},
  {Almeida}, {Anders}, {Anderson}, {Andrews}, {Anguiano}, {Aquino-Ort{\'\i}z},
  {Arag{\'o}n-Salamanca}, {Argudo-Fern{\'a}ndez}, {Ata}, {Aubert},
  {Avila-Reese}, {Badenes}, {Barb{\'a}}, {Barger}, {Barrera-Ballesteros},
  {Beaton}, {Beers}, {Belfiore}, {Bender}, {Bernardi}, {Bershady}, {Beutler},
  {Bidin}, {Bird}, {Bizyaev}, {Blanc}, {Blanton}, {Boardman}, {Bolton},
  {Boquien}, {Borissova}, {Bovy}, {Brandt}, {Brown}, {Brownstein}, {Brusa},
  {Buchner}, {Bundy}, {Burchett}, {Bureau}, {Burgasser}, {Cabang}, {Campbell},
  {Cappellari}, {Carlberg}, {Wanderley}, {Carrera}, {Cash}, {Chen}, {Chen},
  {Cherinka}, {Chiappini}, {Choi}, {Chojnowski}, {Chung}, {Clerc}, {Cohen},
  {Comerford}, {Comparat}, {da Costa}, {Covey}, {Crane}, {Cruz-Gonzalez},
  {Culhane}, {Cunha}, {Dai}, {Damke}, {Darling}, {Davidson}, {Davies},
  {Dawson}, {De Lee}, {Diamond-Stanic}, {Cano-D{\'\i}az}, {S{\'a}nchez},
  {Donor}, {Duckworth}, {Dwelly}, {Eisenstein}, {Elsworth}, {Emsellem},
  {Eracleous}, {Escoffier}, {Fan}, {Farr}, {Feng}, {Fern{\'a}ndez-Trincado},
  {Feuillet}, {Filipp}, {Fillingham}, {Frinchaboy}, {Fromenteau}, {Galbany},
  {Garc{\'\i}a}, {Garc{\'\i}a-Hern{\'a}ndez}, {Ge}, {Geisler}, {Gelfand},
  {G{\'e}ron}, {Gibson}, {Goddy}, {Godoy-Rivera}, {Grabowski}, {Green},
  {Greener}, {Grier}, {Griffith}, {Guo}, {Guy}, {Hadjara}, {Harding},
  {Hasselquist}, {Hayes}, {Hearty}, {Hern{\'a}ndez}, {Hill}, {Hogg},
  {Holtzman}, {Horta}, {Hsieh}, {Hsu}, {Hsu}, {Huber}, {Huertas-Company},
  {Hutchinson}, {Hwang}, {Ibarra-Medel}, {Chitham}, {Ilha}, {Imig}, {Jaekle},
  {Jayasinghe}, {Ji}, {Johnson}, {Jones}, {J{\"o}nsson}, {Katkov}, {Khalatyan},
  {Kinemuchi}, {Kisku}, {Knapen}, {Kneib}, {Kollmeier}, {Kong}, {Kounkel},
  {Kreckel}, {Krishnarao}, {Lacerna}, {Lane}, {Langgin}, {Lavender}, {Law},
  {Lazarz}, {Leung}, {Leung}, {Lewis}, {Li}, {Li}, {Lian}, {Liang}, {Lin},
  {Lin}, {Lin}, {Lintott}, {Long}, {Longa-Pe{\~n}a}, {L{\'o}pez-Cob{\'a}},
  {Lu}, {Lundgren}, {Luo}, {Mackereth}, {de la Macorra}, {Mahadevan},
  {Majewski}, {Manchado}, {Mandeville}, {Maraston}, {Margalef-Bentabol},
  {Masseron}, {Masters}, {Mathur}, {McDermid}, {Mckay}, {Merloni},
  {Merrifield}, {Meszaros}, {Miglio}, {Di Mille}, {Minniti}, {Minsley},
  {Monachesi}, {Moon}, {Mosser}, {Mulchaey}, {Muna}, {Mu{\~n}oz}, {Myers},
  {Myers}, {Nadathur}, {Nair}, {Nandra}, {Neumann}, {Newman}, {Nidever},
  {Nikakhtar}, {Nitschelm}, {O'Connell}, {Garma-Oehmichen}, {Luan Souza de
  Oliveira}, {Olney}, {Oravetz}, {Ortigoza-Urdaneta}, {Osorio}, {Otter},
  {Pace}, {Padilla}, {Pan}, {Pan}, {Parikh}, {Parker}, {Peirani}, {Pe{\~n}a
  Ram{\'\i}rez}, {Penny}, {Percival}, {Perez-Fournon}, {Pinsonneault},
  {Poidevin}, {Poovelil}, {Price-Whelan}, {B{\'a}rbara de Andrade Queiroz},
  {Raddick}, {Ray}, {Rembold}, {Riddle}, {Riffel}, {Riffel}, {Rix}, {Robin},
  {Rodr{\'\i}guez-Puebla}, {Roman-Lopes}, {Rom{\'a}n-Z{\'u}{\~n}iga}, {Rose},
  {Ross}, {Rossi}, {Rubin}, {Salvato}, {S{\'a}nchez}, {S{\'a}nchez-Gallego},
  {Sanderson}, {Santana Rojas}, {Sarceno}, {Sarmiento}, {Sayres}, {Sazonova},
  {Schaefer}, {Schiavon}, {Schlegel}, {Schneider}, {Schultheis}, {Schwope},
  {Serenelli}, {Serna}, {Shao}, {Shapiro}, {Sharma}, {Shen}, {Shetrone}, {Shu},
  {Simon}, {Skrutskie}, {Smethurst}, {Smith}, {Sobeck}, {Spoo}, {Sprague},
  {Stark}, {Stassun}, {Steinmetz}, {Stello}, {Stone-Martinez},
  {Storchi-Bergmann}, {Stringfellow}, {Stutz}, {Su}, {Taghizadeh-Popp},
  {Talbot}, {Tayar}, {Telles}, {Teske}, {Thakar}, {Theissen}, {Tkachenko},
  {Thomas}, {Tojeiro}, {Hernandez Toledo}, {Troup}, {Trump}, {Trussler},
  {Turner}, {Tuttle}, {Unda-Sanzana}, {V{\'a}zquez-Mata}, {Valentini},
  {Valenzuela}, {Vargas-Gonz{\'a}lez}, {Vargas-Maga{\~n}a}, {Alfaro},
  {Villanova}, {Vincenzo}, {Wake}, {Warfield}, {Washington}, {Weaver},
  {Weijmans}, {Weinberg}, {Weiss}, {Westfall}, {Wild}, {Wilde}, {Wilson},
  {Wilson}, {Wilson}, {Wolf}, {Wood-Vasey}, {Yan}, {Zamora}, {Zasowski},
  {Zhang}, {Zhao}, {Zheng}, {Zheng}, \& {Zhu}}]{Abdurrouf22}
{Abdurro'uf}, {Accetta}, K., {Aerts}, C., {et~al.} 2022, \apjs, 259, 35

\bibitem[{Anderson \& Darling(1952)}]{Anderson-Darling52}
Anderson, T.~W. \& Darling, D.~A. 1952, The Annals of Mathematical Statistics,
  23, 193

\bibitem[{Angelis {et~al.}(2023)Angelis, Sofos, \& Karakasidis}]{Angelis23}
Angelis, D., Sofos, F., \& Karakasidis, T.~E. 2023, Archives of Computational
  Methods in Engineering, 30, 3845

\bibitem[{{Astropy Collaboration} {et~al.}(2022){Astropy Collaboration},
  {Price-Whelan}, {Lim}, {Earl}, {Starkman}, {Bradley}, {Shupe}, {Patil},
  {Corrales}, {Brasseur}, {N{"o}the}, {Donath}, {Tollerud}, {Morris},
  {Ginsburg}, {Vaher}, {Weaver}, {Tocknell}, {Jamieson}, {van Kerkwijk},
  {Robitaille}, {Merry}, {Bachetti}, {G{"u}nther}, {Aldcroft},
  {Alvarado-Montes}, {Archibald}, {B{'o}di}, {Bapat}, {Barentsen}, {Baz{'a}n},
  {Biswas}, {Boquien}, {Burke}, {Cara}, {Cara}, {Conroy}, {Conseil}, {Craig},
  {Cross}, {Cruz}, {D'Eugenio}, {Dencheva}, {Devillepoix}, {Dietrich},
  {Eigenbrot}, {Erben}, {Ferreira}, {Foreman-Mackey}, {Fox}, {Freij}, {Garg},
  {Geda}, {Glattly}, {Gondhalekar}, {Gordon}, {Grant}, {Greenfield}, {Groener},
  {Guest}, {Gurovich}, {Handberg}, {Hart}, {Hatfield-Dodds}, {Homeier},
  {Hosseinzadeh}, {Jenness}, {Jones}, {Joseph}, {Kalmbach}, {Karamehmetoglu},
  {Ka{l}uszy{'n}ski}, {Kelley}, {Kern}, {Kerzendorf}, {Koch}, {Kulumani},
  {Lee}, {Ly}, {Ma}, {MacBride}, {Maljaars}, {Muna}, {Murphy}, {Norman},
  {O'Steen}, {Oman}, {Pacifici}, {Pascual}, {Pascual-Granado}, {Patil},
  {Perren}, {Pickering}, {Rastogi}, {Roulston}, {Ryan}, {Rykoff}, {Sabater},
  {Sakurikar}, {Salgado}, {Sanghi}, {Saunders}, {Savchenko}, {Schwardt},
  {Seifert-Eckert}, {Shih}, {Jain}, {Shukla}, {Sick}, {Simpson},
  {Singanamalla}, {Singer}, {Singhal}, {Sinha}, {Sip{H{o}}cz}, {Spitler},
  {Stansby}, {Streicher}, {{{S}}umak}, {Swinbank}, {Taranu}, {Tewary},
  {Tremblay}, {Val-Borro}, {Van Kooten}, {Vasovi{'c}}, {Verma}, {de Miranda
  Cardoso}, {Williams}, {Wilson}, {Winkel}, {Wood-Vasey}, {Xue}, {Yoachim},
  {Zhang}, {Zonca}, \& {Astropy Project Contributors}}]{astropy:2022}
{Astropy Collaboration}, {Price-Whelan}, A.~M., {Lim}, P.~L., {et~al.} 2022,
  \apj, 935, 167

\bibitem[{{Astropy Collaboration} {et~al.}(2018){Astropy Collaboration},
  {Price-Whelan}, {Sip{\H{o}}cz}, {G{\"u}nther}, {Lim}, {Crawford}, {Conseil},
  {Shupe}, {Craig}, {Dencheva}, {Ginsburg}, {Vand erPlas}, {Bradley},
  {P{\'e}rez-Su{\'a}rez}, {de Val-Borro}, {Aldcroft}, {Cruz}, {Robitaille},
  {Tollerud}, {Ardelean}, {Babej}, {Bach}, {Bachetti}, {Bakanov}, {Bamford},
  {Barentsen}, {Barmby}, {Baumbach}, {Berry}, {Biscani}, {Boquien}, {Bostroem},
  {Bouma}, {Brammer}, {Bray}, {Breytenbach}, {Buddelmeijer}, {Burke},
  {Calderone}, {Cano Rodr{\'\i}guez}, {Cara}, {Cardoso}, {Cheedella}, {Copin},
  {Corrales}, {Crichton}, {D'Avella}, {Deil}, {Depagne}, {Dietrich}, {Donath},
  {Droettboom}, {Earl}, {Erben}, {Fabbro}, {Ferreira}, {Finethy}, {Fox},
  {Garrison}, {Gibbons}, {Goldstein}, {Gommers}, {Greco}, {Greenfield},
  {Groener}, {Grollier}, {Hagen}, {Hirst}, {Homeier}, {Horton}, {Hosseinzadeh},
  {Hu}, {Hunkeler}, {Ivezi{\'c}}, {Jain}, {Jenness}, {Kanarek}, {Kendrew},
  {Kern}, {Kerzendorf}, {Khvalko}, {King}, {Kirkby}, {Kulkarni}, {Kumar},
  {Lee}, {Lenz}, {Littlefair}, {Ma}, {Macleod}, {Mastropietro}, {McCully},
  {Montagnac}, {Morris}, {Mueller}, {Mumford}, {Muna}, {Murphy}, {Nelson},
  {Nguyen}, {Ninan}, {N{\"o}the}, {Ogaz}, {Oh}, {Parejko}, {Parley}, {Pascual},
  {Patil}, {Patil}, {Plunkett}, {Prochaska}, {Rastogi}, {Reddy Janga},
  {Sabater}, {Sakurikar}, {Seifert}, {Sherbert}, {Sherwood-Taylor}, {Shih},
  {Sick}, {Silbiger}, {Singanamalla}, {Singer}, {Sladen}, {Sooley},
  {Sornarajah}, {Streicher}, {Teuben}, {Thomas}, {Tremblay}, {Turner},
  {Terr{\'o}n}, {van Kerkwijk}, {de la Vega}, {Watkins}, {Weaver}, {Whitmore},
  {Woillez}, {Zabalza}, \& {Astropy Contributors}}]{astropy:2018}
{Astropy Collaboration}, {Price-Whelan}, A.~M., {Sip{\H{o}}cz}, B.~M., {et~al.}
  2018, \aj, 156, 123

\bibitem[{{Astropy Collaboration} {et~al.}(2013){Astropy Collaboration},
  {Robitaille}, {Tollerud}, {Greenfield}, {Droettboom}, {Bray}, {Aldcroft},
  {Davis}, {Ginsburg}, {Price-Whelan}, {Kerzendorf}, {Conley}, {Crighton},
  {Barbary}, {Muna}, {Ferguson}, {Grollier}, {Parikh}, {Nair}, {Unther},
  {Deil}, {Woillez}, {Conseil}, {Kramer}, {Turner}, {Singer}, {Fox}, {Weaver},
  {Zabalza}, {Edwards}, {Azalee Bostroem}, {Burke}, {Casey}, {Crawford},
  {Dencheva}, {Ely}, {Jenness}, {Labrie}, {Lim}, {Pierfederici}, {Pontzen},
  {Ptak}, {Refsdal}, {Servillat}, \& {Streicher}}]{astropy:2013}
{Astropy Collaboration}, {Robitaille}, T.~P., {Tollerud}, E.~J., {et~al.} 2013,
  \aap, 558, A33

\bibitem[{{Barbary}(2016)}]{Barbary16-ext}
{Barbary}, K. 2016, {extinction v0.3.0}

\bibitem[{Barbary(2016)}]{Barbary16-sep}
Barbary, K. 2016, Journal of Open Source Software, 1, 58

\bibitem[{{Barbon} {et~al.}(2010){Barbon}, {Buondi}, {Cappellaro}, \&
  {Turatto}}]{Barbon10}
{Barbon}, R., {Buondi}, V., {Cappellaro}, E., \& {Turatto}, M. 2010, {VizieR
  Online Data Catalog: Asiago Supernova Catalogue (Barbon et al., 1999-)},
  VizieR On-line Data Catalog: B/sn. Originally published in: Astron.
  Astrophys. Suppl. Ser. 139, 531 (1999)

\bibitem[{{Bell} \& {de Jong}(2001)}]{Bell01}
{Bell}, E.~F. \& {de Jong}, R.~S. 2001, \apj, 550, 212

\bibitem[{{Bertin} \& {Arnouts}(1996)}]{Bertin96-sex}
{Bertin}, E. \& {Arnouts}, S. 1996, \aaps, 117, 393

\bibitem[{{Bianchi} {et~al.}(2014){Bianchi}, {Conti}, \& {Shiao}}]{GALEX-DR7}
{Bianchi}, L., {Conti}, A., \& {Shiao}, B. 2014, Advances in Space Research,
  53, 900

\bibitem[{{Bianchi}(2007)}]{Bianchi07}
{Bianchi}, S. 2007, \aap, 471, 765

\bibitem[{{Blondin} {et~al.}(2009){Blondin}, {Prieto}, {Patat}, {Challis},
  {Hicken}, {Kirshner}, {Matheson}, \& {Modjaz}}]{Blondin09}
{Blondin}, S., {Prieto}, J.~L., {Patat}, F., {et~al.} 2009, \apj, 693, 207

\bibitem[{{Boiss{\'e}} {et~al.}(2015){Boiss{\'e}}, {Bergeron}, {Prochaska},
  {P{\'e}roux}, \& {York}}]{Boisse15}
{Boiss{\'e}}, P., {Bergeron}, J., {Prochaska}, J.~X., {P{\'e}roux}, C., \&
  {York}, D.~G. 2015, \aap, 581, A109

\bibitem[{Bonferroni(1936)}]{Bonferroni36}
Bonferroni, C. 1936, Teoria statistica delle classi e calcolo delle
  probabilit{\`a}, Pubblicazioni del R. Istituto superiore di scienze
  economiche e commerciali di Firenze (Seeber)

\bibitem[{{Boquien} {et~al.}(2014){Boquien}, {Buat}, \& {Perret}}]{Boquien14}
{Boquien}, M., {Buat}, V., \& {Perret}, V. 2014, \aap, 571, A72

\bibitem[{{Bottinelli} {et~al.}(1983){Bottinelli}, {Gouguenheim}, {Paturel}, \&
  {de Vaucouleurs}}]{Bottinelli83}
{Bottinelli}, L., {Gouguenheim}, L., {Paturel}, G., \& {de Vaucouleurs}, G.
  1983, \aap, 118, 4

\bibitem[{Bradley {et~al.}(2023)Bradley, Sip{\H o}cz, Robitaille, Tollerud,
  Vin{\'{\i}}cius, Deil, Barbary, Wilson, Busko, Donath, G{\"u}nther, Cara,
  Lim, Me{\ss}linger, Conseil, Bostroem, Droettboom, Bray, Bratholm, Barentsen,
  Craig, Rathi, Pascual, Perren, Georgiev, de~Val-Borro, Kerzendorf, Bach,
  Quint, \& Souchereau}]{bradley23-photutils}
Bradley, L., Sip{\H o}cz, B., Robitaille, T., {et~al.} 2023, astropy/photutils:
  1.8.0

\bibitem[{{Bresolin}(2019)}]{Bresolin19}
{Bresolin}, F. 2019, \mnras, 488, 3826

\bibitem[{{Brinchmann} {et~al.}(2004){Brinchmann}, {Charlot}, {White},
  {Tremonti}, {Kauffmann}, {Heckman}, \& {Brinkmann}}]{Brinchman04}
{Brinchmann}, J., {Charlot}, S., {White}, S.~D.~M., {et~al.} 2004, \mnras, 351,
  1151

\bibitem[{{Brout} \& {Scolnic}(2021)}]{Brout21}
{Brout}, D. \& {Scolnic}, D. 2021, \apj, 909, 26

\bibitem[{{Bulla} {et~al.}(2018){Bulla}, {Goobar}, \& {Dhawan}}]{Bulla18}
{Bulla}, M., {Goobar}, A., \& {Dhawan}, S. 2018, \mnras, 479, 3663

\bibitem[{{Byrne} {et~al.}(2023){Byrne}, {Fraser}, {Cai}, {Reguitti}, \&
  {Valerin}}]{Byrne23}
{Byrne}, R.~A., {Fraser}, M., {Cai}, Y.~Z., {Reguitti}, A., \& {Valerin}, G.
  2023, \mnras, 524, 2978

\bibitem[{{Calzetti} {et~al.}(2000){Calzetti}, {Armus}, {Bohlin}, {Kinney},
  {Koornneef}, \& {Storchi-Bergmann}}]{Calzetti00}
{Calzetti}, D., {Armus}, L., {Bohlin}, R.~C., {et~al.} 2000, \apj, 533, 682

\bibitem[{{Casasola} {et~al.}(2017){Casasola}, {Cassar{\`a}}, {Bianchi},
  {Verstocken}, {Xilouris}, {Magrini}, {Smith}, {De Looze}, {Galametz},
  {Madden}, {Baes}, {Clark}, {Davies}, {De Vis}, {Evans}, {Fritz}, {Galliano},
  {Jones}, {Mosenkov}, {Viaene}, \& {Ysard}}]{Casasola17}
{Casasola}, V., {Cassar{\`a}}, L.~P., {Bianchi}, S., {et~al.} 2017, \aap, 605,
  A18

\bibitem[{{Cazzoli} {et~al.}(2022){Cazzoli}, {Hermosa Mu{\~n}oz},
  {M{\'a}rquez}, {Masegosa}, {Castillo-Morales}, {Gil de Paz},
  {Hern{\'a}ndez-Garc{\'\i}a}, {La Franca}, \& {Ramos Almeida}}]{Cazzoli22}
{Cazzoli}, S., {Hermosa Mu{\~n}oz}, L., {M{\'a}rquez}, I., {et~al.} 2022, \aap,
  664, A135

\bibitem[{{Chen} {et~al.}(2010){Chen}, {Tremonti}, {Heckman}, {Kauffmann},
  {Weiner}, {Brinchmann}, \& {Wang}}]{Chen10}
{Chen}, Y.-M., {Tremonti}, C.~A., {Heckman}, T.~M., {et~al.} 2010, \aj, 140,
  445

\bibitem[{{Chevalier} \& {Fransson}(1994)}]{Chevalier94}
{Chevalier}, R.~A. \& {Fransson}, C. 1994, \apj, 420, 268

\bibitem[{{Chevance} {et~al.}(2020){Chevance}, {Kruijssen}, {Hygate},
  {Schruba}, {Longmore}, {Groves}, {Henshaw}, {Herrera}, {Hughes}, {Jeffreson},
  {Lang}, {Leroy}, {Meidt}, {Pety}, {Razza}, {Rosolowsky}, {Schinnerer},
  {Bigiel}, {Blanc}, {Emsellem}, {Faesi}, {Glover}, {Haydon}, {Ho}, {Kreckel},
  {Lee}, {Liu}, {Querejeta}, {Saito}, {Sun}, {Usero}, \& {Utomo}}]{Chevance20}
{Chevance}, M., {Kruijssen}, J.~M.~D., {Hygate}, A. P.~S., {et~al.} 2020,
  \mnras, 493, 2872

\bibitem[{{Choi} {et~al.}(2016){Choi}, {Dotter}, {Conroy}, {Cantiello},
  {Paxton}, \& {Johnson}}]{Choi16}
{Choi}, J., {Dotter}, A., {Conroy}, C., {et~al.} 2016, \apj, 823, 102

\bibitem[{{Chugai} \& {Danziger}(1994)}]{Chugai94}
{Chugai}, N.~N. \& {Danziger}, I.~J. 1994, \mnras, 268, 173

\bibitem[{{Clark} {et~al.}(2021){Clark}, {Maguire}, {Bulla}, {Galbany},
  {Sullivan}, {Anderson}, \& {Smartt}}]{Clark21}
{Clark}, P., {Maguire}, K., {Bulla}, M., {et~al.} 2021, \mnras, 507, 4367

\bibitem[{{Conroy} \& {Gunn}(2010)}]{Conroy10}
{Conroy}, C. \& {Gunn}, J.~E. 2010, \apj, 712, 833

\bibitem[{{Conroy} {et~al.}(2009){Conroy}, {Gunn}, \& {White}}]{Conroy09}
{Conroy}, C., {Gunn}, J.~E., \& {White}, M. 2009, \apj, 699, 486

\bibitem[{Cranmer(2023)}]{PySR}
Cranmer, M. 2023, Interpretable Machine Learning for Science with PySR and
  SymbolicRegression.jl

\bibitem[{{Cross} {et~al.}(2012){Cross}, {Collins}, {Mann}, {Read}, {Sutorius},
  {Blake}, {Holliman}, {Hambly}, {Emerson}, {Lawrence}, \&
  {Noddle}}]{VISTA-archive}
{Cross}, N.~J.~G., {Collins}, R.~S., {Mann}, R.~G., {et~al.} 2012, \aap, 548,
  A119

\bibitem[{{Davies} {et~al.}(2019){Davies}, {Nersesian}, {Baes}, {Bianchi},
  {Casasola}, {Cassar{\`a}}, {Clark}, {De Looze}, {De Vis}, {Evans}, {Fritz},
  {Galametz}, {Galliano}, {Jones}, {Lianou}, {Madden}, {Mosenkov}, {Smith},
  {Verstocken}, {Viaene}, {Vika}, {Xilouris}, \& {Ysard}}]{Davies19}
{Davies}, J.~I., {Nersesian}, A., {Baes}, M., {et~al.} 2019, \aap, 626, A63

\bibitem[{{de Vaucouleurs}(1959)}]{deVaucouleurs59}
{de Vaucouleurs}, G. 1959, Handbuch der Physik, 53, 275

\bibitem[{{Dessart} {et~al.}(2023){Dessart}, {Guti{\'e}rrez}, {Kuncarayakti},
  {Fox}, \& {Filippenko}}]{Dessart23a}
{Dessart}, L., {Guti{\'e}rrez}, C.~P., {Kuncarayakti}, H., {Fox}, O.~D., \&
  {Filippenko}, A.~V. 2023, \aap, 675, A33

\bibitem[{{Dotter}(2016)}]{Dotter16}
{Dotter}, A. 2016, \apjs, 222, 8

\bibitem[{{Draine}(2011)}]{Draine11}
{Draine}, B.~T. 2011, {Physics of the Interstellar and Intergalactic Medium}

\bibitem[{{Duarte} {et~al.}(2023){Duarte}, {Gonz{\'a}lez-Gait{\'a}n},
  {Mour{\~a}o}, {Paulino-Afonso}, {Guilherme-Garcia}, {{\'A}guas}, {Galbany},
  {Kelsey}, {Scolnic}, {Sullivan}, {Brout}, {Palmese}, {Wiseman}, {Aguena},
  {Alves}, {Bacon}, {Bertin}, {Bocquet}, {Brooks}, {Burke}, {Carnero Rosell},
  {Carrasco Kind}, {Carretero}, {Costanzi}, {Pereira}, {Davis}, {De Vicente},
  {Desai}, {Diehl}, {Doel}, {Everett}, {Ferrero}, {Friedel}, {Frieman},
  {Garc{\'\i}a-Bellido}, {Gatti}, {Gerdes}, {Gruen}, {Gruendl}, {Gutierrez},
  {Hinton}, {Hollowood}, {Honscheid}, {James}, {Kuehn}, {Kuropatkin},
  {Melchior}, {Miquel}, {Paz-Chinch{\'o}n}, {Pieres}, {Plazas Malag{\'o}n},
  {Raveri}, {Rodriguez-Monroy}, {Sanchez}, {Scarpine}, {Sevilla-Noarbe},
  {Smith}, {Suchyta}, {Tarle}, {To}, \& {Weaverdyck}}]{Duarte23}
{Duarte}, J., {Gonz{\'a}lez-Gait{\'a}n}, S., {Mour{\~a}o}, A., {et~al.} 2023,
  \aap, 680, A56

\bibitem[{{Duarte} {et~al.}(2025){Duarte}, {Gonz{\'a}lez-Gait{\'a}n},
  {Mour{\~a}o}, {Rino-Silvestre}, {Baes}, {Anderson}, {Galbany}, \&
  {Stalevski}}]{Duarte25}
{Duarte}, J., {Gonz{\'a}lez-Gait{\'a}n}, S., {Mour{\~a}o}, A., {et~al.} 2025,
  arXiv e-prints, arXiv:2503.04906

\bibitem[{{Easeman} {et~al.}(2022){Easeman}, {Schady}, {Wuyts}, \&
  {Yates}}]{Easeman22}
{Easeman}, B., {Schady}, P., {Wuyts}, S., \& {Yates}, R.~M. 2022, \mnras, 511,
  371

\bibitem[{{Falc{\'o}n-Barroso} {et~al.}(2011){Falc{\'o}n-Barroso},
  {S{\'a}nchez-Bl{\'a}zquez}, {Vazdekis}, {Ricciardelli}, {Cardiel}, {Cenarro},
  {Gorgas}, \& {Peletier}}]{Falcon-Barroso11-miles}
{Falc{\'o}n-Barroso}, J., {S{\'a}nchez-Bl{\'a}zquez}, P., {Vazdekis}, A.,
  {et~al.} 2011, \aap, 532, A95

\bibitem[{{Fathi} {et~al.}(2010){Fathi}, {Allen}, {Boch}, {Hatziminaoglou}, \&
  {Peletier}}]{Fathi10}
{Fathi}, K., {Allen}, M., {Boch}, T., {Hatziminaoglou}, E., \& {Peletier},
  R.~F. 2010, \mnras, 406, 1595

\bibitem[{{Feldmann}(2020)}]{Feldmann20}
{Feldmann}, R. 2020, Communications Physics, 3, 226

\bibitem[{{Ferretti} {et~al.}(2016){Ferretti}, {Amanullah}, {Goobar},
  {Johansson}, {Vreeswijk}, {Butler}, {Cao}, {Cenko}, {Doran}, {Filippenko},
  {Freeland }, {Hosseinzadeh}, {Howell}, {Lundqvist}, {Mattila}, {Nordin},
  {Nugent}, {Petrushevska}, {Valenti}, {Vogt}, \& {Wozniak}}]{Ferretti16}
{Ferretti}, R., {Amanullah}, R., {Goobar}, A., {et~al.} 2016, \aap, 592, A40

\bibitem[{{Ferretti} {et~al.}(2017){Ferretti}, {Amanullah}, {Goobar},
  {Petrushevska}, {Borthakur}, {Bulla}, {Fox}, {Freeland}, {Fremling},
  {Hangard}, \& {Hayes}}]{Ferretti17}
{Ferretti}, R., {Amanullah}, R., {Goobar}, A., {et~al.} 2017, \aap, 606, A111

\bibitem[{{Fielding} \& {Bryan}(2022)}]{Fielding22}
{Fielding}, D.~B. \& {Bryan}, G.~L. 2022, \apj, 924, 82

\bibitem[{{Flaugher} {et~al.}(2015){Flaugher}, {Diehl}, {Honscheid}, {Abbott},
  {Alvarez}, {Angstadt}, {Annis}, {Antonik}, {Ballester}, {Beaufore},
  {Bernstein}, {Bernstein}, {Bigelow}, {Bonati}, {Boprie}, {Brooks},
  {Buckley-Geer}, {Campa}, {Cardiel-Sas}, {Castander}, {Castilla}, {Cease},
  {Cela-Ruiz}, {Chappa}, {Chi}, {Cooper}, {da Costa}, {Dede}, {Derylo},
  {DePoy}, {de Vicente}, {Doel}, {Drlica-Wagner}, {Eiting}, {Elliott}, {Emes},
  {Estrada}, {Fausti Neto}, {Finley}, {Flores}, {Frieman}, {Gerdes},
  {Gladders}, {Gregory}, {Gutierrez}, {Hao}, {Holland}, {Holm}, {Huffman},
  {Jackson}, {James}, {Jonas}, {Karcher}, {Karliner}, {Kent}, {Kessler},
  {Kozlovsky}, {Kron}, {Kubik}, {Kuehn}, {Kuhlmann}, {Kuk}, {Lahav}, {Lathrop},
  {Lee}, {Levi}, {Lewis}, {Li}, {Mandrichenko}, {Marshall}, {Martinez},
  {Merritt}, {Miquel}, {Mu{\~n}oz}, {Neilsen}, {Nichol}, {Nord}, {Ogando},
  {Olsen}, {Palaio}, {Patton}, {Peoples}, {Plazas}, {Rauch}, {Reil}, {Rheault},
  {Roe}, {Rogers}, {Roodman}, {Sanchez}, {Scarpine}, {Schindler}, {Schmidt},
  {Schmitt}, {Schubnell}, {Schultz}, {Schurter}, {Scott}, {Serrano}, {Shaw},
  {Smith}, {Soares-Santos}, {Stefanik}, {Stuermer}, {Suchyta}, {Sypniewski},
  {Tarle}, {Thaler}, {Tighe}, {Tran}, {Tucker}, {Walker}, {Wang}, {Watson},
  {Weaverdyck}, {Wester}, {Woods}, {Yanny}, \& {DES Collaboration}}]{DECam}
{Flaugher}, B., {Diehl}, H.~T., {Honscheid}, K., {et~al.} 2015, \aj, 150, 150

\bibitem[{{Flewelling} {et~al.}(2020){Flewelling}, {Magnier}, {Chambers},
  {Heasley}, {Holmberg}, {Huber}, {Sweeney}, {Waters}, {Calamida}, {Casertano},
  {Chen}, {Farrow}, {Hasinger}, {Henderson}, {Long}, {Metcalfe}, {Narayan},
  {Nieto-Santisteban}, {Norberg}, {Rest}, {Saglia}, {Szalay}, {Thakar},
  {Tonry}, {Valenti}, {Werner}, {White}, {Denneau}, {Draper}, {Hodapp},
  {Jedicke}, {Kaiser}, {Kudritzki}, {Price}, {Wainscoat}, {Chastel}, {McLean},
  {Postman}, \& {Shiao}}]{Flewelling20-PS1}
{Flewelling}, H.~A., {Magnier}, E.~A., {Chambers}, K.~C., {et~al.} 2020, \apjs,
  251, 7

\bibitem[{{Foreman-Mackey} {et~al.}(2013){Foreman-Mackey}, {Hogg}, {Lang}, \&
  {Goodman}}]{emcee}
{Foreman-Mackey}, D., {Hogg}, D.~W., {Lang}, D., \& {Goodman}, J. 2013, \pasp,
  125, 306

\bibitem[{{F{\"o}rster} {et~al.}(2013){F{\"o}rster}, {Gonz{\'a}lez-Gait{\'a}n},
  {Folatelli}, \& {Morrell}}]{Forster13}
{F{\"o}rster}, F., {Gonz{\'a}lez-Gait{\'a}n}, S., {Folatelli}, G., \&
  {Morrell}, N. 2013, \apj, 772, 19

\bibitem[{{Fouque} {et~al.}(1990){Fouque}, {Bottinelli}, {Gouguenheim}, \&
  {Paturel}}]{Fouque90}
{Fouque}, P., {Bottinelli}, L., {Gouguenheim}, L., \& {Paturel}, G. 1990, \apj,
  349, 1

\bibitem[{{Fransson} {et~al.}(2002){Fransson}, {Chevalier}, {Filippenko},
  {Leibundgut}, {Barth}, {Fesen}, {Kirshner}, {Leonard}, {Li}, {Lundqvist},
  {Sollerman}, \& {Van Dyk}}]{Fransson02}
{Fransson}, C., {Chevalier}, R.~A., {Filippenko}, A.~V., {et~al.} 2002, \apj,
  572, 350

\bibitem[{{Fraser-McKelvie} {et~al.}(2022){Fraser-McKelvie}, {Cortese},
  {Groves}, {Brough}, {Bryant}, {Catinella}, {Croom}, {D'Eugenio},
  {L{\'o}pez-S{\'a}nchez}, {van de Sande}, {Sweet}, {Vaughan},
  {Bland-Hawthorn}, {Lawrence}, {Lorente}, \& {Owers}}]{Fraser-McKelvie22}
{Fraser-McKelvie}, A., {Cortese}, L., {Groves}, B., {et~al.} 2022, \mnras, 510,
  320

\bibitem[{{Frerking} {et~al.}(1982){Frerking}, {Langer}, \&
  {Wilson}}]{Frerking82}
{Frerking}, M.~A., {Langer}, W.~D., \& {Wilson}, R.~W. 1982, \apj, 262, 590

\bibitem[{{Gagliano} {et~al.}(2021){Gagliano}, {Narayan}, {Engel}, {Carrasco
  Kind}, \& {LSST Dark Energy Science Collaboration}}]{Gagliano21}
{Gagliano}, A., {Narayan}, G., {Engel}, A., {Carrasco Kind}, M., \& {LSST Dark
  Energy Science Collaboration}. 2021, \apj, 908, 170

\bibitem[{{Gaia Collaboration} {et~al.}(2016){Gaia Collaboration}, {Prusti},
  {de Bruijne}, {Brown}, {Vallenari}, {Babusiaux}, {Bailer-Jones}, {Bastian},
  {Biermann}, {Evans}, {Eyer}, {Jansen}, {Jordi}, {Klioner}, {Lammers},
  {Lindegren}, {Luri}, {Mignard}, {Milligan}, {Panem}, {Poinsignon},
  {Pourbaix}, {Randich}, {Sarri}, {Sartoretti}, {Siddiqui}, {Soubiran},
  {Valette}, {van Leeuwen}, {Walton}, {Aerts}, {Arenou}, {Cropper}, {Drimmel},
  {H{\o}g}, {Katz}, {Lattanzi}, {O'Mullane}, {Grebel}, {Holland}, {Huc},
  {Passot}, {Bramante}, {Cacciari}, {Casta{\~n}eda}, {Chaoul}, {Cheek}, {De
  Angeli}, {Fabricius}, {Guerra}, {Hern{\'a}ndez}, {Jean-Antoine-Piccolo},
  {Masana}, {Messineo}, {Mowlavi}, {Nienartowicz}, {Ord{\'o}{\~n}ez-Blanco},
  {Panuzzo}, {Portell}, {Richards}, {Riello}, {Seabroke}, {Tanga},
  {Th{\'e}venin}, {Torra}, {Els}, {Gracia-Abril}, {Comoretto},
  {Garcia-Reinaldos}, {Lock}, {Mercier}, {Altmann}, {Andrae}, {Astraatmadja},
  {Bellas-Velidis}, {Benson}, {Berthier}, {Blomme}, {Busso}, {Carry},
  {Cellino}, {Clementini}, {Cowell}, {Creevey}, {Cuypers}, {Davidson}, {De
  Ridder}, {de Torres}, {Delchambre}, {Dell'Oro}, {Ducourant}, {Fr{\'e}mat},
  {Garc{\'\i}a-Torres}, {Gosset}, {Halbwachs}, {Hambly}, {Harrison}, {Hauser},
  {Hestroffer}, {Hodgkin}, {Huckle}, {Hutton}, {Jasniewicz}, {Jordan},
  {Kontizas}, {Korn}, {Lanzafame}, {Manteiga}, {Moitinho}, {Muinonen},
  {Osinde}, {Pancino}, {Pauwels}, {Petit}, {Recio-Blanco}, {Robin}, {Sarro},
  {Siopis}, {Smith}, {Smith}, {Sozzetti}, {Thuillot}, {van Reeven}, {Viala},
  {Abbas}, {Abreu Aramburu}, {Accart}, {Aguado}, {Allan}, {Allasia},
  {Altavilla}, {{\'A}lvarez}, {Alves}, {Anderson}, {Andrei}, {Anglada Varela},
  {Antiche}, {Antoja}, {Ant{\'o}n}, {Arcay}, {Atzei}, {Ayache}, {Bach},
  {Baker}, {Balaguer-N{\'u}{\~n}ez}, {Barache}, {Barata}, {Barbier}, {Barblan},
  {Baroni}, {Barrado y Navascu{\'e}s}, {Barros}, {Barstow}, {Becciani},
  {Bellazzini}, {Bellei}, {Bello Garc{\'\i}a}, {Belokurov}, {Bendjoya},
  {Berihuete}, {Bianchi}, {Bienaym{\'e}}, {Billebaud}, {Blagorodnova},
  {Blanco-Cuaresma}, {Boch}, {Bombrun}, {Borrachero}, {Bouquillon}, {Bourda},
  {Bouy}, {Bragaglia}, {Breddels}, {Brouillet}, {Br{\"u}semeister},
  {Bucciarelli}, {Budnik}, {Burgess}, {Burgon}, {Burlacu}, {Busonero}, {Buzzi},
  {Caffau}, {Cambras}, {Campbell}, {Cancelliere}, {Cantat-Gaudin}, {Carlucci},
  {Carrasco}, {Castellani}, {Charlot}, {Charnas}, {Charvet}, {Chassat},
  {Chiavassa}, {Clotet}, {Cocozza}, {Collins}, {Collins}, {Costigan}, {Crifo},
  {Cross}, {Crosta}, {Crowley}, {Dafonte}, {Damerdji}, {Dapergolas}, {David},
  {David}, {De Cat}, {de Felice}, {de Laverny}, {De Luise}, {De March}, {de
  Martino}, {de Souza}, {Debosscher}, {del Pozo}, {Delbo}, {Delgado},
  {Delgado}, {di Marco}, {Di Matteo}, {Diakite}, {Distefano}, {Dolding}, {Dos
  Anjos}, {Drazinos}, {Dur{\'a}n}, {Dzigan}, {Ecale}, {Edvardsson}, {Enke},
  {Erdmann}, {Escolar}, {Espina}, {Evans}, {Eynard Bontemps}, {Fabre},
  {Fabrizio}, {Faigler}, {Falc{\~a}o}, {Farr{\`a}s Casas}, {Faye}, {Federici},
  {Fedorets}, {Fern{\'a}ndez-Hern{\'a}ndez}, {Fernique}, {Fienga}, {Figueras},
  {Filippi}, {Findeisen}, {Fonti}, {Fouesneau}, {Fraile}, {Fraser}, {Fuchs},
  {Furnell}, {Gai}, {Galleti}, {Galluccio}, {Garabato}, {Garc{\'\i}a-Sedano},
  {Gar{\'e}}, {Garofalo}, {Garralda}, {Gavras}, {Gerssen}, {Geyer}, {Gilmore},
  {Girona}, {Giuffrida}, {Gomes}, {Gonz{\'a}lez-Marcos},
  {Gonz{\'a}lez-N{\'u}{\~n}ez}, {Gonz{\'a}lez-Vidal}, {Granvik}, {Guerrier},
  {Guillout}, {Guiraud}, {G{\'u}rpide}, {Guti{\'e}rrez-S{\'a}nchez}, {Guy},
  {Haigron}, {Hatzidimitriou}, {Haywood}, {Heiter}, {Helmi}, {Hobbs},
  {Hofmann}, {Holl}, {Holland}, {Hunt}, {Hypki}, {Icardi}, {Irwin}, {Jevardat
  de Fombelle}, {Jofr{\'e}}, {Jonker}, {Jorissen}, {Julbe}, {Karampelas},
  {Kochoska}, {Kohley}, {Kolenberg}, {Kontizas}, {Koposov}, {Kordopatis},
  {Koubsky}, {Kowalczyk}, {Krone-Martins}, {Kudryashova}, {Kull}, {Bachchan},
  {Lacoste-Seris}, {Lanza}, {Lavigne}, {Le Poncin-Lafitte}, {Lebreton},
  {Lebzelter}, {Leccia}, {Leclerc}, {Lecoeur-Taibi}, {Lemaitre}, {Lenhardt},
  {Leroux}, {Liao}, {Licata}, {Lindstr{\o}m}, {Lister}, {Livanou}, {Lobel},
  {L{\"o}ffler}, {L{\'o}pez}, {Lopez-Lozano}, {Lorenz}, {Loureiro},
  {MacDonald}, {Magalh{\~a}es Fernandes}, {Managau}, {Mann}, {Mantelet},
  {Marchal}, {Marchant}, {Marconi}, {Marie}, {Marinoni}, {Marrese},
  {Marschalk{\'o}}, {Marshall}, {Mart{\'\i}n-Fleitas}, {Martino}, {Mary},
  {Matijevi{\v{c}}}, {Mazeh}, {McMillan}, {Messina}, {Mestre}, {Michalik},
  {Millar}, {Miranda}, {Molina}, {Molinaro}, {Molinaro}, {Moln{\'a}r},
  {Moniez}, {Montegriffo}, {Monteiro}, {Mor}, {Mora}, {Morbidelli}, {Morel},
  {Morgenthaler}, {Morley}, {Morris}, {Mulone}, {Muraveva}, {Musella},
  {Narbonne}, {Nelemans}, {Nicastro}, {Noval}, {Ord{\'e}novic},
  {Ordieres-Mer{\'e}}, {Osborne}, {Pagani}, {Pagano}, {Pailler}, {Palacin},
  {Palaversa}, {Parsons}, {Paulsen}, {Pecoraro}, {Pedrosa}, {Pentik{\"a}inen},
  {Pereira}, {Pichon}, {Piersimoni}, {Pineau}, {Plachy}, {Plum}, {Poujoulet},
  {Pr{\v{s}}a}, {Pulone}, {Ragaini}, {Rago}, {Rambaux}, {Ramos-Lerate},
  {Ranalli}, {Rauw}, {Read}, {Regibo}, {Renk}, {Reyl{\'e}}, {Ribeiro},
  {Rimoldini}, {Ripepi}, {Riva}, {Rixon}, {Roelens}, {Romero-G{\'o}mez},
  {Rowell}, {Royer}, {Rudolph}, {Ruiz-Dern}, {Sadowski}, {Sagrist{\`a}
  Sell{\'e}s}, {Sahlmann}, {Salgado}, {Salguero}, {Sarasso}, {Savietto},
  {Schnorhk}, {Schultheis}, {Sciacca}, {Segol}, {Segovia}, {Segransan},
  {Serpell}, {Shih}, {Smareglia}, {Smart}, {Smith}, {Solano}, {Solitro},
  {Sordo}, {Soria Nieto}, {Souchay}, {Spagna}, {Spoto}, {Stampa}, {Steele},
  {Steidelm{\"u}ller}, {Stephenson}, {Stoev}, {Suess}, {S{\"u}veges}, {Surdej},
  {Szabados}, {Szegedi-Elek}, {Tapiador}, {Taris}, {Tauran}, {Taylor},
  {Teixeira}, {Terrett}, {Tingley}, {Trager}, {Turon}, {Ulla}, {Utrilla},
  {Valentini}, {van Elteren}, {Van Hemelryck}, {van Leeuwen}, {Varadi},
  {Vecchiato}, {Veljanoski}, {Via}, {Vicente}, {Vogt}, {Voss}, {Votruba},
  {Voutsinas}, {Walmsley}, {Weiler}, {Weingrill}, {Werner}, {Wevers},
  {Whitehead}, {Wyrzykowski}, {Yoldas}, {{\v{Z}}erjal}, {Zucker}, {Zurbach},
  {Zwitter}, {Alecu}, {Allen}, {Allende Prieto}, {Amorim},
  {Anglada-Escud{\'e}}, {Arsenijevic}, {Azaz}, {Balm}, {Beck}, {Bernstein},
  {Bigot}, {Bijaoui}, {Blasco}, {Bonfigli}, {Bono}, {Boudreault}, {Bressan},
  {Brown}, {Brunet}, {Bunclark}, {Buonanno}, {Butkevich}, {Carret}, {Carrion},
  {Chemin}, {Ch{\'e}reau}, {Corcione}, {Darmigny}, {de Boer}, {de Teodoro}, {de
  Zeeuw}, {Delle Luche}, {Domingues}, {Dubath}, {Fodor}, {Fr{\'e}zouls},
  {Fries}, {Fustes}, {Fyfe}, {Gallardo}, {Gallegos}, {Gardiol}, {Gebran},
  {Gomboc}, {G{\'o}mez}, {Grux}, {Gueguen}, {Heyrovsky}, {Hoar}, {Iannicola},
  {Isasi Parache}, {Janotto}, {Joliet}, {Jonckheere}, {Keil}, {Kim},
  {Klagyivik}, {Klar}, {Knude}, {Kochukhov}, {Kolka}, {Kos}, {Kutka}, {Lainey},
  {LeBouquin}, {Liu}, {Loreggia}, {Makarov}, {Marseille}, {Martayan},
  {Martinez-Rubi}, {Massart}, {Meynadier}, {Mignot}, {Munari}, {Nguyen},
  {Nordlander}, {Ocvirk}, {O'Flaherty}, {Olias Sanz}, {Ortiz}, {Osorio},
  {Oszkiewicz}, {Ouzounis}, {Palmer}, {Park}, {Pasquato}, {Peltzer}, {Peralta},
  {P{\'e}turaud}, {Pieniluoma}, {Pigozzi}, {Poels}, {Prat}, {Prod'homme},
  {Raison}, {Rebordao}, {Risquez}, {Rocca-Volmerange}, {Rosen}, {Ruiz-Fuertes},
  {Russo}, {Sembay}, {Serraller Vizcaino}, {Short}, {Siebert}, {Silva},
  {Sinachopoulos}, {Slezak}, {Soffel}, {Sosnowska}, {Strai{\v{z}}ys}, {ter
  Linden}, {Terrell}, {Theil}, {Tiede}, {Troisi}, {Tsalmantza}, {Tur},
  {Vaccari}, {Vachier}, {Valles}, {Van Hamme}, {Veltz}, {Virtanen}, {Wallut},
  {Wichmann}, {Wilkinson}, {Ziaeepour}, \& {Zschocke}}]{Gaia-mission}
{Gaia Collaboration}, {Prusti}, T., {de Bruijne}, J.~H.~J., {et~al.} 2016,
  \aap, 595, A1

\bibitem[{{Gaia Collaboration} {et~al.}(2023){Gaia Collaboration}, {Vallenari},
  {Brown}, {Prusti}, {de Bruijne}, {Arenou}, {Babusiaux}, {Biermann},
  {Creevey}, {Ducourant}, {Evans}, {Eyer}, {Guerra}, {Hutton}, {Jordi},
  {Klioner}, {Lammers}, {Lindegren}, {Luri}, {Mignard}, {Panem}, {Pourbaix},
  {Randich}, {Sartoretti}, {Soubiran}, {Tanga}, {Walton}, {Bailer-Jones},
  {Bastian}, {Drimmel}, {Jansen}, {Katz}, {Lattanzi}, {van Leeuwen}, {Bakker},
  {Cacciari}, {Casta{\~n}eda}, {De Angeli}, {Fabricius}, {Fouesneau},
  {Fr{\'e}mat}, {Galluccio}, {Guerrier}, {Heiter}, {Masana}, {Messineo},
  {Mowlavi}, {Nicolas}, {Nienartowicz}, {Pailler}, {Panuzzo}, {Riclet}, {Roux},
  {Seabroke}, {Sordo}, {Th{\'e}venin}, {Gracia-Abril}, {Portell}, {Teyssier},
  {Altmann}, {Andrae}, {Audard}, {Bellas-Velidis}, {Benson}, {Berthier},
  {Blomme}, {Burgess}, {Busonero}, {Busso}, {C{\'a}novas}, {Carry}, {Cellino},
  {Cheek}, {Clementini}, {Damerdji}, {Davidson}, {de Teodoro}, {Nu{\~n}ez
  Campos}, {Delchambre}, {Dell'Oro}, {Esquej}, {Fern{\'a}ndez-Hern{\'a}ndez},
  {Fraile}, {Garabato}, {Garc{\'\i}a-Lario}, {Gosset}, {Haigron}, {Halbwachs},
  {Hambly}, {Harrison}, {Hern{\'a}ndez}, {Hestroffer}, {Hodgkin}, {Holl},
  {Jan{\ss}en}, {Jevardat de Fombelle}, {Jordan}, {Krone-Martins}, {Lanzafame},
  {L{\"o}ffler}, {Marchal}, {Marrese}, {Moitinho}, {Muinonen}, {Osborne},
  {Pancino}, {Pauwels}, {Recio-Blanco}, {Reyl{\'e}}, {Riello}, {Rimoldini},
  {Roegiers}, {Rybizki}, {Sarro}, {Siopis}, {Smith}, {Sozzetti}, {Utrilla},
  {van Leeuwen}, {Abbas}, {{\'A}brah{\'a}m}, {Abreu Aramburu}, {Aerts},
  {Aguado}, {Ajaj}, {Aldea-Montero}, {Altavilla}, {{\'A}lvarez}, {Alves},
  {Anders}, {Anderson}, {Anglada Varela}, {Antoja}, {Baines}, {Baker},
  {Balaguer-N{\'u}{\~n}ez}, {Balbinot}, {Balog}, {Barache}, {Barbato},
  {Barros}, {Barstow}, {Bartolom{\'e}}, {Bassilana}, {Bauchet}, {Becciani},
  {Bellazzini}, {Berihuete}, {Bernet}, {Bertone}, {Bianchi}, {Binnenfeld},
  {Blanco-Cuaresma}, {Blazere}, {Boch}, {Bombrun}, {Bossini}, {Bouquillon},
  {Bragaglia}, {Bramante}, {Breedt}, {Bressan}, {Brouillet}, {Brugaletta},
  {Bucciarelli}, {Burlacu}, {Butkevich}, {Buzzi}, {Caffau}, {Cancelliere},
  {Cantat-Gaudin}, {Carballo}, {Carlucci}, {Carnerero}, {Carrasco},
  {Casamiquela}, {Castellani}, {Castro-Ginard}, {Chaoul}, {Charlot}, {Chemin},
  {Chiaramida}, {Chiavassa}, {Chornay}, {Comoretto}, {Contursi}, {Cooper},
  {Cornez}, {Cowell}, {Crifo}, {Cropper}, {Crosta}, {Crowley}, {Dafonte},
  {Dapergolas}, {David}, {David}, {de Laverny}, {De Luise}, {De March}, {De
  Ridder}, {de Souza}, {de Torres}, {del Peloso}, {del Pozo}, {Delbo},
  {Delgado}, {Delisle}, {Demouchy}, {Dharmawardena}, {Di Matteo}, {Diakite},
  {Diener}, {Distefano}, {Dolding}, {Edvardsson}, {Enke}, {Fabre}, {Fabrizio},
  {Faigler}, {Fedorets}, {Fernique}, {Fienga}, {Figueras}, {Fournier},
  {Fouron}, {Fragkoudi}, {Gai}, {Garcia-Gutierrez}, {Garcia-Reinaldos},
  {Garc{\'\i}a-Torres}, {Garofalo}, {Gavel}, {Gavras}, {Gerlach}, {Geyer},
  {Giacobbe}, {Gilmore}, {Girona}, {Giuffrida}, {Gomel}, {Gomez},
  {Gonz{\'a}lez-N{\'u}{\~n}ez}, {Gonz{\'a}lez-Santamar{\'\i}a},
  {Gonz{\'a}lez-Vidal}, {Granvik}, {Guillout}, {Guiraud},
  {Guti{\'e}rrez-S{\'a}nchez}, {Guy}, {Hatzidimitriou}, {Hauser}, {Haywood},
  {Helmer}, {Helmi}, {Sarmiento}, {Hidalgo}, {Hilger}, {H{\l}adczuk}, {Hobbs},
  {Holland}, {Huckle}, {Jardine}, {Jasniewicz}, {Jean-Antoine Piccolo},
  {Jim{\'e}nez-Arranz}, {Jorissen}, {Juaristi Campillo}, {Julbe}, {Karbevska},
  {Kervella}, {Khanna}, {Kontizas}, {Kordopatis}, {Korn}, {K{\'o}sp{\'a}l},
  {Kostrzewa-Rutkowska}, {Kruszy{\'n}ska}, {Kun}, {Laizeau}, {Lambert},
  {Lanza}, {Lasne}, {Le Campion}, {Lebreton}, {Lebzelter}, {Leccia}, {Leclerc},
  {Lecoeur-Taibi}, {Liao}, {Licata}, {Lindstr{\o}m}, {Lister}, {Livanou},
  {Lobel}, {Lorca}, {Loup}, {Madrero Pardo}, {Magdaleno Romeo}, {Managau},
  {Mann}, {Manteiga}, {Marchant}, {Marconi}, {Marcos}, {Marcos Santos},
  {Mar{\'\i}n Pina}, {Marinoni}, {Marocco}, {Marshall}, {Martin Polo},
  {Mart{\'\i}n-Fleitas}, {Marton}, {Mary}, {Masip}, {Massari},
  {Mastrobuono-Battisti}, {Mazeh}, {McMillan}, {Messina}, {Michalik}, {Millar},
  {Mints}, {Molina}, {Molinaro}, {Moln{\'a}r}, {Monari}, {Mongui{\'o}},
  {Montegriffo}, {Montero}, {Mor}, {Mora}, {Morbidelli}, {Morel}, {Morris},
  {Muraveva}, {Murphy}, {Musella}, {Nagy}, {Noval}, {Oca{\~n}a}, {Ogden},
  {Ordenovic}, {Osinde}, {Pagani}, {Pagano}, {Palaversa}, {Palicio},
  {Pallas-Quintela}, {Panahi}, {Payne-Wardenaar}, {Pe{\~n}alosa Esteller},
  {Penttil{\"a}}, {Pichon}, {Piersimoni}, {Pineau}, {Plachy}, {Plum}, {Poggio},
  {Pr{\v{s}}a}, {Pulone}, {Racero}, {Ragaini}, {Rainer}, {Raiteri}, {Rambaux},
  {Ramos}, {Ramos-Lerate}, {Re Fiorentin}, {Regibo}, {Richards}, {Rios Diaz},
  {Ripepi}, {Riva}, {Rix}, {Rixon}, {Robichon}, {Robin}, {Robin}, {Roelens},
  {Rogues}, {Rohrbasser}, {Romero-G{\'o}mez}, {Rowell}, {Royer}, {Ruz Mieres},
  {Rybicki}, {Sadowski}, {S{\'a}ez N{\'u}{\~n}ez}, {Sagrist{\`a} Sell{\'e}s},
  {Sahlmann}, {Salguero}, {Samaras}, {Sanchez Gimenez}, {Sanna},
  {Santove{\~n}a}, {Sarasso}, {Schultheis}, {Sciacca}, {Segol}, {Segovia},
  {S{\'e}gransan}, {Semeux}, {Shahaf}, {Siddiqui}, {Siebert}, {Siltala},
  {Silvelo}, {Slezak}, {Slezak}, {Smart}, {Snaith}, {Solano}, {Solitro},
  {Souami}, {Souchay}, {Spagna}, {Spina}, {Spoto}, {Steele},
  {Steidelm{\"u}ller}, {Stephenson}, {S{\"u}veges}, {Surdej}, {Szabados},
  {Szegedi-Elek}, {Taris}, {Taylor}, {Teixeira}, {Tolomei}, {Tonello}, {Torra},
  {Torra}, {Torralba Elipe}, {Trabucchi}, {Tsounis}, {Turon}, {Ulla}, {Unger},
  {Vaillant}, {van Dillen}, {van Reeven}, {Vanel}, {Vecchiato}, {Viala},
  {Vicente}, {Voutsinas}, {Weiler}, {Wevers}, {Wyrzykowski}, {Yoldas}, {Yvard},
  {Zhao}, {Zorec}, {Zucker}, \& {Zwitter}}]{Gaia-DR3}
{Gaia Collaboration}, {Vallenari}, A., {Brown}, A.~G.~A., {et~al.} 2023, \aap,
  674, A1

\bibitem[{{Galazutdinov} {et~al.}(2004){Galazutdinov}, {Manic{\`o}},
  {Pirronello}, \& {Kre{\l}owski}}]{Galazutdinov04}
{Galazutdinov}, G.~A., {Manic{\`o}}, G., {Pirronello}, V., \& {Kre{\l}owski},
  J. 2004, \mnras, 355, 169

\bibitem[{{Gallazzi} {et~al.}(2005){Gallazzi}, {Charlot}, {Brinchmann},
  {White}, \& {Tremonti}}]{Gallazzi05}
{Gallazzi}, A., {Charlot}, S., {Brinchmann}, J., {White}, S. D.~M., \&
  {Tremonti}, C.~A. 2005, \mnras, 362, 41

\bibitem[{{Garn} \& {Best}(2010)}]{Garn10}
{Garn}, T. \& {Best}, P.~N. 2010, \mnras, 409, 421

\bibitem[{{Ginsburg} {et~al.}(2019){Ginsburg}, {Sip{\H{o}}cz}, {Brasseur},
  {Cowperthwaite}, {Craig}, {Deil}, {Guillochon}, {Guzman}, {Liedtke}, {Lian
  Lim}, {Lockhart}, {Mommert}, {Morris}, {Norman}, {Parikh}, {Persson},
  {Robitaille}, {Segovia}, {Singer}, {Tollerud}, {de Val-Borro}, {Valtchanov},
  {Woillez}, {Astroquery Collaboration}, \& {a subset of astropy
  Collaboration}}]{Ginsburg19-astroquery}
{Ginsburg}, A., {Sip{\H{o}}cz}, B.~M., {Brasseur}, C.~E., {et~al.} 2019, \aj,
  157, 98

\bibitem[{{Gonz{\'a}lez-Gait{\'a}n} {et~al.}(2021){Gonz{\'a}lez-Gait{\'a}n},
  {de Jaeger}, {Galbany}, {Mour{\~a}o}, {Paulino-Afonso}, \&
  {Filippenko}}]{Gonzalez21}
{Gonz{\'a}lez-Gait{\'a}n}, S., {de Jaeger}, T., {Galbany}, L., {et~al.} 2021,
  \mnras, 508, 4656

\bibitem[{{Gonz{\'a}lez-Gait{\'a}n} {et~al.}(2024){Gonz{\'a}lez-Gait{\'a}n},
  {Guti{\'e}rrez}, {Anderson}, {Morales-Garoffolo}, {Galbany}, {Goswami},
  {Mour{\~a}o}, {Mattila}, \& {Sullivan}}]{GG24}
{Gonz{\'a}lez-Gait{\'a}n}, S., {Guti{\'e}rrez}, C.~P., {Anderson}, J.~P.,
  {et~al.} 2024, \aap, 687, A108

\bibitem[{{Goodwin} {et~al.}(1998){Goodwin}, {Gribbin}, \&
  {Hendry}}]{Goodwin98}
{Goodwin}, S.~P., {Gribbin}, J., \& {Hendry}, M.~A. 1998, The Observatory, 118,
  201

\bibitem[{{Graham} {et~al.}(2014){Graham}, {Plante}, {Tody}, \&
  {Fitzpatrick}}]{Graham14-pyvo}
{Graham}, M., {Plante}, R., {Tody}, D., \& {Fitzpatrick}, M. 2014, {PyVO:
  Python access to the Virtual Observatory}, Astrophysics Source Code Library,
  record ascl:1402.004

\bibitem[{{Graham} {et~al.}(2015){Graham}, {Valenti}, {Fulton}, {Weiss},
  {Shen}, {Kelly}, {Zheng}, {Filippenko}, {Marcy}, {Howell}, {Burt}, \&
  {Rivera}}]{Graham15}
{Graham}, M.~L., {Valenti}, S., {Fulton}, B.~J., {et~al.} 2015, \apj, 801, 136

\bibitem[{{Gupta} {et~al.}(2016){Gupta}, {Kuhlmann}, {Kovacs}, {Spinka},
  {Kessler}, {Goldstein}, {Liotine}, {Pomian}, {D'Andrea}, {Sullivan},
  {Carretero}, {Castander}, {Nichol}, {Finley}, {Fischer}, {Foley}, {Kim},
  {Papadopoulos}, {Sako}, {Scolnic}, {Smith}, {Tucker}, {Uddin}, {Wolf},
  {Yuan}, {Abbott}, {Abdalla}, {Benoit-L{\'e}vy}, {Bertin}, {Brooks}, {Carnero
  Rosell}, {Carrasco Kind}, {Cunha}, {da Costa}, {Desai}, {Doel}, {Eifler},
  {Evrard}, {Flaugher}, {Fosalba}, {Gazta{\~n}aga}, {Gruen}, {Gruendl},
  {James}, {Kuehn}, {Kuropatkin}, {Maia}, {Marshall}, {Miquel}, {Plazas},
  {Romer}, {S{\'a}nchez}, {Schubnell}, {Sevilla-Noarbe}, {Sobreira}, {Suchyta},
  {Swanson}, {Tarle}, {Walker}, \& {Wester}}]{Gupta16}
{Gupta}, R.~R., {Kuhlmann}, S., {Kovacs}, E., {et~al.} 2016, \aj, 152, 154

\bibitem[{{Guti{\'e}rrez} {et~al.}(2017){Guti{\'e}rrez}, {Anderson}, {Hamuy},
  {Gonz{\'a}lez-Gaitan}, {Galbany}, {Dessart}, {Stritzinger}, {Phillips},
  {Morrell}, \& {Folatelli}}]{Gutierrez17b}
{Guti{\'e}rrez}, C.~P., {Anderson}, J.~P., {Hamuy}, M., {et~al.} 2017, \apj,
  850, 90

\bibitem[{{Guti{\'e}rrez} {et~al.}(2025){Guti{\'e}rrez}, {Gonz{\'a}lez-Gaitan},
  \& et~al.}]{Gutierrez24}
{Guti{\'e}rrez}, C.~P., {Gonz{\'a}lez-Gaitan}, S., \& et~al., o. 2025, in
  preparation

\bibitem[{{Guti{\'e}rrez} {et~al.}(2016){Guti{\'e}rrez},
  {Gonz{\'a}lez-Gait{\'a}n}, {Folatelli}, {Pignata}, {Anderson}, {Hamuy},
  {Morrell}, {Stritzinger}, {Taubenberger}, {Bufano}, {Olivares E.}, {Haislip},
  \& {Reichart}}]{Gutierrez16}
{Guti{\'e}rrez}, C.~P., {Gonz{\'a}lez-Gait{\'a}n}, S., {Folatelli}, G.,
  {et~al.} 2016, \aap, 590, A5

\bibitem[{{Hachinger} {et~al.}(2017){Hachinger}, {R{\"o}pke}, {Mazzali},
  {Gal-Yam}, {Maguire}, {Sullivan}, {Taubenberger}, {Ashall}, {Campbell},
  {Elias-Rosa}, {Feindt}, {Greggio}, {Inserra}, {Miluzio}, {Smartt}, \&
  {Young}}]{Hachinger17}
{Hachinger}, S., {R{\"o}pke}, F.~K., {Mazzali}, P.~A., {et~al.} 2017, \mnras,
  471, 491

\bibitem[{Harris {et~al.}(2020)Harris, Millman, van~der Walt, Gommers,
  Virtanen, Cournapeau, Wieser, Taylor, Berg, Smith, Kern, Picus, Hoyer, van
  Kerkwijk, Brett, Haldane, del Río, Wiebe, Peterson, Gérard-Marchant,
  Sheppard, Reddy, Weckesser, Abbasi, Gohlke, \& Oliphant}]{numpy}
Harris, C.~R., Millman, K.~J., van~der Walt, S.~J., {et~al.} 2020, Nature, 585,
  357–362

\bibitem[{{Heidmann} {et~al.}(1972){Heidmann}, {Heidmann}, \& {de
  Vaucouleurs}}]{Heidmann72}
{Heidmann}, J., {Heidmann}, N., \& {de Vaucouleurs}, G. 1972, \memras, 75, 85

\bibitem[{{Ho} {et~al.}(2011){Ho}, {Li}, {Barth}, {Seigar}, \& {Peng}}]{Ho11}
{Ho}, L.~C., {Li}, Z.-Y., {Barth}, A.~J., {Seigar}, M.~S., \& {Peng}, C.~Y.
  2011, \apjs, 197, 21

\bibitem[{{Hoang} {et~al.}(2019){Hoang}, {Tram}, {Lee}, \& {Ahn}}]{Hoang19}
{Hoang}, T., {Tram}, L.~N., {Lee}, H., \& {Ahn}, S.-H. 2019, Nature Astronomy,
  3, 766

\bibitem[{{Holmberg}(1946)}]{Holmberg46}
{Holmberg}, E. 1946, Meddelanden fran Lunds Astronomiska Observatorium Serie
  II, 117, 3

\bibitem[{{Holmberg}(1975)}]{Holmberg75}
{Holmberg}, E. 1975, in Galaxies and the Universe, ed. A.~{Sandage},
  M.~{Sandage}, \& J.~{Kristian}, 123

\bibitem[{{Hubble}(1926)}]{Hubble26}
{Hubble}, E.~P. 1926, \apj, 64, 321

\bibitem[{Hunter(2007)}]{matplotlib}
Hunter, J.~D. 2007, Computing in Science \& Engineering, 9, 90

\bibitem[{{Irwin} {et~al.}(2004){Irwin}, {Lewis}, {Hodgkin}, {Bunclark},
  {Evans}, {McMahon}, {Emerson}, {Stewart}, \& {Beard}}]{VISTA-pipeline}
{Irwin}, M.~J., {Lewis}, J., {Hodgkin}, S., {et~al.} 2004, in Society of
  Photo-Optical Instrumentation Engineers (SPIE) Conference Series, Vol. 5493,
  Optimizing Scientific Return for Astronomy through Information Technologies,
  ed. P.~J. {Quinn} \& A.~{Bridger}, 411--422

\bibitem[{{Jack} \& {Schr{\"o}der}(2019)}]{Jack19}
{Jack}, D. \& {Schr{\"o}der}, K.~P. 2019, \rmxaa, 55, 141

\bibitem[{Johnson {et~al.}(2023)Johnson, Foreman-Mackey, Sick, Leja, Walmsley,
  Tollerud, Leung, Scott, \& Park}]{johnson23-python-fsps}
Johnson, B., Foreman-Mackey, D., Sick, J., {et~al.} 2023, dfm/python-fsps:
  v0.4.6

\bibitem[{{Johnson} {et~al.}(2021){Johnson}, {Leja}, {Conroy}, \&
  {Speagle}}]{Johnson21-prospector}
{Johnson}, B.~D., {Leja}, J., {Conroy}, C., \& {Speagle}, J.~S. 2021, \apjs,
  254, 22

\bibitem[{{Kacprzak} {et~al.}(2015){Kacprzak}, {Churchill}, {Murphy}, \&
  {Cooke}}]{Kacprzak15}
{Kacprzak}, G.~G., {Churchill}, C.~W., {Murphy}, M.~T., \& {Cooke}, J. 2015,
  \mnras, 446, 2861

\bibitem[{{Kangas} {et~al.}(2016)}]{Kangas16}
{Kangas}, T. {et~al.} 2016, \mnras, 456, 323

\bibitem[{{Kauffmann} {et~al.}(2003){Kauffmann}, {Heckman}, {White}, {Charlot},
  {Tremonti}, {Brinchmann}, {Bruzual}, {Peng}, {Seibert}, {Bernardi},
  {Blanton}, {Brinkmann}, {Castander}, {Cs{\'a}bai}, {Fukugita}, {Ivezic},
  {Munn}, {Nichol}, {Padmanabhan}, {Thakar}, {Weinberg}, \&
  {York}}]{Kauffmann03}
{Kauffmann}, G., {Heckman}, T.~M., {White}, S. D.~M., {et~al.} 2003, \mnras,
  341, 33

\bibitem[{{Kelly}(2007)}]{Kelly07}
{Kelly}, B.~C. 2007, \apj, 665, 1489

\bibitem[{Kolmogorov(1933)}]{ks-test1}
Kolmogorov, A.~N. 1933, Giorn Dell'inst Ital Degli Att, 4, 89

\bibitem[{{Kron}(1980)}]{Kron80}
{Kron}, R.~G. 1980, \apjs, 43, 305

\bibitem[{{Kroupa}(2001)}]{Kroupa01}
{Kroupa}, P. 2001, \mnras, 322, 231

\bibitem[{{Lang}(2014)}]{unWISE}
{Lang}, D. 2014, \aj, 147, 108

\bibitem[{{Leja} {et~al.}(2017){Leja}, {Johnson}, {Conroy}, {van Dokkum}, \&
  {Byler}}]{Leja17}
{Leja}, J., {Johnson}, B.~D., {Conroy}, C., {van Dokkum}, P.~G., \& {Byler}, N.
  2017, \apj, 837, 170

\bibitem[{{Lewis} {et~al.}(2010){Lewis}, {Irwin}, \&
  {Bunclark}}]{VISTA-pipeline2}
{Lewis}, J.~R., {Irwin}, M., \& {Bunclark}, P. 2010, in Astronomical Society of
  the Pacific Conference Series, Vol. 434, Astronomical Data Analysis Software
  and Systems XIX, ed. Y.~{Mizumoto}, K.~I. {Morita}, \& M.~{Ohishi}, 91

\bibitem[{{Lin} {et~al.}(2020){Lin}, {Calzetti}, {Kong}, {Adamo}, {Cignoni},
  {Cook}, {Dale}, {Grasha}, {Grebel}, {Messa}, {Sacchi}, \& {Smith}}]{Lin20}
{Lin}, Z., {Calzetti}, D., {Kong}, X., {et~al.} 2020, \apj, 896, 16

\bibitem[{{Magnier} {et~al.}(2020){Magnier}, {Chambers}, {Flewelling},
  {Hoblitt}, {Huber}, {Price}, {Sweeney}, {Waters}, {Denneau}, {Draper},
  {Hodapp}, {Jedicke}, {Kaiser}, {Kudritzki}, {Metcalfe}, {Stubbs}, \&
  {Wainscoat}}]{Magnier20-PS1}
{Magnier}, E.~A., {Chambers}, K.~C., {Flewelling}, H.~A., {et~al.} 2020, \apjs,
  251, 3

\bibitem[{{Maguire} {et~al.}(2013){Maguire}, {Sullivan}, {Patat}, {Gal-Yam},
  {Hook}, {Dhawan}, {Howell}, {Mazzali}, {Nugent}, {Pan}, {Podsiadlowski},
  {Simon}, {Sternberg}, {Valenti}, {Baltay}, {Bersier}, {Blagorodnova}, {Chen},
  {Ellman}, {Feindt}, {F{\"o}rster}, {Fraser}, {Gonz{\'a}lez-Gait{\'a}n},
  {Graham}, {Guti{\'e}rrez}, {Hachinger}, {Hadjiyska}, {Inserra}, {Knapic},
  {Laher}, {Leloudas}, {Margheim}, {McKinnon}, {Molinaro}, {Morrell}, {Ofek},
  {Rabinowitz}, {Rest}, {Sand}, {Smareglia}, {Smartt}, {Taddia}, {Walker},
  {Walton}, \& {Young}}]{Maguire13}
{Maguire}, K., {Sullivan}, M., {Patat}, F., {et~al.} 2013, \mnras, 436, 222

\bibitem[{{Martin}(2005)}]{Martin05}
{Martin}, C.~L. 2005, \apj, 621, 227

\bibitem[{{McMahon} {et~al.}(2013){McMahon}, {Banerji}, {Gonzalez}, {Koposov},
  {Bejar}, {Lodieu}, {Rebolo}, \& {VHS Collaboration}}]{VISTA}
{McMahon}, R.~G., {Banerji}, M., {Gonzalez}, E., {et~al.} 2013, The Messenger,
  154, 35

\bibitem[{Meisner {et~al.}(2017{\natexlab{a}})Meisner, Lang, \&
  Schlegel}]{unWISE3yr}
Meisner, A.~M., Lang, D., \& Schlegel, D.~J. 2017{\natexlab{a}}, The
  Astronomical Journal, 154, 161

\bibitem[{Meisner {et~al.}(2017{\natexlab{b}})Meisner, Lang, \&
  Schlegel}]{unWISE1yr}
Meisner, A.~M., Lang, D., \& Schlegel, D.~J. 2017{\natexlab{b}}, The
  Astronomical Journal, 153, 38

\bibitem[{{Milisavljevic} {et~al.}(2014){Milisavljevic}, {Margutti},
  {Crabtree}, {Foster}, {Soderberg}, {Fesen}, {Parrent}, {Sanders}, {Drout},
  {Kamble}, {Chakraborti}, {Pickering}, {Cenko}, {Silverman}, {Filippenko},
  {Kirshner}, {Mazzali}, {Maeda}, {Marion}, {Vinko}, \&
  {Wheeler}}]{Milisavljevic14}
{Milisavljevic}, D., {Margutti}, R., {Crabtree}, K.~N., {et~al.} 2014, \apjl,
  782, L5

\bibitem[{Miller(1981)}]{Miller81}
Miller, R.~G. 1981

\bibitem[{{Morganson} {et~al.}(2018){Morganson}, {Gruendl}, {Menanteau},
  {Carrasco Kind}, {Chen}, {Daues}, {Drlica-Wagner}, {Friedel}, {Gower},
  {Johnson}, {Johnson}, {Kessler}, {Paz-Chinch{\'o}n}, {Petravick}, {Pond},
  {Yanny}, {Allam}, {Armstrong}, {Barkhouse}, {Bechtol}, {Benoit-L{\'e}vy},
  {Bernstein}, {Bertin}, {Buckley-Geer}, {Covarrubias}, {Desai}, {Diehl},
  {Goldstein}, {Gruen}, {Li}, {Lin}, {Marriner}, {Mohr}, {Neilsen}, {Ngeow},
  {Paech}, {Rykoff}, {Sako}, {Sevilla-Noarbe}, {Sheldon}, {Sobreira}, {Tucker},
  {Wester}, \& {DES Collaboration}}]{DES_pipeline}
{Morganson}, E., {Gruendl}, R.~A., {Menanteau}, F., {et~al.} 2018, \pasp, 130,
  074501

\bibitem[{{Morokuma-Matsui} \& {Baba}(2015)}]{Morokuma-Matsui15}
{Morokuma-Matsui}, K. \& {Baba}, J. 2015, \mnras, 454, 3792

\bibitem[{Morrissey {et~al.}(2008)Morrissey, Conrow, Barlow, Small, Seibert,
  Wyder, Budavari, Arnouts, Friedman, Forster, Martin, Neff, Schiminovich,
  Bianchi, Donas, Heckman, Lee, Madore, Milliard, \& Yi}]{GALEX}
Morrissey, P., Conrow, T., Barlow, T., {et~al.} 2008, The Astrophysical Journal
  Supplement Series, 173, 682

\bibitem[{{Mu{\~n}oz-Mateos} {et~al.}(2011){Mu{\~n}oz-Mateos}, {Boissier}, {Gil
  de Paz}, {Zamorano}, {Kennicutt}, {Moustakas}, {Prantzos}, \&
  {Gallego}}]{Munoz-Mateos11}
{Mu{\~n}oz-Mateos}, J.~C., {Boissier}, S., {Gil de Paz}, A., {et~al.} 2011,
  \apj, 731, 10

\bibitem[{{M{\"u}ller-Bravo} \& {Galbany}(2022)}]{Muller-Bravo22}
{M{\"u}ller-Bravo}, T. \& {Galbany}, L. 2022, The Journal of Open Source
  Software, 7, 4508

\bibitem[{{Namiki} {et~al.}(2021){Namiki}, {Koyama}, {Koyama}, {Yamashita},
  {Hayashi}, {Haynes}, {Shimakawa}, \& {Onodera}}]{Namiki21}
{Namiki}, S.~V., {Koyama}, Y., {Koyama}, S., {et~al.} 2021, \apj, 918, 68

\bibitem[{{Noll} {et~al.}(2009){Noll}, {Burgarella}, {Giovannoli}, {Buat},
  {Marcillac}, \& {Mu{\~n}oz-Mateos}}]{Noll09}
{Noll}, S., {Burgarella}, D., {Giovannoli}, E., {et~al.} 2009, \aap, 507, 1793

\bibitem[{{Noordermeer} \& {van der Hulst}(2007)}]{Noordermeer07}
{Noordermeer}, E. \& {van der Hulst}, J.~M. 2007, \mnras, 376, 1480

\bibitem[{{Orellana} {et~al.}(2017){Orellana}, {Nagar}, {Elbaz},
  {Calder{\'o}n-Castillo}, {Leiton}, {Ibar}, {Magnelli}, {Daddi}, {Messias},
  {Cerulo}, \& {Slater}}]{Orellana17}
{Orellana}, G., {Nagar}, N.~M., {Elbaz}, D., {et~al.} 2017, \aap, 602, A68

\bibitem[{{Park} {et~al.}(2015){Park}, {Jeong}, \& {Yi}}]{Park15}
{Park}, J., {Jeong}, H., \& {Yi}, S.~K. 2015, \apj, 809, 91

\bibitem[{{Pascucci} {et~al.}(2015){Pascucci}, {Edwards}, {Heyer}, {Rigliaco},
  {Hillenbrand}, {Gorti}, {Hollenbach}, \& {Simon}}]{Pascucci15}
{Pascucci}, I., {Edwards}, S., {Heyer}, M., {et~al.} 2015, \apj, 814, 14

\bibitem[{{Patat} {et~al.}(2007){Patat}, {Chandra}, {Chevalier}, {Justham},
  {Podsiadlowski}, {Wolf}, {Gal-Yam}, {Pasquini}, {Crawford}, {Mazzali},
  {Pauldrach}, {Nomoto}, {Benetti}, {Cappellaro}, {Elias-Rosa}, {Hillebrandt},
  {Leonard}, {Pastorello}, {Renzini}, {Sabbadin}, {Simon}, \&
  {Turatto}}]{Patat07}
{Patat}, F., {Chandra}, P., {Chevalier}, R., {et~al.} 2007, Science, 317, 924

\bibitem[{{Paxton} {et~al.}(2011){Paxton}, {Bildsten}, {Dotter}, {Herwig},
  {Lesaffre}, \& {Timmes}}]{Paxton11}
{Paxton}, B., {Bildsten}, L., {Dotter}, A., {et~al.} 2011, \apjs, 192, 3

\bibitem[{{Paxton} {et~al.}(2013){Paxton}, {Cantiello}, {Arras}, {Bildsten},
  {Brown}, {Dotter}, {Mankovich}, {Montgomery}, {Stello}, {Timmes}, \&
  {Townsend}}]{Paxton13}
{Paxton}, B., {Cantiello}, M., {Arras}, P., {et~al.} 2013, \apjs, 208, 4

\bibitem[{{Paxton} {et~al.}(2015){Paxton}, {Marchant}, {Schwab}, {Bauer},
  {Bildsten}, {Cantiello}, {Dessart}, {Farmer}, {Hu}, {Langer}, {Townsend},
  {Townsley}, \& {Timmes}}]{Paxton15}
{Paxton}, B., {Marchant}, P., {Schwab}, J., {et~al.} 2015, \apjs, 220, 15

\bibitem[{{Phillips} {et~al.}(2013)}]{Phillips13}
{Phillips}, M.~M. {et~al.} 2013, \apj, 779, 38

\bibitem[{{Qin} {et~al.}(2022){Qin}, {Zheng}, {Fang}, {Pan}, {Wuyts}, {Shi},
  {Peng}, {Gonzalez}, {Bian}, {Huang}, {Gu}, {Liu}, {Tan}, {Shi}, {Ren},
  {Zhang}, {Qiao}, {Wen}, \& {Liu}}]{Qin22}
{Qin}, J., {Zheng}, X.~Z., {Fang}, M., {et~al.} 2022, \mnras, 511, 765

\bibitem[{{Ritchey} {et~al.}(2015){Ritchey}, {Welty}, {Dahlstrom}, \&
  {York}}]{Ritchey15}
{Ritchey}, A.~M., {Welty}, D.~E., {Dahlstrom}, J.~A., \& {York}, D.~G. 2015,
  \apj, 799, 197

\bibitem[{{Roberts} \& {Haynes}(1994)}]{Roberts94}
{Roberts}, M.~S. \& {Haynes}, M.~P. 1994, \araa, 32, 115

\bibitem[{{Rodr{\'\i}guez} \& {Padilla}(2013)}]{Rodriguez13}
{Rodr{\'\i}guez}, S. \& {Padilla}, N.~D. 2013, \mnras, 434, 2153

\bibitem[{{Rupke} {et~al.}(2005{\natexlab{a}}){Rupke}, {Veilleux}, \&
  {Sanders}}]{Rupke05}
{Rupke}, D.~S., {Veilleux}, S., \& {Sanders}, D.~B. 2005{\natexlab{a}}, \apjs,
  160, 87

\bibitem[{{Rupke} {et~al.}(2005{\natexlab{b}}){Rupke}, {Veilleux}, \&
  {Sanders}}]{Rupke05b}
{Rupke}, D.~S., {Veilleux}, S., \& {Sanders}, D.~B. 2005{\natexlab{b}}, \apjs,
  160, 115

\bibitem[{{Saintonge} \& {Catinella}(2022)}]{Saintonge22}
{Saintonge}, A. \& {Catinella}, B. 2022, \araa, 60, 319

\bibitem[{{Salim} {et~al.}(2007){Salim}, {Rich}, {Charlot}, {Brinchmann},
  {Johnson}, {Schiminovich}, {Seibert}, {Mallery}, {Heckman}, {Forster},
  {Friedman}, {Martin}, {Morrissey}, {Neff}, {Small}, {Wyder}, {Bianchi},
  {Donas}, {Lee}, {Madore}, {Milliard}, {Szalay}, {Welsh}, \& {Yi}}]{Salim07}
{Salim}, S., {Rich}, R.~M., {Charlot}, S., {et~al.} 2007, \apjs, 173, 267

\bibitem[{{S{\'a}nchez-Menguiano} {et~al.}(2018){S{\'a}nchez-Menguiano},
  {S{\'a}nchez}, {P{\'e}rez}, {Ruiz-Lara}, {Galbany}, {Anderson},
  {Kr{\"u}hler}, {Kuncarayakti}, \& {Lyman}}]{Sanchez-Menguiano18}
{S{\'a}nchez-Menguiano}, L., {S{\'a}nchez}, S.~F., {P{\'e}rez}, I., {et~al.}
  2018, \aap, 609, A119

\bibitem[{{Scargle} {et~al.}(2013){Scargle}, {Norris}, {Jackson}, \&
  {Chiang}}]{Scargle13-bayesianblocks}
{Scargle}, J.~D., {Norris}, J.~P., {Jackson}, B., \& {Chiang}, J. 2013, \apj,
  764, 167

\bibitem[{{Simha} {et~al.}(2014){Simha}, {Weinberg}, {Conroy}, {Dave},
  {Fardal}, {Katz}, \& {Oppenheimer}}]{Simha14}
{Simha}, V., {Weinberg}, D.~H., {Conroy}, C., {et~al.} 2014, arXiv e-prints,
  arXiv:1404.0402

\bibitem[{{Simon} {et~al.}(2009){Simon}, {Gal-Yam}, {Gnat}, {Quimby},
  {Ganeshalingam}, {Silverman}, {Blondin}, {Li}, {Filippenko}, {Wheeler},
  {Kirshner}, {Patat}, {Nugent}, {Foley}, {Vogt}, {Butler}, {Peek},
  {Rosolowsky}, {Herczeg}, {Sauer}, \& {Mazzali}}]{Simon09}
{Simon}, J.~D., {Gal-Yam}, A., {Gnat}, O., {et~al.} 2009, \apj, 702, 1157

\bibitem[{{Skrutskie} {et~al.}(2006){Skrutskie}, {Cutri}, {Stiening},
  {Weinberg}, {Schneider}, {Carpenter}, {Beichman}, {Capps}, {Chester},
  {Elias}, {Huchra}, {Liebert}, {Lonsdale}, {Monet}, {Price}, {Seitzer},
  {Jarrett}, {Kirkpatrick}, {Gizis}, {Howard}, {Evans}, {Fowler}, {Fullmer},
  {Hurt}, {Light}, {Kopan}, {Marsh}, {McCallon}, {Tam}, {Van Dyk}, \&
  {Wheelock}}]{2MASS}
{Skrutskie}, M.~F., {Cutri}, R.~M., {Stiening}, R., {et~al.} 2006, \aj, 131,
  1163

\bibitem[{Smirnov(1939)}]{ks-test2}
Smirnov, N.~V. 1939, Bull. Math. Univ. Moscou, 2, 3

\bibitem[{{Sollerman} {et~al.}(2005){Sollerman}, {Cox}, {Mattila},
  {Ehrenfreund}, {Kaper}, {Leibundgut}, \& {Lundqvist}}]{Sollerman05}
{Sollerman}, J., {Cox}, N., {Mattila}, S., {et~al.} 2005, \aap, 429, 559

\bibitem[{Spearman(1904)}]{Spearman1904}
Spearman, C. 1904, The American Journal of Psychology, 15, 201

\bibitem[{{Sternberg} {et~al.}(2011){Sternberg}, {Gal-Yam}, {Simon}, {Leonard},
  {Quimby}, {Phillips}, {Morrell}, {Thompson}, {Ivans}, {Marshall},
  {Filippenko}, {Marcy}, {Bloom}, {Patat}, {Foley}, {Yong}, {Penprase},
  {Beeler}, {Allende Prieto}, \& {Stringfellow}}]{Sternberg11}
{Sternberg}, A., {Gal-Yam}, A., {Simon}, J.~D., {et~al.} 2011, Science, 333,
  856

\bibitem[{{Sternberg} {et~al.}(2013){Sternberg}, {Gal Yam}, {Simon}, {Patat},
  {Hillebrandt}, {Phillips}, {Foley}, {Thompson}, {Morrell}, {Chomiuk},
  {Soderberg}, {Yong}, {Kraus}, {Herczeg}, {Hsiao}, {Raskutti}, {Cohen},
  {Mazzali}, \& {Nomoto}}]{Sternberg13}
{Sternberg}, A., {Gal Yam}, A., {Simon}, J.~D., {et~al.} 2013, ArXiv e-prints
  [\eprint[arXiv]{1311.3645}]

\bibitem[{{Sullivan} {et~al.}(2006)}]{Sullivan06}
{Sullivan}, M. {et~al.} 2006, \apj, 648, 868

\bibitem[{{The pandas development team}(2020)}]{pandas}
{The pandas development team}. 2020, pandas-dev/pandas: Pandas

\bibitem[{Tonidandel \& LeBreton(2015)}]{kda}
Tonidandel, S. \& LeBreton, J. 2015, Journal of Business and Psychology, 30,
  207, publisher Copyright: {\textcopyright} 2014, Springer Science+Business
  Media New York.

\bibitem[{{Toy} {et~al.}(2025){Toy}, {Wiseman}, {Sullivan}, {Scolnic},
  {Vincenzi}, {Brout}, {Davis}, {Frohmaier}, {Galbany}, {Lidman}, {Lee},
  {Kelsey}, {Kessler}, {M{\"o}ller}, {Popovic}, {S{\'a}nchez}, {Shah}, {Smith},
  {Allam}, {Aguena}, {Alves}, {Bacon}, {Brooks}, {Burke}, {Carnero Rosell},
  {Carretero}, {da Costa}, {Pereira}, {Desai}, {Diehl}, {Doel},
  {Drlica-Wagner}, {Everett}, {Ferrero}, {Flaugher}, {Frieman},
  {Garc{\'\i}a-Bellido}, {Gatti}, {Gaztanaga}, {Giannini}, {Gruendl},
  {Gutierrez}, {Hinton}, {Hollowood}, {Honscheid}, {James}, {Lahav}, {Lee},
  {Marshall}, {Mena-Fern{\'a}ndez}, {Miquel}, {Palmese}, {Pieres}, {Plazas
  Malag{\'o}n}, {Romer}, {Samuroff}, {Sanchez}, {Sanchez Cid}, {Schubnell},
  {Suchyta}, {Swanson}, {Tarle}, {Tucker}, {Vikram}, {Walker}, \&
  {Weaverdyck}}]{Toy25}
{Toy}, M., {Wiseman}, P., {Sullivan}, M., {et~al.} 2025, \mnras
  [\eprint[arXiv]{2408.03749}]

\bibitem[{{Tuffs} {et~al.}(2004){Tuffs}, {Popescu}, {Voelk}, {Kylafis}, \&
  {Dopita}}]{Tuffs04}
{Tuffs}, R.~J., {Popescu}, C.~C., {Voelk}, H.~J., {Kylafis}, N., \& {Dopita},
  M.~A. 2004, Astronomische Nachrichten Supplement, 325, 1.118

\bibitem[{Virtanen {et~al.}(2020)Virtanen, Gommers, Oliphant, Haberland, Reddy,
  Cournapeau, Burovski, Peterson, Weckesser, Bright, {van der Walt}, Brett,
  Wilson, Millman, Mayorov, Nelson, Jones, Kern, Larson, Carey, Polat, Feng,
  Moore, {VanderPlas}, Laxalde, Perktold, Cimrman, Henriksen, Quintero, Harris,
  Archibald, Ribeiro, Pedregosa, {van Mulbregt}, \& {SciPy 1.0
  Contributors}}]{scipy}
Virtanen, P., Gommers, R., Oliphant, T.~E., {et~al.} 2020, Nature Methods, 17,
  261

\bibitem[{{Walcher} {et~al.}(2011){Walcher}, {Groves}, {Budav{\'a}ri}, \&
  {Dale}}]{Walcher11}
{Walcher}, J., {Groves}, B., {Budav{\'a}ri}, T., \& {Dale}, D. 2011, \apss,
  331, 1

\bibitem[{{Waters} {et~al.}(2020){Waters}, {Magnier}, {Price}, {Chambers},
  {Burgett}, {Draper}, {Flewelling}, {Hodapp}, {Huber}, {Jedicke}, {Kaiser},
  {Kudritzki}, {Lupton}, {Metcalfe}, {Rest}, {Sweeney}, {Tonry}, {Wainscoat},
  \& {Wood-Vasey}}]{Waters20-PS1}
{Waters}, C.~Z., {Magnier}, E.~A., {Price}, P.~A., {et~al.} 2020, \apjs, 251, 4

\bibitem[{{Welty} {et~al.}(2014){Welty}, {Ritchey}, {Dahlstrom}, \&
  {York}}]{Welty14}
{Welty}, D.~E., {Ritchey}, A.~M., {Dahlstrom}, J.~A., \& {York}, D.~G. 2014,
  \apj, 792, 106

\bibitem[{{W}es {M}c{K}inney(2010)}]{mckinney-pandas}
{W}es {M}c{K}inney. 2010, in {P}roceedings of the 9th {P}ython in {S}cience
  {C}onference, ed. {S}t\'efan van~der {W}alt \& {J}arrod {M}illman, 56 -- 61

\bibitem[{{Xu} {et~al.}(2023){Xu}, {Heckman}, {Yoshida}, {Henry}, \&
  {Ohyama}}]{Xu23}
{Xu}, X., {Heckman}, T., {Yoshida}, M., {Henry}, A., \& {Ohyama}, Y. 2023,
  \apj, 956, 142

\bibitem[{{Yuan} {et~al.}(2021){Yuan}, {Lu}, {Shen}, \& {Boquien}}]{Yuan21}
{Yuan}, F.-T., {Lu}, J., {Shen}, S., \& {Boquien}, M. 2021, \apj, 911, 145

\bibitem[{{Yuan} \& {Zhu}(2004)}]{Yuan04}
{Yuan}, Q.-r. \& {Zhu}, C.-x. 2004, \caa, 28, 127

\end{thebibliography}


\appendix

\section{Relation among different lines}\label{ap:lines}

\begin{figure}
\centering
\includegraphics[width=\columnwidth]{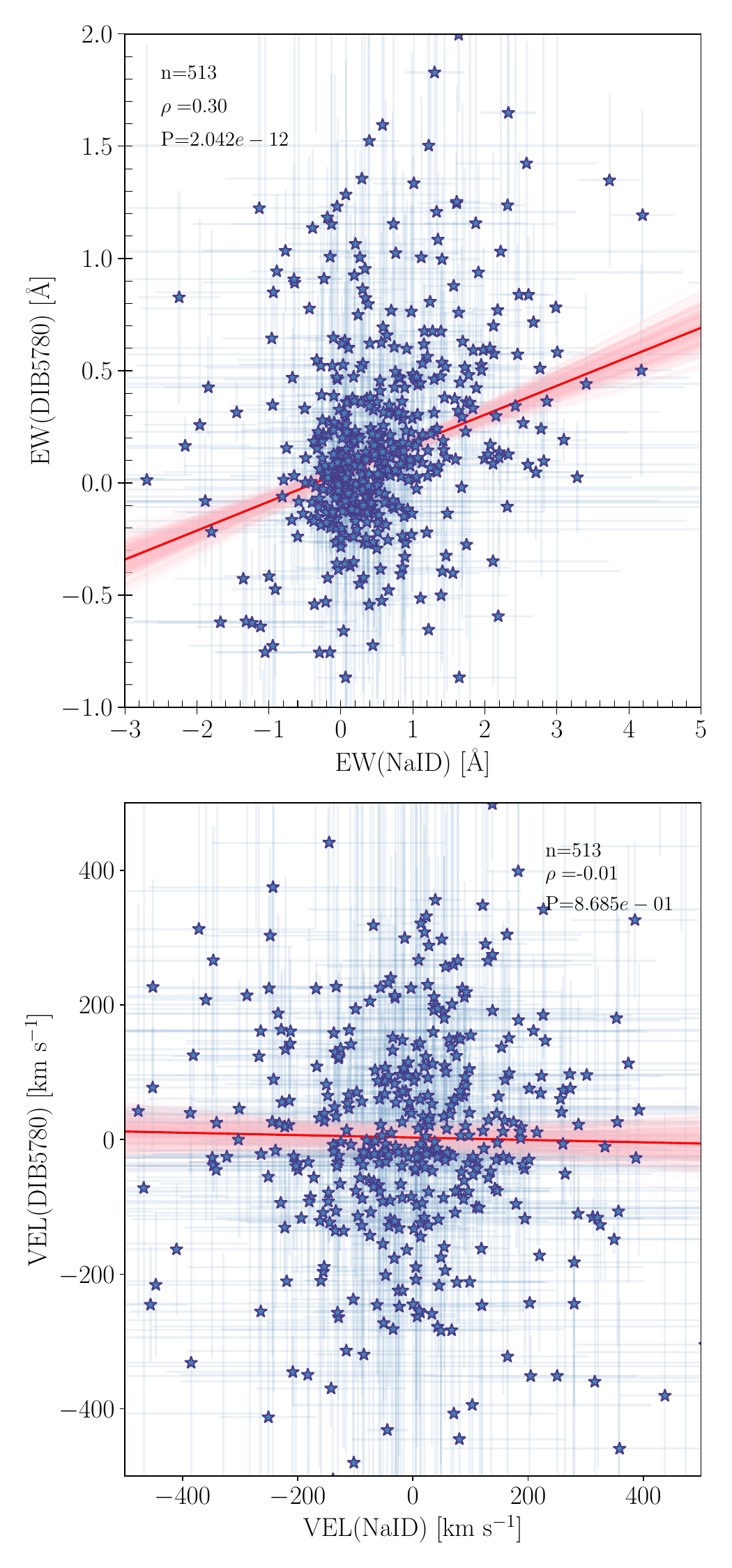}
\caption{Comparison of \naid\ lines with DIB-5780 for EW (top) and VEL (bottom). The median linear fit is shown in red while in pink are various realizations of the MCMC. The correlation is $\rho=0.30$ for the EW and $\rho=-0.01$ for the VEL, while the median slope is $0.127^{+0.015}_{-0.016}$, 8.4$\sigma$ above zero, for EW, and $-0.914^{+1.96}_{-1.87}$ for VEL,  fully consistent with zero.}
\label{fig:EW-NaID-DIB5780}
\end{figure}

In this section, we compare the EW and VEL measured from different narrow lines in SN spectra. We consider the following lines: \naid, \caii H \& K, \ki 1 \& 2, and DIBs 5780, 4428 and 6283.  An example of the relation between \naid\ and DIB-5780 is shown in Figure~\ref{fig:EW-NaID-DIB5780} including a Markov Chain Monte Carlo (MCMC) linear fit to the data that includes both errors in $x$ and $y$ \citep{Kelly07}. Despite the high dispersion and weak correlation ($\rho=0.30$), these two tracers have a significantly positive EW slope ($0.126\pm0.015$). As can be seen for these two lines but also in the comparison with other lines, the sodium lines are stronger and more distinguishable than the rest, especially with low-resolution spectra. Nonetheless, other lines such as the DIB-5780 appear quite clearly in SNe with high extinction. Table~\ref{table:medlines} confirms that the median EW values are largest for the \naid\ lines. 

\begin{figure*}
\centering
\includegraphics[width=\textwidth,clip=True]{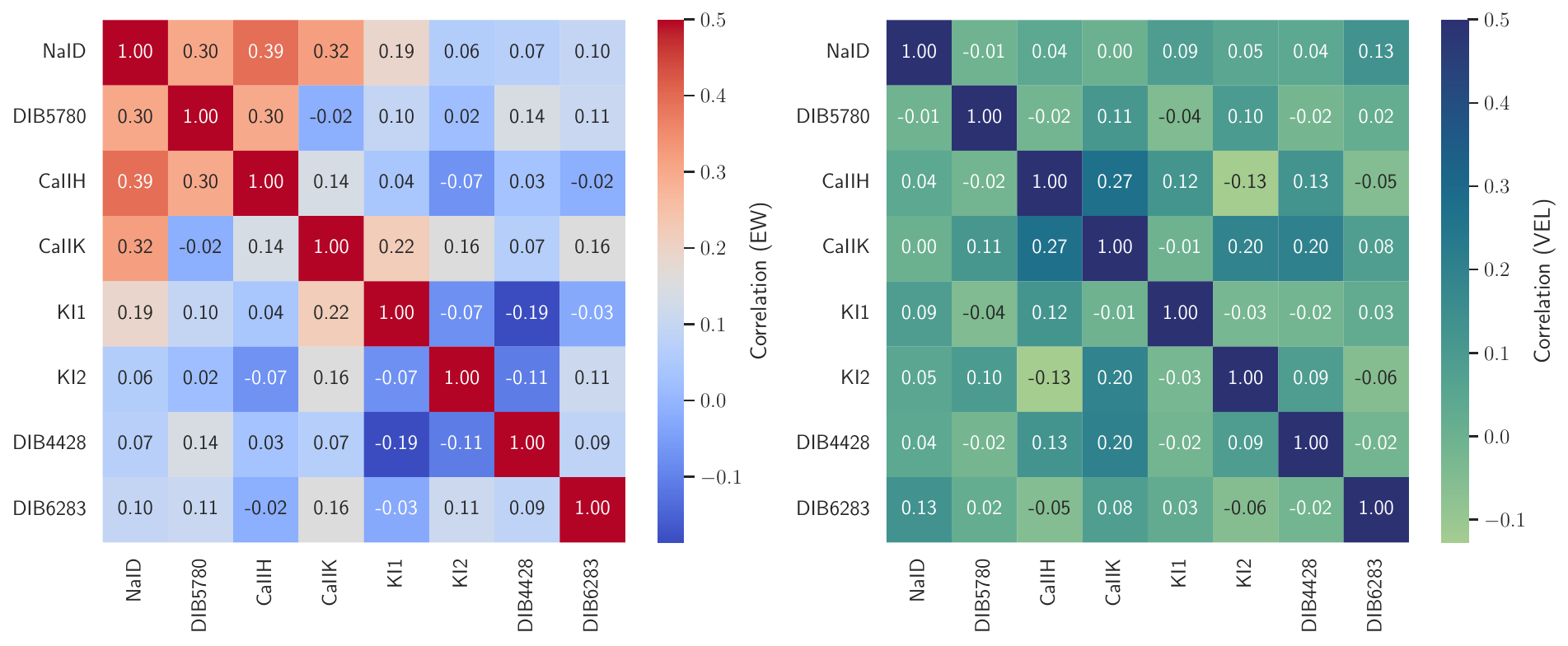}
\caption{Spearman correlation coefficients for the EW (left) and the VEL (right) of various combinations of ISM tracers in SN spectra.}
\label{fig:corr}
\end{figure*}

The correlation between every pair of lines for both EW and VEL is shown in Figure~\ref{fig:corr}. It can be seen that the correlation between the EW lines is generally weak and for the VEL it is rather absent. The negative correlations in EW are driven by low-number statistics whereas the linear fits for these cases give significantly positive slopes. It is interesting that the EW slopes between potassium lines and DIBs are close to one while they differ with lines of ionized calcium and sodium, in agreement with previous findings \citep{Galazutdinov04}.


\begin{table}
\centering
\caption{Median and median absolute deviations of EW and VEL distributions for various lines. 
}
\label{table:medlines}
\renewcommand{\arraystretch}{1.4}
\begin{tabular}{c|c|c}
\hline
Line & $<$EW$>$ (\AA) & $<$VEL$>$ (km/s)  \\
\hline
\hline
\naid\ & $0.35\pm0.50$ & $-3\pm123$ \\
DIB-5780 & $0.11\pm0.23$ & $-4\pm124$ \\
\caii H & $0.23\pm0.46$ & $-39\pm168$ \\
\caii K & $0.21\pm0.33$ & $-60\pm185$ \\
\ki 1 & $0.03\pm0.34$ & $-40\pm165$ \\
\ki 2 & $-0.06\pm0.33$ & $10\pm159$ \\
DIB-4428 & $0.06\pm0.24$ & $2\pm175$ \\
DIB-6283 & $0.02\pm0.24$ & $-1\pm146$ \\
\hline
\end{tabular}
\end{table}

\section{Relation between EW and VEL}\label{ap:EW-VEL}

\begin{figure}
\centering
\includegraphics[width=\columnwidth]{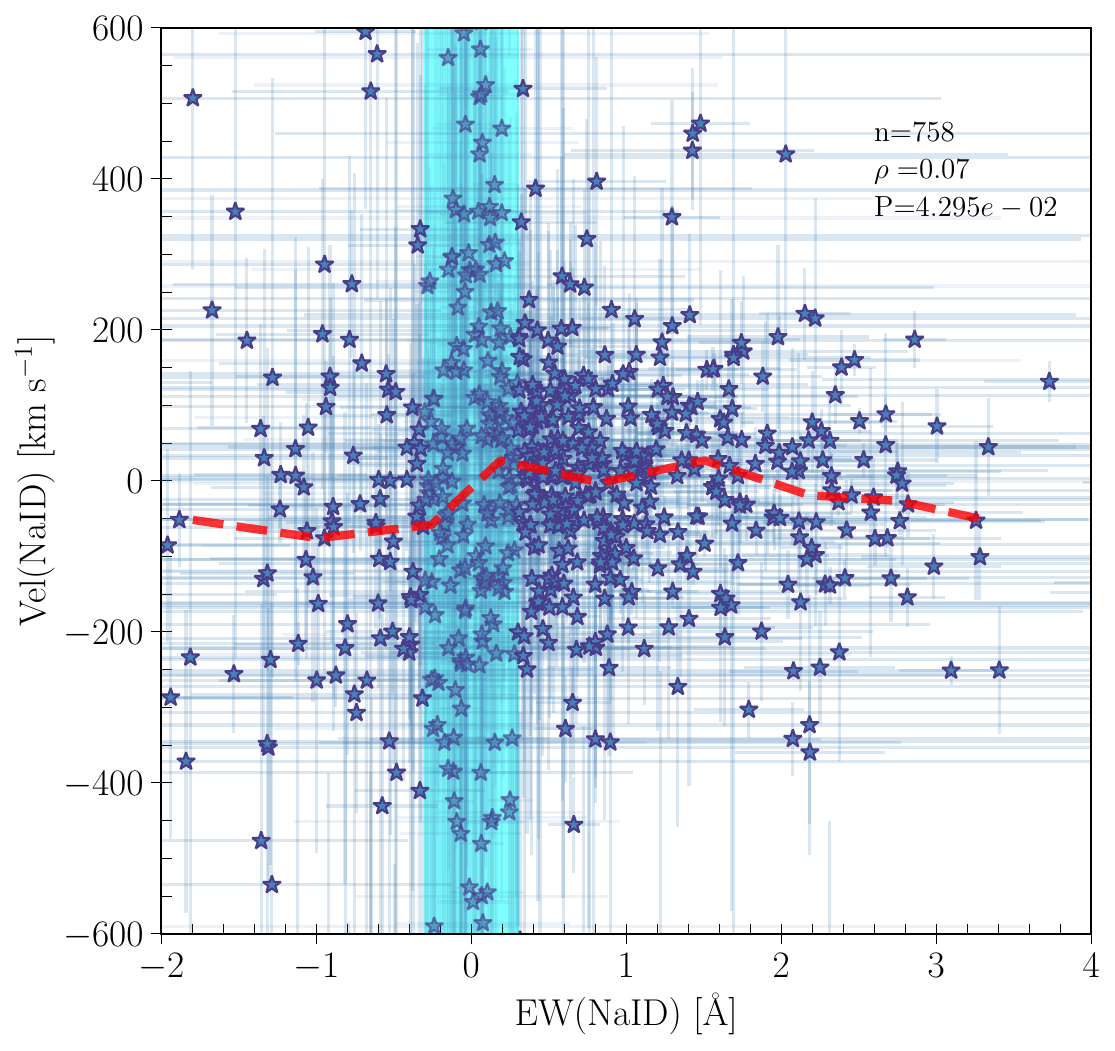}
\caption{Comparison of \naid\ EW and VEL. The red line portrays the rolling median and the vertical cyan region between $\pm0.3$\AA\, represents the values for which the line is very weak and the uncertainty in VEL is rather large.}
\label{fig:EW-VEL}
\end{figure}

We present here the relation between EW and VEL of the narrow lines. In Figure~\ref{fig:EW-VEL} we show the case of \naid. The VEL values are spread across $\pm500$ km/s around zero and there is no correlation with EW ($\rho=0.07$). A linear fit to the data results in a slope of $-0.08\pm5.8$, fully consistent with zero. Obtaining accurate velocity measurements when the line is weak or absent is very difficult or impossible. This is reflected in the large VEL dispersion seen at values between $\pm0.3$\AA\,. Therefore, we do not take SNe in this region into account when doing tests of VEL with galaxy properties in section~\ref{sec:res}. For the other lines, which are weaker, this cut eliminates a higher fraction of SNe. We confirm that we do not find any strong relation between EW and VEL for any of the ISM tracers considered here. 

\section{Robustness of the K-S tests}
\label{ap:KSquart}

The K-S tests performed in this study are very dependent on the location of the split of the two samples being compared. Although in Section~\ref{sec:res} we do a sweep changing the split between 40\% and 60\% of the distribution, and perform a bootstrap simulation, here we do some additional tests. For the global sample, we also take the median split of the MPA/JHU sample (see Figure~\ref{fig:galglobal}), which lies outside the 40-60\% of our sample and find consistent results for the global mass but somewhat lower p-values for SFR ($p=0.035$, $\mathbf{P}=43$\%) and sSFR ($p=0.10$, $\mathbf{P}=24$\%). Moreover, in Table~\ref{table:KSquart}, we perform K-S tests comparing the EW of the lower and upper quartiles of each galaxy parameter distribution. We confirm all the significant p-values found in Table~\ref{table:KS} and also slightly more significant p-values for SFR and sSFR. Finally, although no significant p-values were found for the same sample using the MW lines, we also consider here the look-elsewhere effect (LEE), in which the probability of finding significant differences between populations increases with the number of multiple parameters searched in a given dataset \citep{Miller81}. The LEE can be compensated by testing each individual hypothesis at a significance corrected by the number of parameters tested \citep{Bonferroni36}. So, in Table~\ref{table:KSquart}, we include an extra column to the right, in which we recalculate the probabilities of Table~\ref{table:KS} for a limiting p-value now of $0.05/N=0.005$, where $N=10$ is the number of parameters searched: normalised offset, T-type, inclination and the seven stellar/dust parameters from SED fits (local or global). We leave out other highly correlated parameters (e.g. angular offset). We find that the probabilities still provide significantly high values for all parameters except for the dust index and the local age.

\renewcommand{\arraystretch}{1.3}
\begin{table}
\begin{threeparttable}
\small
\centering
\caption{K-S quartile statistics for \naid EW divided according to galaxy properties.}
\label{table:KSquart}
\renewcommand{\arraystretch}{1.6}
\begin{tabular}{c|c|cccc}
Property & Nr & $<D^{\mathrm{EW}}_{\mathrm{MC}}>$ & $<p_{MC}^{\mathrm{EW}}>$ & $\mathbf{P}$ & $\mathbf{P}_{\mathbf{LEE}}$\\
\hline                                         
\hline
\multicolumn{5}{c}{\textbf{General properties}}\\
\hdashline 
\boldsymbol{$\overline{\Delta\alpha}$} & 334          & 0.39   & $6.1\times10^{-12}$ & {\bf 100\%} & {\bf 100\%} \\ 
{\bf T-type}                           & 330          & 0.19   & $5.0\times10^{-4}$  & {\bf 95\%}  & {\bf 92\%}  \\
\boldsymbol{$i(^{\circ})$}             & 333          & 0.20   & $1.7\times10^{-3}$  & {\bf 86\%}  & {\bf 96\%}  \\
\hline                      
\multicolumn{5}{c}{\textbf{Local properties}}\\
\hdashline     
\bf{M$_*^L$}                           & $179$        & 0.32   & $1.7\times10^{-4}$  & {\bf 98\%}  & {\bf 99\%}  \\
\bf{SFR\boldsymbol{$_0^L$}}            & $179$        & 0.35   & $4.0\times10^{-5}$  & {\bf 100\%} & {\bf 100\%} \\
\bf{sSFR\boldsymbol{$_0^L$}}           & 179          & 0.30   & $7.3\times10^{-4}$  & \bf{94\%}   & \bf{91\%}   \\
\boldsymbol{$A_V^L$}                   & 179          & 0.34   & $6.0\times10^{-5}$  & \bf{99\%}   & \bf{100\%}  \\
\boldsymbol{$n^L$}                     & 179          & 0.25   & $5.8\times10^{-3}$  & {\bf 74\%}  & 24\%        \\
\boldsymbol{$t_{\mathrm{age}}^L$}      & 179          & 0.27   & $3.2\times10^{-3}$  & \bf{77\%}   & 46\%        \\
$\tau^L$                               & 179          & $0.17$ & $0.14$              & 16\%        & 2\%         \\
$Z_*^L$                                & 179          & 0.13   & $0.44$              & 4\%         & 1\%         \\
\hline                                 
\multicolumn{5}{c}{\textbf{Global properties}}\\
\hdashline              
M$_*^G$                                & 307          & 0.18   & 0.019                & 44\%       &  4\%        \\
SFR$_0^G$                              & 307          & 0.14   & 0.092                & 21\%       &  0\%        \\
sSFR$_0^G$                             & 307          & 0.15   & 0.069                & 26\%       & 0\%         \\
$A_V^G$                                & 307          & 0.14   & 0.087                & 21\%       & 2\%         \\
$n^G$                                  & 307          & 0.11   & 0.32                 & 6\%        & 0\%         \\
$t_{\mathrm{age}}^G$                   & 307          & 0.10   & 0.42                 & 5\%        & 1\%         \\
$\tau^G$                               & 307          & 0.11   & 0.38                 & 5\%        & 1\%         \\
$Z_*^G$                                & 307          & 0.12   & 0.25                 & 9\%        & 1\%         \\
\hline
\end{tabular}
\tablefoot{Similar to Table~\ref{table:KS} with the two samples corresponding to the lower and upper quartiles of the galaxy property. The last column is the regular K-S test (with division in 40-60\%) with the probability of the look-elsewhere-effect p-value being smaller than 0.005. 
}
\end{threeparttable}
\end{table}

\section{Semi-major axes}\label{ap:NED-HP}

\begin{figure}
\centering
\includegraphics[width=\columnwidth]{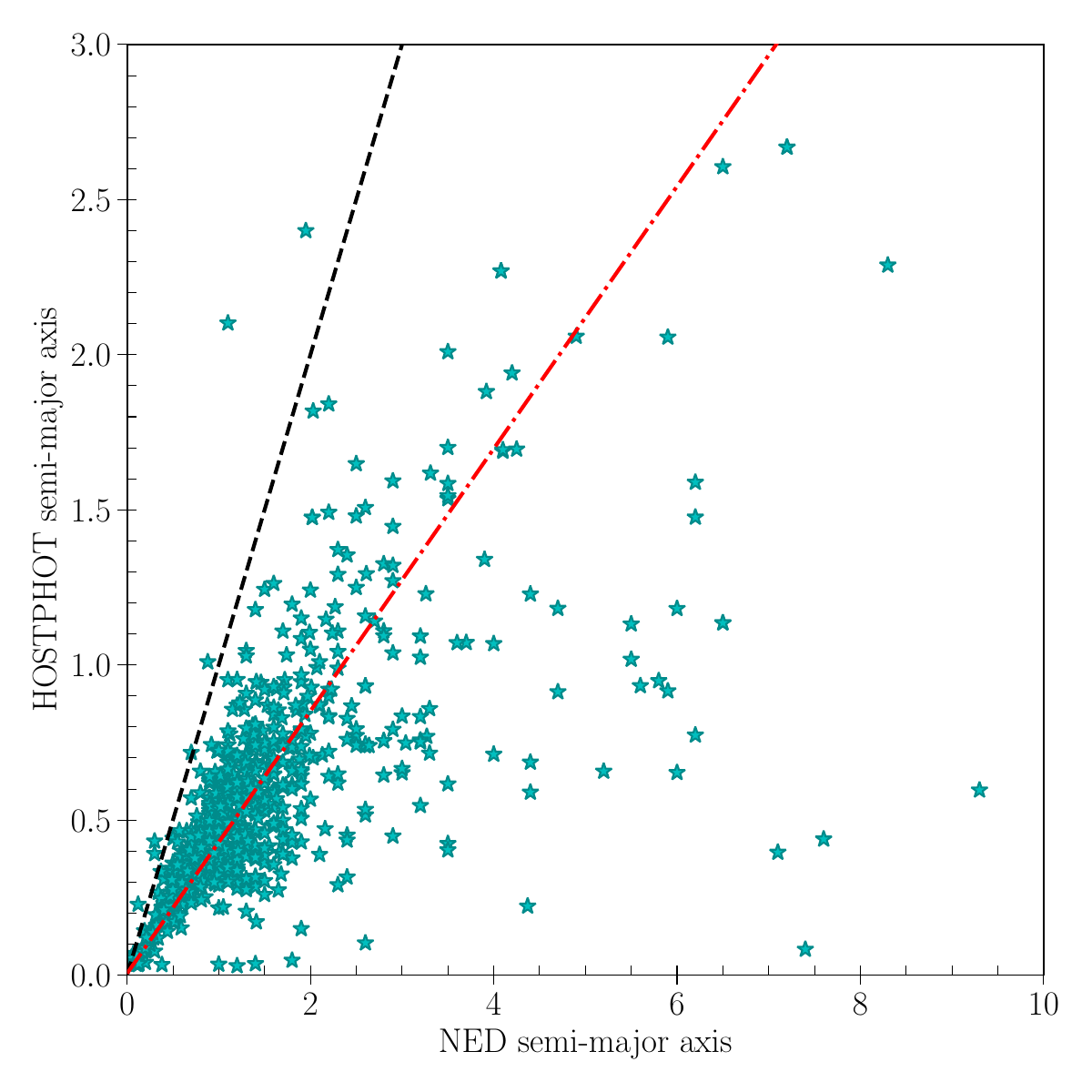}
\caption{Comparison of semi-major axes from NED and {\sc hostphot}. The black dashed line is a 1:1 relation, whereas the red dot-dashed line shows a fit between the two.}
\label{fig:semimajor}
\end{figure}

We compare here the semi-major axes obtained from the homogenised NED database and those extracted from the Kron apertures of {\sc hostphot}. Figure~\ref{fig:semimajor} shows that the NED values are at least twice as high (with a slope of 0.42 or a median ratio difference of 0.44). This explains why in Figure~\ref{fig:offnorm}, the range of normalised offsets is mostly below 1, i.e. at smaller distances than the semi-major axis, whereas other studies show the normalised offset or DLR for SNe below and above 1 \citep[e.g.][]{Toy25}. Since our results are more significant when using NED values (both for offset and inclination), perhaps due to an increased number of objects, we focus our work on the normalisation with NED semi-major axes.

\label{lastpage}

\end{document}